%% file: MAIN.tex
\documentclass[twocolumn]{aastex63}

\input{header.tex}

\received{Month Day, Year}
\revised{Month Day, Year}
\accepted{Month Day, Year}
\submitjournal{ApJS}

\shorttitle{Breaking the degeneracy in MCV X-ray spectral modeling with LC}
\shortauthors{D. Belloni et al.}

\begin{document}

\title{\bf \Large Breaking the degeneracy in magnetic cataclysmic variable X-ray spectral modeling using X-ray light curves}


\correspondingauthor{Diogo Belloni}
\email{diogobellonizorzi@gmail.com}

\author[0000-0003-1535-0866]{Diogo Belloni~\includegraphics[width=9pt]{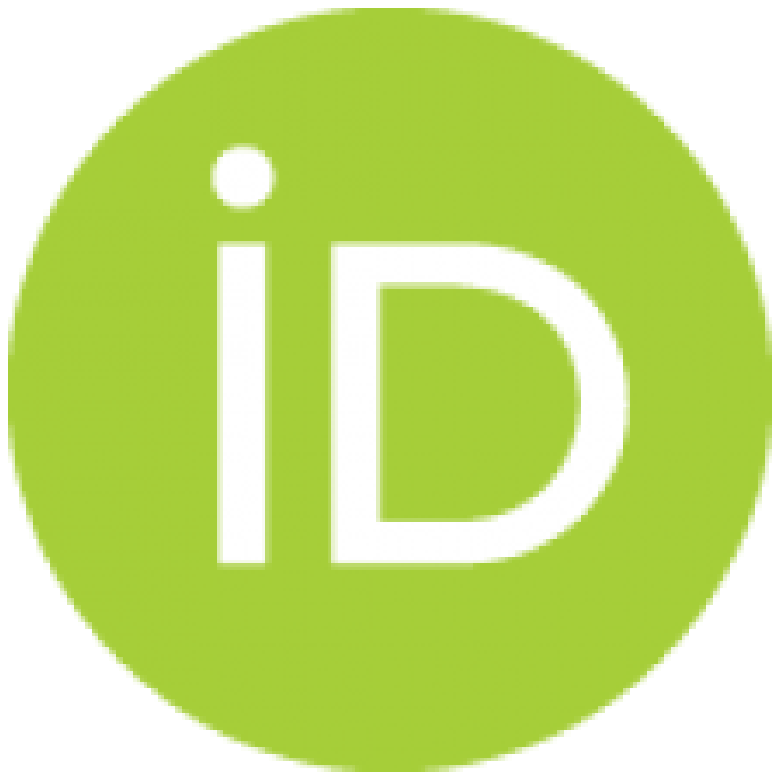}}
\affiliation{Instituto Nacional de Pesquisas Espaciais, Av. dos Astronautas, 1758, 12227-010, S\~ao Jos\'e dos Campos, SP, Brazil}
\affiliation{Departamento de F\'isica, Universidad T\'ecnica Federico Santa Mar\'ia, Av. Espa\~na 1680, Valpara\'iso, Chile}

\author[0000-0002-9459-043X]{Claudia V. Rodrigues~\includegraphics[width=9pt]{FIGURES/Orcid-ID.eps}}
\affiliation{Instituto Nacional de Pesquisas Espaciais, Av. dos Astronautas, 1758, 12227-010, S\~ao Jos\'e dos Campos, SP, Brazil}

\author[0000-0003-3903-8009]{Matthias R. Schreiber~\includegraphics[width=9pt]{FIGURES/Orcid-ID.eps}}
\affiliation{Departamento de F\'isica, Universidad T\'ecnica Federico Santa Mar\'ia, Av. Espa\~na 1680, Valpara\'iso, Chile}
\affiliation{Millenium Nucleus for Planet Formation (NPF), Valpara{\'i}so, Chile}

\author[0000-0003-2162-8393]{Manuel Castro~\includegraphics[width=9pt]{FIGURES/Orcid-ID.eps}}
\affiliation{Instituto Nacional de Pesquisas Espaciais, Av. dos Astronautas, 1758, 12227-010, S\~ao Jos\'e dos Campos, SP, Brazil}

\author[0000-0002-0703-4735]{Joaquim E. R. Costa~\includegraphics[width=9pt]{FIGURES/Orcid-ID.eps}}
\affiliation{Instituto Nacional de Pesquisas Espaciais, Av. dos Astronautas, 1758, 12227-010, S\~ao Jos\'e dos Campos, SP, Brazil}

\author{Takayuki Hayashi}
\affiliation{NASA Goddard Space Flight Center, Code 662, Greenbelt, MD 20771, U.S.A.}
\affiliation{University of Maryland, Baltimore County, 1000 Hilltop Cir, Baltimore, MD 21250 USA}

\author[0000-0001-6013-1772]{Isabel J. Lima~\includegraphics[width=9pt]{FIGURES/Orcid-ID.eps}}
\affiliation{Instituto Nacional de Pesquisas Espaciais, Av. dos Astronautas, 1758, 12227-010, S\~ao Jos\'e dos Campos, SP, Brazil}

\author[0000-0002-2647-4373]{Gerardo J. M. Luna~\includegraphics[width=9pt]{FIGURES/Orcid-ID.eps}}
\affiliation{CONICET-Universidad de Buenos Aires, Instituto de Astronom\'ia y F\'isica del Espacio (IAFE), Av. Inte. G\"uiraldes 2620, C1428ZAA, Buenos Aires, Argentina}
\affiliation{Universidad de Buenos Aires, Facultad de Ciencias Exactas y Naturales, Buenos Aires, Argentina}
\affiliation{Universidad Nacional de Hurlingham, Av. Gdor. Vergara 2222, Villa Tesei, Buenos Aires, Argentina}

\author[0000-0003-3014-7989]{Murilo Martins~\includegraphics[width=9pt]{FIGURES/Orcid-ID.eps}}
\affil{IP\&D, Universidade do Vale do Para\'iba, 12244-000, S\~ao Jos\'e dos Campos, SP, Brazil}

\author[0000-0001-6422-9486]{Alexandre S. Oliveira~\includegraphics[width=9pt]{FIGURES/Orcid-ID.eps}}
\affil{IP\&D, Universidade do Vale do Para\'iba, 12244-000, S\~ao Jos\'e dos Campos, SP, Brazil}

\author[0000-0002-2695-2654]{Steven G. Parsons~\includegraphics[width=9pt]{FIGURES/Orcid-ID.eps}}
\affiliation{Department of Physics and Astronomy, University of Sheffield, Sheffield S3 7RH, UK}

\author[0000-0003-1949-4621]{Karleyne M. G. Silva~\includegraphics[width=9pt]{FIGURES/Orcid-ID.eps}}
\affiliation{Gemini Observatory, Casilla 603, La Serena, Chile}
\affiliation{European Southern Observatory,  Alonso  de  Cordova 3107, Vitacura, Casilla 19001, Santiago, Chile}

\author[0000-0002-2503-2434]{Paulo E. Stecchini~\includegraphics[width=9pt]{FIGURES/Orcid-ID.eps}}
\affiliation{Instituto Nacional de Pesquisas Espaciais, Av. dos Astronautas, 1758, 12227-010, S\~ao Jos\'e dos Campos, SP, Brazil}

\author[0000-0003-0700-6205]{Teresa J. Stuchi~\includegraphics[width=9pt]{FIGURES/Orcid-ID.eps}}
\affiliation{Universidade Federal do Rio de Janeiro, Cidade Universit\'aria, 21941-972, Rio de Janeiro, RJ, Brazil}

\author[0000-0002-4526-0469]{Monica Zorotovic~\includegraphics[width=9pt]{FIGURES/Orcid-ID.eps}}
\affiliation{Instituto de F\'sica y Astronom\'ia, Universidad de Valpara\'iso, Av. Gran Breta\~na 1111, 2360102, Valpara\'iso, Chile}


\input{abstract.tex}


\vspace{0.2cm}
\section{Introduction} 
\label{intro}

\input{CONTENT/Section01.tex}

\section{\cyclops~code}
\label{cyclopscode}

\input{CONTENT/Section02.tex}

\section{PSR structure}
\label{influPSR}

\input{CONTENT/Section03.tex}

\vspace{2.5cm}
\section{X-ray spectra and the degeneracy problem}
\label{xrayobs}

\input{CONTENT/Section04.tex}

\vspace{2.5cm}
\section{Methods of breaking the degeneracy in the parameter space}
\label{degeneracymethods}

\input{CONTENT/Section05.tex}

\section{Solving the degeneracy problem}
\label{degeneracysolution}

\input{CONTENT/Section06.tex}

\section{Future developments of the \cyclops~code}
\label{future}

\input{CONTENT/Section07.tex}

\vspace{2.5cm}
\section{Summary and Conclusions}
\label{conclusions}

\input{CONTENT/Section08.tex}

\section*{Acknowledgements}

\input{Acknowledgements.tex}

\bibliographystyle{aasjournal}
\bibliography{references}

\appendix

\section{Post-shock region (PSR)}
\label{app_cyclopspsr}

\input{CONTENT/AppendixA.tex}

\section{Consistency with previous works}
\label{app_comppre}

\input{CONTENT/AppendixB.tex}

\section{Comparison with \xspec}
\label{app_compxspec}

\input{CONTENT/AppendixC.tex}

\end{document}

%% file: header.tex
\usepackage{tikz,siunitx,mwe}
\usepackage{booktabs}
\usepackage{tabto}
\usepackage{float}
\usepackage{color}
\usepackage{amssymb} 
\usepackage{amsfonts,amsmath,amscd} 
\usepackage[subnum]{cases}
\usepackage{mathtools}
\usepackage{verbatim}
\usepackage{t1enc}
\usepackage[T1]{fontenc}
\usepackage{ae,aecompl}
\usepackage{graphics}
\usepackage{euscript}
\usepackage[normalem]{ulem} 
\usepackage{pstricks}
\usepackage{epsfig}
\usepackage{pst-grad} 
\usepackage{pst-plot} 
\usepackage{multirow} 
\usepackage{times} 
\usepackage{indentfirst} 
\usepackage{natbib}
\usepackage{threeparttable}
\usepackage[space]{grffile} 

\newcommand{\Msun}{M$_{\odot}$}

\newcommand{\Rwd}{$R_{\mathrm{WD}}$}

\newcommand{\der}[2]{\frac{{\rm d} #1}{{\rm d} #2}} 

\newcommand{\cyclops}{{\sc cyclops}}
\newcommand{\ipolar}{{\sc ipolar}}
\newcommand{\spex}{{\sc spex}}
\newcommand{\xspec}{{\sc xspec}}
\newcommand{\bremss}{{\sc bremss}}
\newcommand{\apec}{{\sc apec}}
\newcommand{\mekal}{{\sc mekal}}

\newcommand{\mkcflow}{{\sc mkcflow}}
\newcommand{\phabs}{{\sc phabs}}
\newcommand{\pcfabs}{{\sc pcfabs}}

\newcommand{\brem}{bremsstrahlung}

\newcommand{\nustar}{\emph{NuSTAR}}

\newcommand{\chandra}{\emph{Chandra}}
\newcommand{\gaia}{\emph{Gaia}}

\newcommand{\gscm}{${\rm g\,s^{-1}\,cm^{-2}}$}
\newcommand{\Msunyr}{M$_{\odot}$~yr$^{-1}$}
\definecolor{smalt(darkpowderblue)}{rgb}{0.0, 0.2, 0.6}
\definecolor{forestgreen(traditional)}{rgb}{0.0, 0.5, 0.0}

\usepackage{soulutf8} 

\newcommand{\RNum}[1]{\uppercase\expandafter{\romannumeral #1\relax}}

\defcitealias{MD17}{MD17}

\newcommand{\setwindow}[5]{\def\xmin{#1}%
                            \def\ymin{#2}%
                            \def\xmax{#3}%
                            \def\ymax{#4}%
                            \pstFPsub\viewingwidth{#3}{#1}%
         		 \pstFPdiv\result{\strip@pt#5}{\viewingwidth}%
			 \psset{unit=\result pt}}

\graphicspath{{./}{FIGURES/}}

%% file: abstract.tex
\begin{abstract}
We present an analysis of mock X-ray spectra and light curves of magnetic cataclysmic variables using an upgraded version of the 3D \cyclops~code.
This 3D representation of the accretion flow allows us to properly model total and partial occultation of the post-shock region by the white dwarf as well as the modulation of the X-ray light curves due to the phase-dependent extinction of the pre-shock region.
We carried out detailed post-shock region modeling in a four-dimensional parameter space by varying the white dwarf mass and magnetic field strength as well as the magnetosphere radius and the specific accretion rate.
To calculate the post-shock region temperature and density profiles, we assumed equipartition between ions and electrons, took into account the white dwarf gravitational potential, the finite size of the magnetosphere and a dipole-like magnetic field geometry, and considered cooling by both \brem~and cyclotron radiative processes.
By investigating the impact of the parameters on the resulting X-ray continuum spectra, we show that there is an inevitable degeneracy in the four-dimensional parameter space investigated here, which compromises X-ray continuum spectral fitting strategies and can lead to incorrect parameter estimates.
However, the inclusion of X-ray light curves in different energy ranges can break this degeneracy, and it therefore remains, in principle, possible to use X-ray data to derive fundamental parameters of magnetic cataclysmic variables, which represents an essential step toward understanding their formation and evolution.
\end{abstract}

\keywords{
methods: numerical --
novae, cataclysmic variables --
stars: fundamental parameters -- 
stars: magnetic field --
white dwarf
}

%% file: CONTENT/Section01.tex

Magnetic cataclysmic variables (CVs) are interacting binaries, in which a strongly magnetized white dwarf (WD) accretes matter from a low-mass star \citep[e.g.,][]{Warner_1995,Hellier_2001}.
In magnetic CVs, WD magnetic fields are strong enough to play a role in the dynamics of the accretion flow, and they are generally separated into two main classes, namely intermediate polars (IPs) and polars \citep[see, e.g.,][for comprehensive reviews on magnetic CVs]{Cropper_1990,Patterson_1994,Wickramasinghe_2000,Ferrario_2015,Mukai_2017,Ferrario_2020}.


These two types of CVs differ by the impact the WD magnetic field has on the accretion process. 
In polars, the WD spin is usually synchronized  with the orbital revolution, due to the torque exerted by the donor magnetic field on the WD \citep[e.g.,][]{Hameury_1987}, and its magnetic field is sufficiently strong such that the magnetic pressure exceeds the gas ram pressure outside the circularization radius, which prevents the formation of an accretion disk.
In IPs, on the other hand, given their on average weaker fields, the WD spin is not synchronized with the orbit and a truncated accretion disk is allowed to form, because the magnetic pressure exceeds the gas ram pressure at a radius greater than the WD radius, but smaller than the circularization radius.
\citet{NatAst} proposed a rotation- and crystallization-driven dynamo to be responsible for the generation of WD magnetic fields in CVs. According to this scenario, which successfully explains the observed relative numbers of magnetic WDs in close binaries, the occurrence of strong magnetic fields is intrinsically related to close binary evolution \citep[see also][]{bellonietal21-1}.


In both polars and IPs, matter is accreted onto the WD in a field-channeled accretion flow, starting at the threading region, which is the region where the magnetic field captures the mass flow from the secondary star, and extending to the WD surface.
Such an accretion flow is supersonic when it reaches the region close to the WD surface where a shock is formed.
The matter in the post-shock region (PSR), which is the region between the shock and the WD surface, is compressed and heated to temperatures up to a few tens of keV, and is usually the dominant emission component in magnetic CVs.


Polars are characterized by the strong circular and linear polarization of the optical and near-infrared thermal cyclotron emission.
IPs, on the other hand, usually do not exhibit measurable polarization and the PSR emission at optical wavelengths is diluted by the radiation emitted by the accretion disk.
Additionally, due to the high temperature in the PSR, most magnetic CVs are strong X-rays emitters and have been discovered by high-energy surveys.

The X-ray emission in polars and IPs is mostly produced by \brem~in the PSR. 
However, there might be contributions from the WD surface and from the pre-shock region in soft X-rays. 
For energies greater than ${\sim10}$~keV, Compton scattering can also contribute to the observed flux \citep[e.g.,][]{Mukai_2015}.
In addition, line emission provides significant emission at soft X-rays in these systems.
At energies smaller than ${\sim1}$~keV, X-ray emission from the irradiated/heated WD photosphere has been observed in many systems \citep[e.g.,][]{Ramsay_Cropper_2004,Bernardini_2017} and from the photoionizioned pre-shock region in EX~Hya \citep[][]{Luna_2010}.


Among all magnetic CV parameters, four deserve special attention: the WD mass, the WD magnetic field strength, the specific accretion rate and the threading region radius (the magnetosphere boundary), as they are the main parameters characterizing the emission from the PSR, especially bremsstrahlung \citep[e.g.,][]{Wu_1994,Cropper_1998,Wu_2000,Hayashi_2014,Suleimanov_2019}.
This is because the PSR temperature and density profiles are mainly determined by these four parameters.

Even though the determination of these parameters is generally not straightforward, X-ray emission provides a tool to estimate them in a relatively simple way \citep[e.g.,][]{Cropper_1998,Cropper_1999,Ramsay_2000,Suleimanov_2005,Yuasa_2010,Hayashi_2014b,Suleimanov_2016,Suleimanov_2019}.
Inspired by the early works of \citet{Aizu_1973} and \citet{Hoshi_1973}, modeling of the X-ray emission from the PSR has been improved by several groups with the aim 
to derive strong constraints on crucial parameters of magnetic CVs.
These models take into account the influence of cyclotron emission \citep[e.g.,][]{Wu_1994}, the WD gravity \citep[e.g.,][]{Cropper_1999},
dipole geometry of the WD magnetic field \citep[e.g.,][]{Canalle_2005}, and the difference between the electron and ion temperatures \citep[e.g.,][]{Imamura_1987,Saxton_2007}.


In particular, \citet{Hayashi_2014} investigated the influence of the specific accretion rate and the WD mass on predicted PSR and X-ray spectrum properties by considering a dipole-like geometry.
However, these authors did not take into account the effects of cyclotron emission in their analysis.
That said, it is still not clear how their results would change in the presence of strong cyclotron emission, which is the case for polars and some IPs (e.g., V405~Aur, PG~Gem, V2400~Oph).
In order to address this issue, we thoroughly analyze here PSR properties by varying four parameters, namely the WD mass, the WD magnetic field, the magnetosphere/threading region radius and the specific accretion rate.
To do so, we upgraded the 3D \cyclops~code \citep{Costa_Rodrigues_2009,Silva_2013} such that the PSR is consistently built based on the model parameters.
An example of fitting with this new version of the code has been recently performed by \citet{Oliveira_2019}, who investigated the polar V348~Pav in optical wavelengths.
In addition, we analyze the properties of the X-ray spectra resulting from the PSR modeling and discuss how model parameters affect them.


Despite the fact that X-ray emission modeling became a quite convenient technique to estimate magnetic CV parameters, there is one major difficulty with this approach, namely the \textit{degeneracy problem}.
Within a fitting scheme, many combinations of the parameters naturally lead to virtually identical X-ray spectra, despite the PSRs being substantially different.
This implies that fitting X-ray continuum spectra alone does not necessarily provide unambiguous estimates for magnetic CV parameters, even in simplified schemes.

In previous works, several assumptions had to be made to break the degeneracy in the complex parameter space of magnetic CVs.
\citet{Yuasa_2010} used an improved version of the approach of \citet{Suleimanov_2005}, who assumed cyclotron emission to be negligible and the magnetosphere to be infinite.
A similar approach was used more recently by \citet{Hayashi_2014}.
In both cases, the additional assumptions reduce the parameter space, leaving only the WD mass and the specific accretion rate to be fitted.

Even though \citet{Yuasa_2010} broke the degeneracy by assuming a constant specific accretion rate of $1$~\gscm, in a more general situation of unknown specific accretion rate, both \citet{Yuasa_2010} and \citet{Hayashi_2014} would still have problems with the degeneracy of the WD mass and the specific accretion rate.
For instance, \citet{Hayashi_2014b} applied their model to investigate the IPs EX~Hya and V1223~Sgr and found a strong degeneracy in the plane of these two parameters.
These authors managed to estimate both parameters for EX~Hya with their method, but needed additional constraints, related to the PSR height, in order to properly estimate the properties of V1223~Sgr.
This is not a desirable solution to the degeneracy problem, though, since detailed information, which provides such additional `external' constraints, is not usually available for most magnetic CVs.

The last example we discuss is the approach by \citet{Suleimanov_2016,Suleimanov_2019}.
In addition to assuming negligible cyclotron emission, in their short-column method they fixed the accretion rate and the fraction of the WD surface occupied by the PSR, and calculated the specific accretion rate for given WD mass and radius.
On ther other hand, their tall-column method consists of fixing the PSR height, so that the specific accretion rate in the fitting scheme is simply adjusted to match the desired height, for a given combination of magnetosphere radius and WD mass.
In their model, the WD mass and the magnetosphere radius are the free parameters, which turned out to be degenerate.
Therefore, should only X-ray spectra be used in their fitting scheme, the degeneracy would still remain, even in a 2D parameter space.

In order to break this degeneracy, these authors added information about the break frequency in the power spectra of X-ray light curves, which corresponds to the Keplerian frequency at the magnetospheric boundary.
By doing so, these authors managed to infer the magnetosphere radius together with the WD mass.
Despite the fact that this method seems to work, it is usually not possible to extract the break frequency from observations \citep[e.g.,][]{Shaw_2020}.
More importantly, it does not solve the degeneracy problem in a four-dimensional fitting scheme, as investigated here.
Thus, a new method is required to break the degeneracy without oversimplifying the problem.


A popular tool to fit X-ray spectra is \xspec\ \citep{Arnaud_1996,Dorman_Arnaud_2001}.
The basic \xspec\ recipe consists of a combination of additive and multiplicative models.
In the context of magnetic CVs, the PSR emission should be represented by an additive model while the absorption is a multiplicative model.
Examples of additive models are: 
\bremss, which represents the thermal bremsstrahlung emission of a hot gas at a given temperature; \mekal, which adds line emission to the continuum bremsstrahlung of a hot gas at a given temperature; \apec, which provides an emission spectrum from collisionally-ionized diffuse gas calculated from the AtomDB atomic database; and \mkcflow, which is a cooling flow model, in which a multitemperature hot gas emits following the \mekal~recipe for each temperature.
In addition, the \ipolar~model, which can be loaded as a table model, represents the X-ray emission ($3-200$~keV) from PSRs obtained from a grid following \citet{Suleimanov_2016}.

The PSR emission can be modified by the photoelectric absorption coming from two astrophysical origins: (i) the interstellar gas between the magnetic CV and the Earth and (ii) the material in the binary.
The interstellar absorption is well represented by the \phabs~multiplicative model of \xspec.
The modulation of the flux with the WD rotation seen in many systems indicates the presence of absorbing material in the magnetic accretion column above the PSR, the pre-shock region, that affects the PSR emission differently with the rotation phase.
To account for the effect of this variable absorption in the spectrum, it is usually adopted the partial covering fraction absorption model (\pcfabs) of \xspec.
It basically assumes that only a fraction of the additive model is absorbed.
That approach allows a quick and easy fitting of X-ray spectra of magnetic CVs.
However, it does not necessarily represent a consistent view of the system.
As we show later in this paper, a 3D representation of the entire magnetic accretion structure, PSR and pre-shock region, is required for a correct modeling of the observed X-ray emission of magnetic CVs.


In this paper, we address the degeneracy of the X-ray continuum spectral modeling of magnetic CVs using four parameters: WD mass, WD magnetic field, specific accretion rate, and magnetosphere radius.
In addition to investigating in detail this degeneracy problem, we also discuss additional observational constraints that could be incorporated in magnetic CV fitting strategies thereby allowing to break the degeneracy.
More specifically, we show how X-ray light curves allow to disentangle models with similar X-ray spectra.
We will present applications of this new approach proposed in this work to observations in forthcoming papers.

%% file: CONTENT/Section02.tex
The {\sc cyclops} code developed by \citet{Costa_Rodrigues_2009} and improved by \citet{Silva_2013} is a tool that enables modeling of cyclotron and bremsstrahlung emissions, in optical and X-rays, from PSRs in magnetic CVs.


In the \cyclops~code a 3D grid is used to represent the entire magnetic CV accretion structure, which is defined by the lines of a dipolar magnetic field, from the threading region to the WD surface.
For a given rotational phase, the accretion structure is represented by cells in a 3D Cartesian grid having one axis parallel to the line-of-sight. 
Despite the fact that the linear dimension of a cell in the plane of sky can assume any value, we enforce the cell dimension in the line of sight to have at least 0.1 of the minimum geometrical depth considering all rotation phases.
This allows us to properly sample the PSR density and temperature profiles.
Even though \cyclops~can account for PSRs either in only one hemisphere or in both hemispheres simultaneously, we consider in this paper only one PSR per system.

\subsection{Radiative transfer in the post-shock region}

The radiative coefficients are calculated for each cell of the PSR according to its physical properties. The magnetic field magnitude and direction follow a dipolar magnetic field parametrized by its axis direction and magnitude at the magnetic pole.
In the optical regime, the cyclotron emissivities of the four Stokes parameters are calculated according to the WD magnetic field and plasma density and temperature. \cyclops~adopts the radiative transport solution of \citet{Pacholczyk_1975} and \citet{Meggitt_1982}. The free-free absorption is also considered in the transport.
In X-rays, the bremsstrahlung emissivity is computed by assuming a fully ionized magnetized hydrogen plasma, following \citet{Gronenschild_1978} and \citet{Mewe_1986}.

The radiative transfer in the PSR is computed, from the bottom to the top of the region, along the line-of-sight. 
So the PSR emission is represented by a 2D array of fluxes, each flux being the result of the radiative transfer is one line-of-sight.
Therefore, the outcome is an image in each rotation phase for each Stokes parameter, which are integrated to obtain fluxes and polarization as a function of the rotation phase.
An X-ray spectrum is calculated combining the fluxes in all rotation phases. 
Routines from the PINTofALE package \citep{Kashyap_2000} are used to convolve the model with the X-ray instrumental files allowing us to compare the models with high-energy observations.

\subsection{Interstellar and pre-shock region extinction}
\label{sec_extinction}

\cyclops\ takes into account two sources of extinction, namely the interstellar medium and the pre-shock region.
In X-rays, Compton scattering by electrons and photoelectric absorption are considered.
Compton scattering cross section is calculated using the Klein-Nishina formula \citep[e.g.,][]{1986rpa..book.....R}.
The photoelectric cross-section is calculated using the {\sc bamabs} routine from the PINTofALE package, an implementation for cross-sections presented by \citet{1992ApJ...400..699B}.
This calculation has an upper energy limit of 10~keV, above which no photoelectric absorption is considered.
Only photoelectric absorption is taken into account for the interstellar extinction, since this material is considered neutral. 
However, since we assume that the pre-shock material is partially ionized, both Compton scattering and photoelectric absorption are included in its extinction.
In optical wavelengths, Thomson scattering in the pre-shock column scatters the PSR emission out of the line of sight.
The reddening of the interstellar medium is calculated using the gas-to-dust ratio of \citet{2017MNRAS.471.3494Z}, ${R=3.1}$, and the extinction law from \citet{1989ApJ...345..245C}.

Each line-of-sight that composes the emission of the PSR crosses a given number of cells of the pre-shock region. This allows us to calculate the pre-shock length in each line-of-sight and consequently its optical depth.
For simplicity, and in the absence of a proper model for the pre-shock properties, the density is assumed constant and equals to the density in the shock front multiplied by a factor that can assume values from 0 to 0.25.
This maximum limit comes from the Rankine-Hugoniot conditions in the shock front (Eqs.~\ref{vff} and \ref{rhoff}).
We arbitrarily assume that pre-shock region is partially ionized, with a fraction of 0.5 of the mass ionized.

\begin{figure*}[htb!]
\centering
\includegraphics[width=0.895\linewidth]{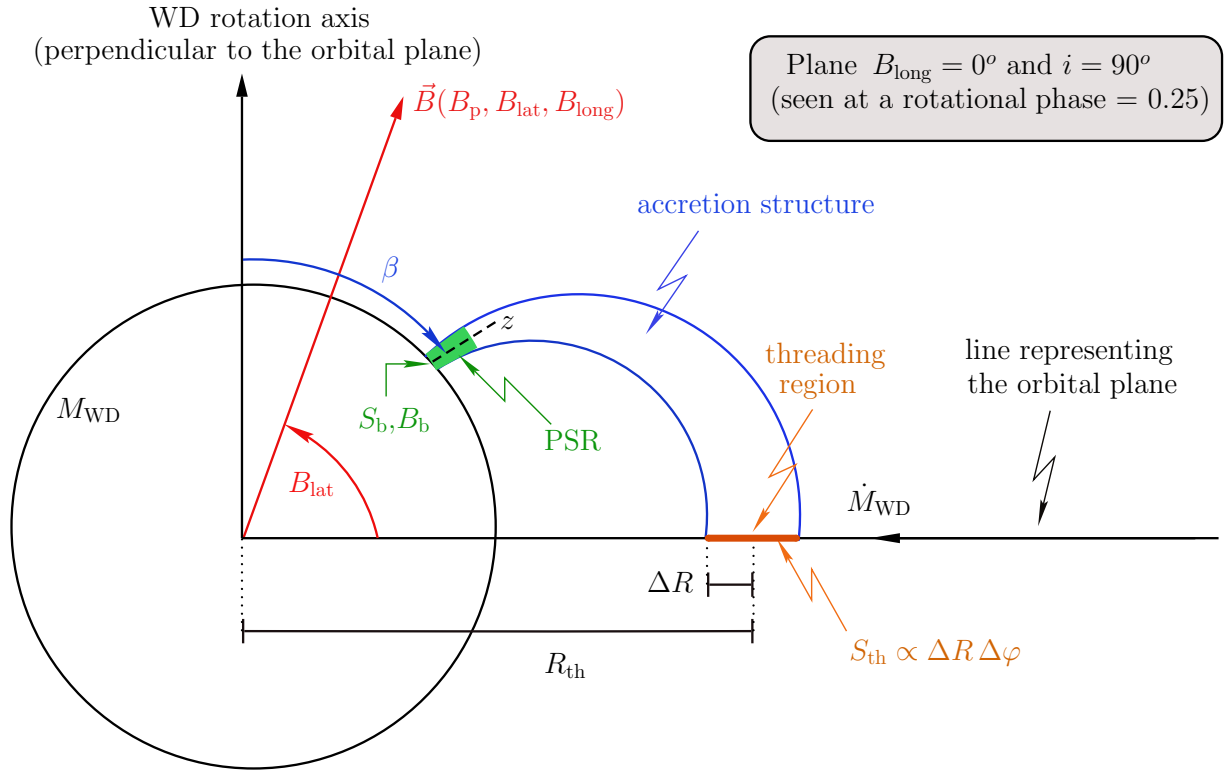}
\caption{Geometrical aspects in the \cyclops~code. 
The WD is assumed to be an opaque sphere with mass $M_{\rm WD}$, which accretes matter at a rate $\dot{M}_{\rm WD}$.
The WD rotation axis (vertical axis) is assumed to be parallel to the donor rotation axis and both are orthogonal to the orbital plane (horizontal axis).
The magnetic field is assumed to be a centered dipolar field with a given intensity at the pole $(B_{\rm p})$ and the magnetic axis is defined by a latitude ($B_{\rm lat}$, from the orbital plane) and a longitude ($B_{\rm long}$, from the line connecting the WD and the donor, and having counterclockwise direction with respect to the WD rotation axis).
Notice that the system can have any inclination $i$, which is defined by the angle between the WD rotation axis and the observer. However, in the figure, we show the case in which $i=90^o$ and $B_{\rm long}=0^o$, for simplicity, for a system seen at a rotation phase of $0.25$.
After the magnetic axis is set in the code, the colatitude of the PSR $(\beta)$ has to be set, which is the angle between the WD rotation axis and the PSR center.
By providing $B_{\rm p}$, $B_{\rm lat}$, $M_{\rm WD}$ and $\dot{M}_{\rm WD}$, the threading region position $(R_{\rm th})$ in the orbital plane is directly set, which is the position at which the magnetic field captures the material from the donor.
In order to define the size of the PSR, the threading region area $(S_{\rm th})$ needs to be defined.
This is done by means of a radial extension $(\Delta R)$ and an angular extension $(\Delta \varphi)$.
Having constructed the threading region, the entire accretion column is built, from the threading region to the WD surface.
In this way, the accretion area on the WD surface, i.e., at the PSR bottom, $(S_{\rm b})$ is directly provided, and the magnetic field strength at the PSR bottom $(B_{\rm b})$ also comes from the adopted geometry.
See \citet{Costa_Rodrigues_2009} for more details.
}
\label{FigPSR0}
\end{figure*}

\begin{figure*}[htb!]
\centering
\includegraphics[width=0.865\linewidth]{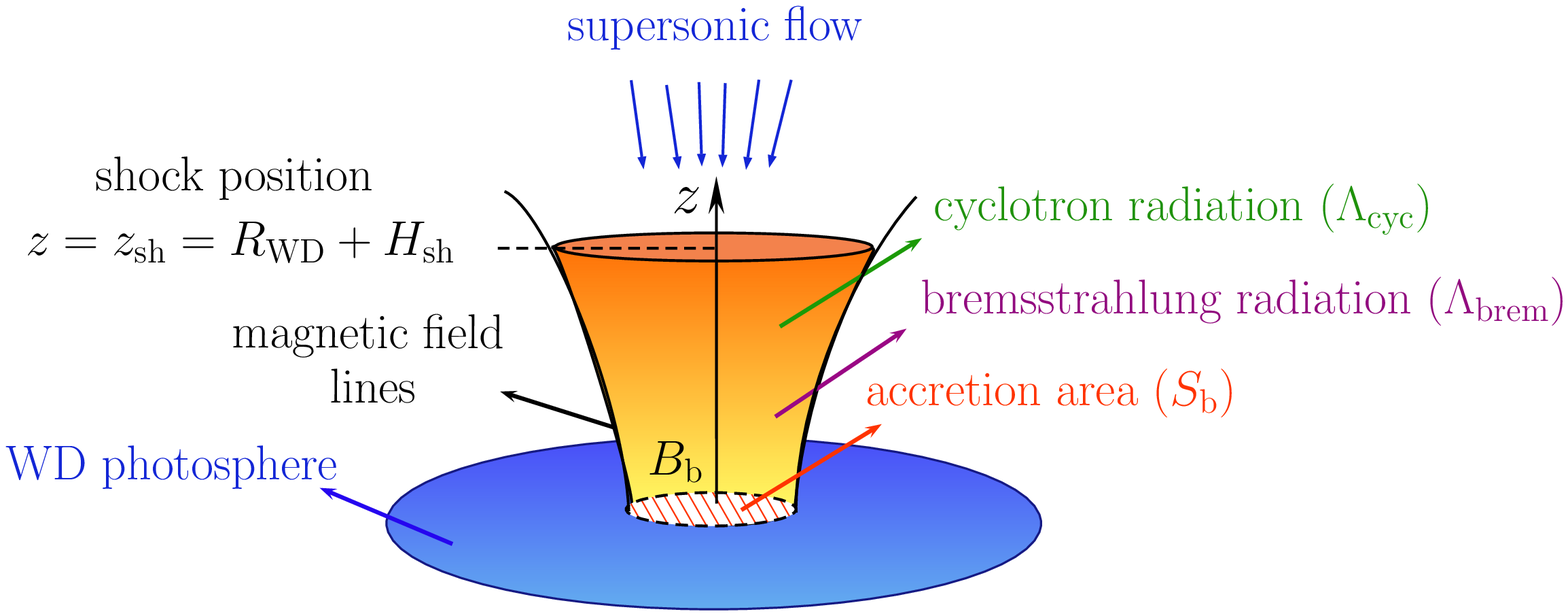}
\caption{Geometry of the PSR assumed here. Both cyclotron cooling $(\Lambda_{\rm cyc})$ and bremsstrahlung cooling $(\Lambda_{\rm brem})$ are considered in our modeling.
The region is defined by the accretion area $(S_{\rm b})$ and the magnetic field strength $(B_{\rm b})$, both at the PSR bottom.
The spatial coordinate $z$ is perpendicular to the WD surface and its starting point is the WD center.
In this way, the shock position is $z_{\rm sh}=R_{\rm WD}+H_{\rm sh}$, where $H_{\rm sh}$ is the shock height with respect to the WD surface.
Both cross-section $S(z)$ and magnetic field $B(z)$ vary through the PSR in the $z$ direction.
}
\label{FigPSR1}
\end{figure*}

\subsection{Model parameters}

In order to have a model in the \cyclops~code, it is needed to specify some geometrical properties.
They are the inclination of the magnetic CV orbital plane with respect to the observer $(i)$, the angular position of the PSR center with respect to the WD rotation axis $(\beta)$, the radial $(\Delta R)$ and the angular $(\Delta \varphi)$ sizes of the threading region, and the dipolar magnetic field parameters, i.e. its intensity at the pole ($B_{\rm p}$), its latitude ($B_{\rm lat}$) with respect to the orbital plane, and its longitude ($B_{\rm long}$) with respect to the line connecting the WD and the donor star.

In addition to the above-mentioned parameters, two other physical parameters have to be set, namely the WD mass $(M_{\rm WD})$ and the accretion rate ($\dot{M}_{\rm WD}$).
Moreover, the distance to the investigated source can be included as a fixed parameter, in case it is known (e.g., from \gaia~accurate parallaxes). 
Finally, an important quantity, which is not a parameter in the code, is the specific accretion rate at the PSR bottom ($\dot{m}_{\rm b}$) defined as $\dot{M}_{\rm WD}/S_{\rm b}$, where $S_{\rm b}$ is the accretion area on the WD surface.
The geometry adopted in the \cyclops~code as well as its parameters are illustrated in Fig.~\ref{FigPSR0} \citep[see also fig.~1 in][]{Costa_Rodrigues_2009}.

\subsection{Post-shock region modeling}
\label{sec_psr_modelling}

In previous versions of the \cyclops~code, the adopted PSR electronic density and temperature profiles were represented by simple fixed analytic expressions \citep[][their equations~1 and 2]{Silva_2013}, which did not depend on the physical parameters of the system and, in addition, did not consistently relate to one other.
We upgraded here the modeling of the PSR structure, which is now obtained from the solution of the stationary one-dimensional hydro-thermodynamic differential equations describing the accreting plasma.
In what follows we briefly discuss the major changes, and a detailed description of our PSR modeling as well as comparisons with previous works are provided in Appendices~\ref{app_cyclopspsr} and \ref{app_comppre}.

In our modeling, we consider the WD gravitational potential \citep[e.g.,][]{Cropper_1999} and assume equipartition between ions and electrons \citep[e.g.,][]{Wu_1994,Som_2018}.
Additionally, we adopt a dipole-like magnetic field geometry \citep[i.e. cubic cross-section variation, e.g.,][]{Hayashi_2014,Suleimanov_2016}, allow the WD magnetic field to decay as the distance from the WD surface increases \citep[e.g.,][]{Canalle_2005,Saxton_2007} and take into account the fact that the threading region is not at infinite \citep[e.g.,][]{Suleimanov_2016}.
Moreover, we assume that \brem~and cyclotron radiative processes are the dominant mechanisms responsible for the cooling of the gas, from the shock to the WD surface \citep{Canalle_2005,Som_2018}.

We illustrate the geometrical and physical aspects of our PSR model in Fig.~\ref{FigPSR1}. 
The spatial coordinate $z$ defining the PSR is perpendicular to the WD surface, and has its origin in the WD center, which implies that the shock position is $z_{\rm sh}=R_{\rm WD}+H_{\rm sh}$,  where $H_{\rm sh}$ is the shock height with respect to the WD surface.
The PSR cross-section increase and the magnetic field strength decreases as $z$ increases, which causes variations in the cooling efficiency.

\subsection{General remarks on the \cyclops~code}

Before proceeding further, a few comments are worth making.
First, the current version of the \cyclops~code substantially differs from previous versions.
In the first version of the code \citep{Costa_Rodrigues_2009}, only cyclotron emission could be modeled, which means that only optical light curves could be used as constraints in any fitting strategy.
In the second version of the code \citep{Silva_2013}, \brem~emission was included, which allowed us to also model X-ray spectra. 
However, in both versions, the post-shock region temperature and density profiles were not consistently obtained from the physical parameters of the model since simple analytical expressions had been used to calculate them \citep[][their equations~1 and 2]{Silva_2013}.
In the version we present here, in addition to include X-ray light curves, the post-shock region properties are consistently obtained from the physical parameters of the system, through solution of the hydro-thermodynamic equations describing the accreting  plasma, as described in detail in Appendix~\ref{app_cyclopspsr}.

Second, the \cyclops~code defines the whole accretion structure from the threading region to the bottom of the PSR region at the WD surface.
That said, the accretion area at the bottom of the PSR ($S_{\rm b}$) comes directly from the adopted magnetic field geometry and the threading region area $S_{\rm th} \propto \Delta R \Delta \varphi$.

Third, in Fig.~\ref{FigPSR0}, we illustrate the particular case in which ${B_{\rm long}=0^{\rm o}}$.
In such a situation, the central line of the accretion structure is in the plane formed by the WD axis and the line connecting the WD center and the threading region center.
Consequently, the PSR footprint on the WD surface forms into an arc `parallel' to the WD latitude circles.
However, if ${B_{\rm long}>0^{\rm o}}$, then the central line of the accretion structure is bent to this plane, and the PSR arc is no longer `parallel' to WD latitude circles.

Fourth, we shall emphasize that unlike other codes \citep[e.g.,][]{Fischer_2001,Canalle_2005,Saxton_2007,Yuasa_2010,Hayashi_2014,Suleimanov_2019},  \cyclops\ is a 3D code, which allows us to properly model optical polarization and light curves as well as X-ray spectra and light curves using the same tool.
In addition, since the WD is treated as 3D body in the code, \cyclops~is able to also model the so called self-eclipse of the PSR by the WD.
In other words, the code takes into account the partial or total occultation of the PSR by the WD, which might occur depending on the geometry of the system, producing a variation in the observed flux as a function of the WD rotation phase.

Fifth, the pre-shock accretion structure is also represented as a 3D structure in \cyclops, which works as a partial covering absorber.
Its absorption of the PSR emission is included in the radiative transport and allows us to consistently calculate the variation of the X-ray emission along the WD rotation and its effect on X-ray light curves and spectra.
All of this makes \cyclops\ a powerful tool to understand magnetic accretion in CVs. In particular, the consistent geometrical approach of the PSR and pre-shock region allows us to model light curves in an unprecedented way.

Sixth, the properties of the PSR are calculated considering an 1D approach, similarly to what is done in most previous studies.
In particular, the magnetic field values are those of the central line of the PSR.
On the other hand, the radiative transfer is performed in a 3D approach.
As a compromise, we assume that the density and temperature radial profiles are the same along the entire PSR and given by the 1D PSR modeling.

Seventh, \cyclops~X-ray spectra are in very good agreement with those generated by the \xspec~code, which is a widely used X-ray package to fit X-ray spectra.
A detailed comparison between both codes, for several cases of uniform PSR distributions, is provided in Appendix~\ref{app_compxspec}.

Finally, despite the fact that the \cyclops~code can handle data in virtually any frequency range, including optical polarized emission, we focus in this paper on X-ray data.
However, a discussion on how one could break the degeneracy using optical light and polarization curves is planned to be hold elsewhere.

%% file: CONTENT/Section03.tex

From now on, we will address properties of PSRs as well as magnetic CV X-ray spectra and light curves.
A detailed description of our modeling, including the physical/geometrical assumptions, the hydro-thermodynamic differential equations and the numerical method to solve them, as well as comparisons with other works, can be found in Appendices~\ref{app_cyclopspsr} and \ref{app_comppre}.
For simplicity, we define a \textit{standard} model, which will be widely used hereafter, as follows.
The standard model has the geometry illustrated in Fig.~\ref{FigPSR0}, i.e. ${B_{\rm long}=0^{\rm o}}$ and ${i=90^{\rm o}}$.
In addition, we set ${\beta=5^{\rm o}}$, ${M_{\rm WD}=0.8}$~\Msun~and ${B_{\rm p}=1}$~MG.
We further set $B_{\rm lat}$ such that ${R_{\rm th}=130~R_{\rm WD}}$.
Moreover, we assume that the standard model corresponds to a source at a distance of $100$~pc.
Finally, we set $\dot{M}_{\rm WD}$, ${\Delta \varphi}$ and ${\Delta R}$, such that ${\dot{m}_{\rm b}=1~{\rm g~s^{-1}~cm^2}}$.
The properties of the standard model are summarized in Table~\ref{TabSTAN}.


\input{TABLES/Table_STANDARD.tex}


\begin{figure*}[tb!]
\begin{center}
\includegraphics[width=0.99\linewidth]{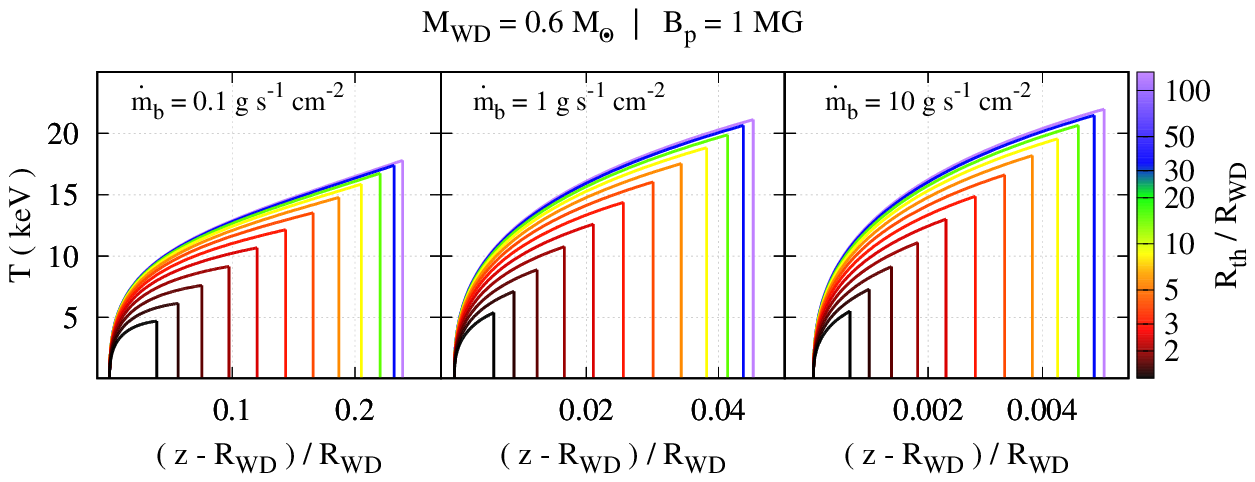}
\includegraphics[width=0.99\linewidth]{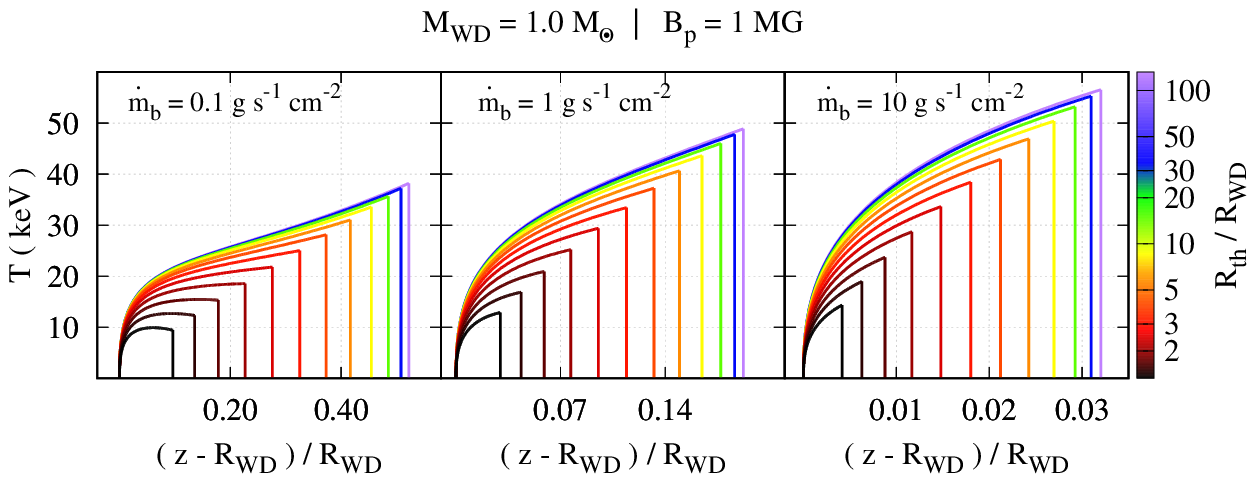}
\end{center}
\caption{Temperature profiles in PSRs defined by particular choices of the model parameters.
We fixed the WD magnetic field strength {$B_{\rm p}=1$~MG} in all panels and compare the solutions for two values of the WD mass ($M_{\rm WD}$), namely $0.6$~\Msun (top row) and $1.0$~\Msun (bottom row).
We further assume that the specific accretion rate $\dot{m}_{\rm b}$ is $0.1$, $1$ and $10$~\gscm~in the left, middle and right panels, respectively.
Finally, in each panel, the profiles are color-coded according to different values of the threading region radius ($R_{\rm th}$).
In all profiles, for simplicity, we set the other \cyclops~parameters as in the standard model (Table~\ref{TabSTAN}).
The $M_{\rm WD}$, $\dot{m}_{\rm b}$ and $R_{\rm th}$ play a sufficiently important role in shaping the $T$ distributions and determining the shock heights ($H_{\rm sh}$) and temperatures ($T_{\rm sh}$).
In particular, the greater $M_{\rm WD}$ and/or the greater $\dot{m}_{\rm b}$ and/or the greater $R_{\rm th}$, the greater $T_{\rm sh}$.
Similarly, the greater $M_{\rm WD}$ and/or the greater $R_{\rm th}$, the greater $H_{\rm sh}$.
On the other hand, $\dot{m}_{\rm b}$ can correlate or anti-correlate with $H_{\rm sh}$, depending on $B_{\rm p}$, as discussed in details in Section~\ref{influCOOL}.
The correlations above are a direct consequence of the interplay between the cooling efficiency and physical conditions at the shock position.
}
  \label{FigPSR4.1}
\end{figure*}


\begin{figure*}[tb!]
\begin{center}\includegraphics[width=0.99\linewidth]{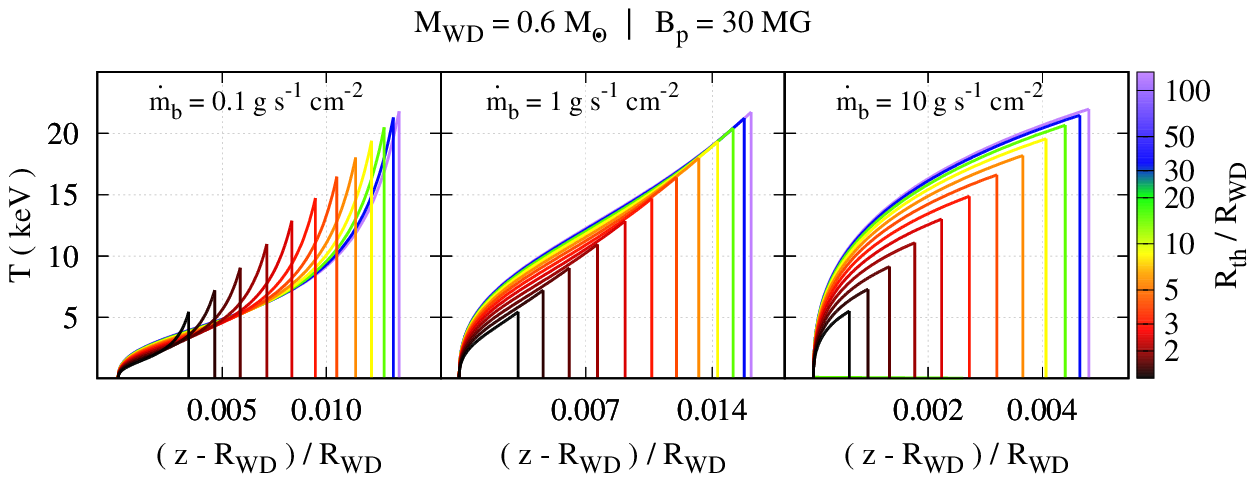}
\includegraphics[width=0.99\linewidth]{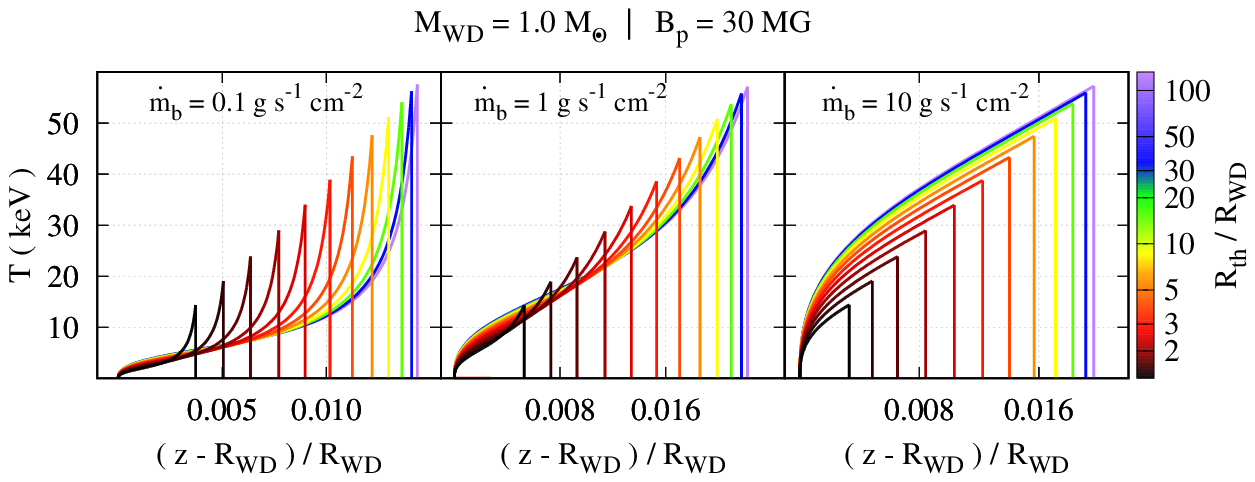}
\end{center}
\caption{Temperature profiles in PSRs defined by the same choices the model parameters as in Fig.~\ref{FigPSR4.1}, with exception of the WD magnetic field strength ($B_{\rm p}$), which is $30$~MG here, while, in Fig.~\ref{FigPSR4.1}, it is $1$~MG.
The impact of increasing $B_{\rm p}$ is seen in the shock height ($H_{\rm sh}$), which is smaller, as well as in the shock temperature ($T_{\rm sh}$), which is greater. 
The decrease in $H_{\rm sh}$ takes place because, for stronger $B_{\rm p}$, the total cooling becomes more efficient, due to an enhanced cyclotron radiation.
This results in a more energetic gas at the shock position, which leads to an increase in $T_{\rm sh}$.
}
  \label{FigPSR4.2}
\end{figure*}


\begin{figure*}[htb!]
\begin{center}\includegraphics[width=0.98\linewidth]{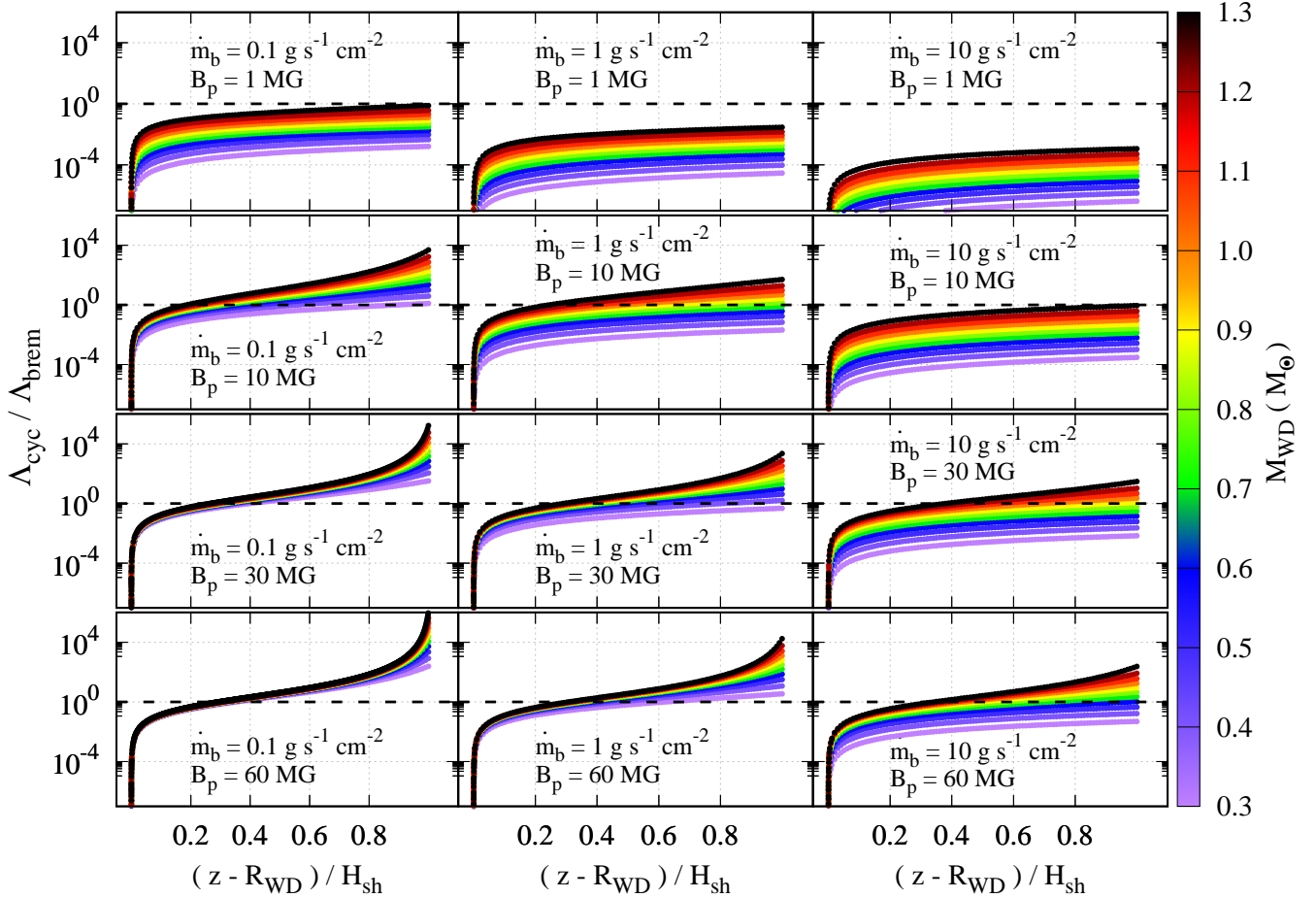} 
\end{center}
\caption{Cyclotron cooling over \brem~cooling $(\Lambda_{\rm cyc}/\Lambda_{\rm brem})$ in the $z$ direction from the WD surface until the shock height $(H_{\rm sh})$, for different combinations of  $B_{\rm p}$, $\dot{m}_{\rm b}$ and $M_{\rm WD}$.
In all cases, we set the non-specified parameters as in the standard model (Table~\ref{TabSTAN}).
We set three values for $\dot{m}_{\rm b}$, namely $0.1$ (left column), $1$ (middle column), and $10$ (right column), in units of \gscm, four values for $B_{\rm p}$, namely $1$ (first row), $10$ (second row), $30$ (third row) and $60$ (fourth row), in units of MG.
In each panel, the color bar corresponds to $M_{\rm WD}$, which has values, in \Msun, from $0.3$ to $1.3$, in steps of $0.1$.
Additionally, the horizontal dashed line corresponds to ${\Lambda_{\rm cyc}/\Lambda_{\rm brem}=1}$, which indicates the region where both radiative processes contribute equally for the cooling.
Notice that the three parameters play a key role in the balance.
In particular, the greater the $B_{\rm p}$ and/or the greater the $M_{\rm WD}$ and/or the smaller the $\dot{m}_{\rm b}$, the more efficient the cooling due to cyclotron radiation.
}
  \label{FigPSR5.1}
\end{figure*}


\begin{figure*}[htb!]
\begin{center}
\includegraphics[width=0.99\linewidth]{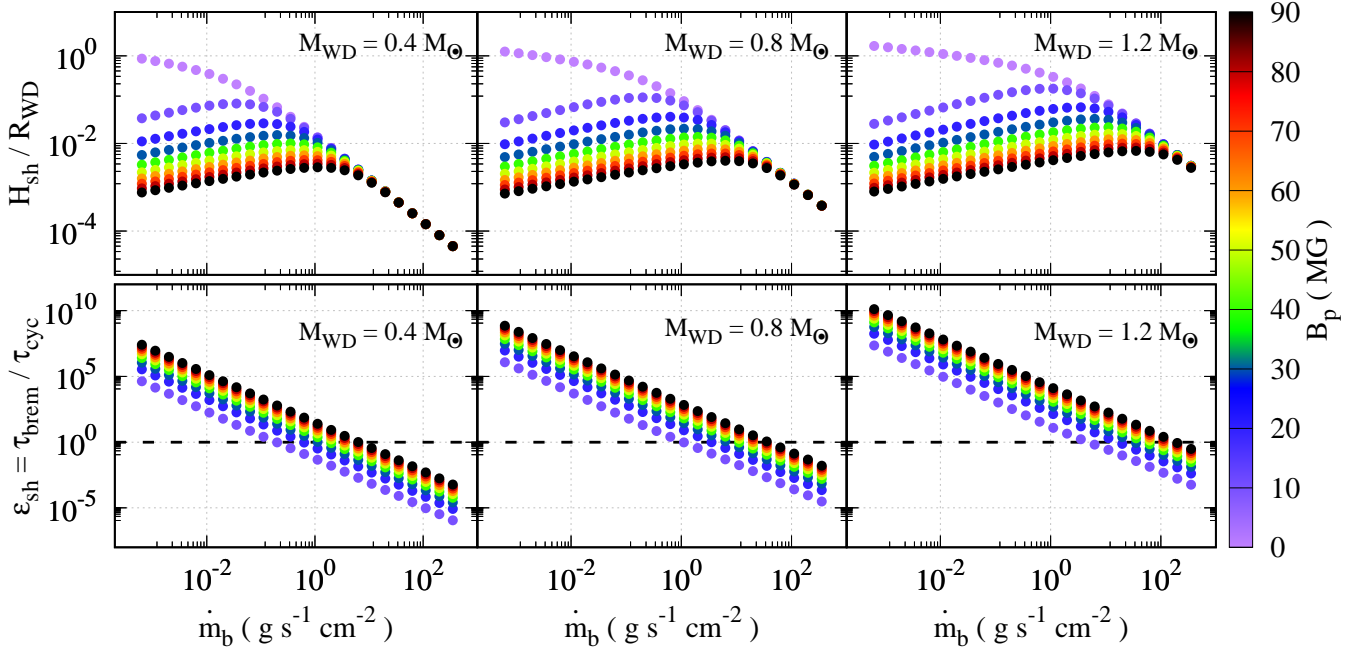} 
\end{center}
\caption{Shock height $(H_{\rm sh})$ in units of the WD radius $(R_{\rm WD})$, in top row, and cooling ratio $(\epsilon_{\rm sh})$, i.e., the ratio between the bremsstrahlung and cyclotron cooling time-scales $(\tau_{\rm brem}/\tau_{\rm cyc})$, in bottom row, for different combinations of model parameters: $M_{\rm WD}$, $B_{\rm p}$ and $\dot{m}_{\rm b}$.
In all cases, for simplicity, we set the remaining parameters as in the standard model (Table~\ref{TabSTAN}).
We took several values for $\dot{m}_{\rm b}$ and $B_{\rm p}$, and three values for $M_{\rm WD}$, namely $0.4$ (left panel), $0.8$ (middle panel), and $1.2$ (right panel), in units of~\Msun.
In each panel, the $x$ axis corresponds to $\dot{m}_{\rm b}$ and the color bar to $B_{\rm p}$.
Since $B_{\rm p}$ has to be non-null for a consistent definition of $\epsilon_{\rm sh}$, the case $B_{\rm p}=0$ is only shown in the top row.
Notice that when cyclotron cooling is negligible, the shock height always decreases with increasing specific accretion rates.
However, for non-negligible cyclotron cooling, there is always a maximum shock height for a given combination of $M_{\rm WD}$ and $B_{\rm p}$, which is caused by the balance between the two physical processes responsible for the cooling, i.e. bremsstrahlung and cyclotron, and this takes place when ${\epsilon_{\rm sh}\sim1}$.
In particular, for a given $M_{\rm WD}$ (or $B_{\rm p}$), the greater the $B_{\rm p}$ (or $M_{\rm WD}$), the greater the critical specific accretion rate such that the balance between bremsstrahlung and cyclotron takes place.
}
  \label{FigPSR5.2}
\end{figure*}


\begin{figure}[htb!]
\begin{center}
\includegraphics[width=0.999\linewidth]{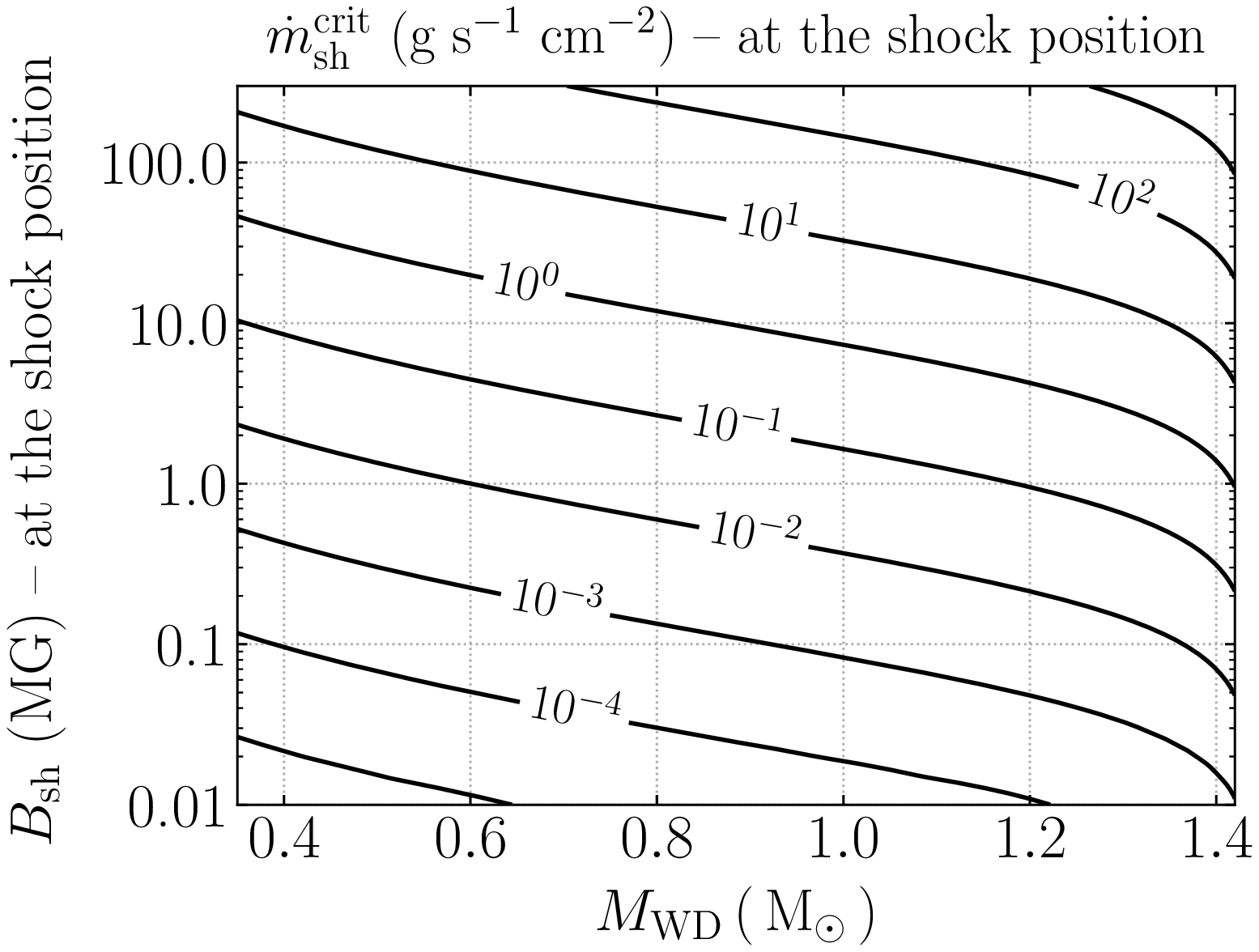}
\end{center}
\caption{Critical specific accretion rate at the shock position $(\dot{m}_{\rm sh}^{\rm crit})$ such that for ${\dot{m}_{\rm sh}<\dot{m}_{\rm sh}^{\rm crit}}$ cyclotron cooling cannot be neglected, as a function of the WD mass $(M_{\rm WD})$ and the magnetic field strength at the shock position $(B_{\rm sh})$, assuming ${S_{\rm sh}=10^{17}}$~cm$^2$.
For a given $\dot{m}_{\rm sh}$, there is a unique line in the plane defined by $M_{\rm WD}$ and $B_{\rm sh}$ so that all values in this line lead to the balance between \brem~and cyclotron, i.e. ${\epsilon_{\rm sh}=1}$.
In this way, for a given $\dot{m}_{\rm sh}$, all combinations of $M_{\rm WD}$ and $B_{\rm sh}$ above the line defined by such a specific accretion rate have non-negligible cyclotron emission. For values of $M_{\rm WD}$ and $B_{\rm sh}$ below such a line, cyclotron cooling can be neglected.
}
\label{FigPSR6.1}
\end{figure}


\subsection{Dependence on main parameters}
\label{influPAR}

In order to illustrate how the PSR profiles are affected by the \cyclops~input parameters, we show in Figs.~\ref{FigPSR4.1} and \ref{FigPSR4.2} temperature profiles built with different values of $B_{\rm p}$, $M_{\rm WD}$, $\dot{m}_{\rm b}$ and $R_{\rm th}$.
For simplicity, the remaining parameters are set as in the standard model (Table~\ref{TabSTAN}).
We fixed  $B_{\rm p}$ in each figure, i.e. 1~MG in Fig.~\ref{FigPSR4.1} and 30 MG in Fig.~\ref{FigPSR4.2}, and $M_{\rm WD}$ in each row, i.e. 0.6~\Msun~(top rows) and 1.0~\Msun~(bottom rows).
With respect to $\dot{m}_{\rm b}$, we set three values in each figure, namely $0.1$~g~s$^{-1}$~cm$^{-2}$ (left panels), $1$~g~s$^{-1}$~cm$^{-2}$ (middle panels) and $10$~g~s$^{-1}$~cm$^{-2}$ (right panels).
Finally, the values used for $R_{\rm th}$ are indicated by the colorbars.
The four parameters analyzed here ($B_{\rm p}$, $M_{\rm WD}$, $\dot{m}_{\rm b}$, $R_{\rm th}$) play a key role in shaping the profiles as well as in determining the shock heights and temperatures.

%
Starting with the threading region radius, we notice that from the Rankine-Hugoniot jump conditions (Eq. \ref{BVPCon1}), it follows that the greater $R_{\rm th}$, the greater the flow velocity at the shock position and, in turn, the greater  $T_{\rm sh}$.
This is nicely illustrated in Figs.~\ref{FigPSR4.1} and \ref{FigPSR4.2}, where $T_{\rm sh}$ correlates with $R_{\rm th}$.
Additionally, keeping all other parameters fixed, the values of $T_{\rm sh}$ are very sensitive to variations in $R_{\rm th}$, when it is in the range ${1~R_{\rm WD}\lesssim R_{\rm th} \lesssim10~R_{\rm WD}}$.
In particular, $T_{\rm sh}$ can vary by a factor of $\sim4$ when changing $R_{\rm th}$ from low to high values.
Moreover, as $T_{\rm sh}$ increases with $R_{\rm th}$, so does $H_{\rm sh}$, since the higher-energy plasma needs longer time to be cooled down, irrespective of the cooling efficiency.
Finally, the shape of the profiles only negligibly changes with $R_{\rm th}$.

%
Regarding the WD magnetic field strength, comparing the panels for $1$~MG with those for $30$~MG, we notice that $H_{\rm sh}$ is smaller and $T_{\rm sh}$ is greater in the cases of stronger $B_{\rm p}$. 
This is because the cooling efficiency is enhanced by cyclotron radiation and thus the stronger $B_{\rm p}$, the more efficient the cooling.
This implies that for sufficiently strong $B_{\rm p}$, values of $H_{\rm sh}$ might be extremely low.
As $H_{\rm sh}$ becomes smaller for stronger $B_{\rm p}$, the gas hits the shock with larger velocities, which implies that $T_{\rm sh}$ increases.
Despite the above mentioned correlation and anti-correlation, we notice that $T_{\rm sh}$ does not change drastically, unlike $H_{\rm sh}$, which is strongly affected by $B_{\rm p}$ (see Fig.~\ref{FigPSR5.2} in Section~\ref{influCOOL}).
Finally, unlike the above-discussed case of $R_{\rm th}$, the profile shape is hugely affected by $B_{\rm p}$, provided that all other parameters are kept
the same.
In fact, as $B_{\rm p}$ increases, the average temperature in the PSR decreases.
This is because of the cyclotron cooling, which is greater for stronger $B_{\rm p}$ and makes the profiles flatter, leading to a strong reduction of the temperature close to the shock.
In particular, the greater $B_{\rm p}$, the more efficient the cyclotron cooling, and the greater is the difference between $T_{\rm sh}$ and the average temperature in the PSR.
This causes the profiles to more closely resemble those of single-temperature plasmas as $B_{\rm p}$ increases.

%
With respect to the WD mass, we can clearly see the correlation between $M_{\rm WD}$ and $T_{\rm sh}$. 
As $T_{\rm sh}$ correlates with the flow velocity at the shock position, which in turn correlates with $M_{\rm WD}$, the greater $M_{\rm WD}$, the greater the flow velocity at the shock position, and, in turn, the greater the $T_{\rm sh}$.
For more details see Appendix~\ref{app_comppre}.
In addition, we can see that $M_{\rm WD}$ also correlates with $H_{\rm sh}$, since a longer time is needed to cool down the hotter gas down, similarly to the case of $R_{\rm th}$.
The WD mass, as in the case of $R_{\rm th}$, has no (or very little, if at all) impact on shape of the $T$ distribution.

%
Concerning the specific accretion rate, we found that it correlates with $T_{\rm sh}$ so that the greater $\dot{m}_{\rm b}$, the greater $T_{\rm sh}$.
This is because both shock density and pressure are directly proportional to the specific accretion rate, which makes cooling more efficient.
The dependence of $H_{\rm sh}$ on $\dot{m}_{\rm b}$ is a bit more complicated and will be discussed in Section~\ref{influCOOL}.
Finally, we notice that the shape of $T$ distribution is somewhat affected by $\dot{m}_{\rm b}$, becoming flatter as $\dot{m}_{\rm b}$ decreases.

%
After discussing how the four above-mentioned parameters separately affect the PSR structure, we turn to a discussion on their influence when compared together.
For sufficiently large values of $\dot{m}_{\rm b}$, it is shown in Figs.~\ref{FigPSR4.1} and \ref{FigPSR4.2} that $B_{\rm p}$ has little impact on the profiles.
However, as $\dot{m}_{\rm b}$ becomes smaller and smaller, even relatively weak $B_{\rm p}$ might change substantially the $T$ profiles. 
Indeed, comparing the three top panels in Fig.~\ref{FigPSR4.2}, which correspond to $M_{\rm WD}=0.6$~\Msun, the profiles are different for different values of $\dot{m}_{\rm b}$.
In particular, the $T$ distributions for $\dot{m}_{\rm b}=1$~\gscm~are less affected by $B_{\rm p}$ than the profiles for $\dot{m}_{\rm b}=0.1$~\gscm.
However, when comparing the three bottom panels, which correspond to $M_{\rm WD}=1.0$~\Msun, we notice that the profiles for $\dot{m}_{\rm b}=1$~\gscm~are more affected by $B_{\rm p}$ than the corresponding profiles associated with $M_{\rm WD}=0.6$~\Msun.
This means that combinations of $M_{\rm WD}$, $B_{\rm p}$ and $\dot{m}_{\rm b}$ might potentially lead to qualitatively very different PSR temperature structures.
The reason for this behavior is connected with the balance between bremsstrahlung and cyclotron radiative processes and will be discussed in more detail in the next section.

\subsection{Cyclotron cooling versus \brem~cooling}
\label{influCOOL}

As briefly discussed before, the balance between \brem\,and cyclotron radiative processes plays an important role in shaping the PSR structure.
Before exploring this balance a bit further, it is convenient to discuss which parameters mainly contribute to both processes.
From Eq.~\ref{coolbrem}, we can see that the density
in the PSR is the main parameter responsible for the cooling by \brem.
In particular, the greater the density, the stronger the \brem\,emission, and since the density depends on the amount of gas in the PSR, it is not difficult to correlate it with the specific accretion rate (Eq.~\ref{BVPCon1}). 
Thus, it naturally follows that the greater the $\dot{m}_{\rm b}$, the stronger the \brem\,emission and consequently the more efficient the cooling by \brem.
On the other hand, from Eqs.~\ref{coolratio} and \ref{coolcyc}, it is clear that the greater $B_{\rm p}$, the stronger the cyclotron emission and the more efficient is the cooling due the cyclotron radiation.

These correlations can be seen in Fig.~\ref{FigPSR5.1}, where we  show profiles regarding the ratio between cyclotron and \brem\,cooling (i.e. $\Lambda_{\rm cyc}/\Lambda_{\rm brem}$), for different values of $B_{\rm p}$, $\dot{m}_{\rm b}$ and $M_{\rm WD}$.
In all cases, all other parameters are set as in the standard model (Table~\ref{TabSTAN}).
The variation in $R_{\rm th}$ causes negligible differences in $\Lambda_{\rm cyc}/\Lambda_{\rm brem}$, so it is not discussed.

%
For models with high accretion rates {(i.e. ${\dot{m}_{\rm b}\gtrsim10}$~\gscm)} and low WD masses {(i.e. ${M_{\rm WD}\lesssim0.8}$~\Msun)}, bremsstrahlung dominates in the entire PSR, even for relatively high magnetic fields {(${\sim60}$~MG)}.
However, models with high accretion rates and high WD masses {(i.e. $M_{\rm WD}\gtrsim0.8$~\Msun)}, cyclotron contributes up to the half of PSR close to the shock, and \brem\,in at least the other half of the PSR (close to the WD surface).

%
Regarding models with low accretion rates (i.e. ${\dot{m}_{\rm b}\lesssim0.1}$~\gscm), bremsstrahlung dominates in the entire PSR only for relatively weak magnetic fields {(${\lesssim1}$~MG)}.
Otherwise, cyclotron is more important for more than $50$\,\% of the PSR from the shock, and bremsstrahlung is important only close to the WD surface, where density is much higher.
The above-mentioned feature takes place regardless of the
WD mass.
Moreover, should $\dot{m}_{\rm b}$ be even smaller, then magnetic fields weaker than ${\sim1}$~MG would be already enough to affect the PSR structure.

%
Concerning models with moderate accretion rates (i.e. ${\dot{m}_{\rm b}\sim1}$~\gscm), cyclotron becomes important only for moderate to high magnetic fields {(${\gtrsim10}$~MG)}.
In particular, a value of $B_{\rm p}=30$~MG is already enough for cyclotron to play a role in at least half of the PSR irrespective of the WD mass.
For values of $B_{\rm p}$ between ${\sim10}$ and ${\sim30}$~MG, cyclotron is important only for high-mass WDs.

Regardless of the magnetic field strength, bremsstrahlung always dominates the cooling process in PSR bottom, close to the WD surface, i.e. at least $\sim20$\,\% of the PSR.
In addition, the smaller the $B_{\rm p}$, the greater the fraction of the PSR, from the bottom, dominated by bremsstrahlung emission.


By inspecting the panels in Fig.~\ref{FigPSR5.1}, we can see that for all combinations of $B_{\rm p}$ and $\dot{m}_{\rm b}$, the greater $M_{\rm WD}$, the greater $\Lambda_{\rm cyc}/\Lambda_{\rm brem}$.
This is because, by considering all other parameters fixed, the greater $M_{\rm WD}$, the more energetic the flow and in turn the longer the cooling time-scale, which leads to higher shock heights and PSRs with lower densities.
This makes the relative importance of cyclotron cooling greater as $M_{\rm WD}$ increases.

Proceeding further with the competition between cyclotron and \brem\,radiative processes, we show in Fig.~\ref{FigPSR5.2} how the shock height ($H_{\rm sh}$, top row) and the cooling ratio ($\epsilon_{\rm sh}$, Eq.~\ref{coolratio}, bottom row) depend on model parameters, for three different values of $M_{\rm WD}$, namely, $0.4$, $0.8$ and $1.2$~\Msun, and several combinations of $B_{\rm p}$ (from 0 to 90 MG) and $\dot{m}_{\rm b}$ (from $\sim10^{-4}$ to $\sim10^{3}$~\gscm), keeping the remaining parameters as in the standard model (Table~\ref{TabSTAN}).

In the case of negligible cyclotron cooling, the shock height always increases with decreasing the specific accretion rate.
This is because the smaller the specific accretion rate, the smaller the density in the PSR and in turn the weaker the \brem\,emission.
This leads to a decrease in the plasma cooling rate and consequently an increase in the \brem~cooling time-scale as the specific accretion rate decreases, yielding taller PSRs.
However, in the presence of non-negligible cyclotron cooling, the shock height does not always increase as the specific accretion rate decreases.

In fact, there is a maximum shock height, which is defined by the balance between the bremsstrahlung and cyclotron cooling.
For a given combination of $M_{\rm WD}$ and $B_{\rm p}$, there is a critical specific accretion rate $\dot{m}_{\rm b}^{\rm crit}$ such that cyclotron/\brem\,cooling is more important when $\dot{m}_{\rm b}$ is smaller/greater than $\dot{m}_{\rm b}^{\rm crit}$.
This can be seen in the top row of Fig.~\ref{FigPSR5.2}, where an anti-correlation takes place between $H_{\rm sh}$ and $\dot{m}_{\rm b}$ when ${\dot{m}_{\rm b}>\dot{m}_{\rm b}^{\rm crit}}$, and a correlation otherwise.

The values of ${\dot{m}_{\rm b}^{\rm crit}}$ are given by the balance between bremsstrahlung and cyclotron processes, when the cooling ratio ${\epsilon_{\rm sh}\sim1}$, i.e. the ratio between the \brem\,and cyclotron cooling time-scales ${(\tau_{\rm brem}/\tau_{\rm cyc})}$ is ${\sim1}$.
In the bottom row of Fig.~\ref{FigPSR5.2} we show how $\epsilon_{\rm sh}$ depends on $M_{\rm WD}$, $B_{\rm p}$ and $\dot{m}_{\rm b}$.
The greater $M_{\rm WD}$, or $B_{\rm p}$, the greater $\epsilon_{\rm sh}$.
This is because the greater those parameters, the stronger the cyclotron radiation and in turn the cyclotron cooling.
For $\dot{m}_{\rm b}$, on the other hand, there is an anti-correlation with $\epsilon_{\rm sh}$, which is a consequence of the increase in \brem\, radiation (and in turn a decrease in $\epsilon_{\rm sh}$) when $\dot{m}_{\rm b}$ increases.

With respect to the critical specific accretion rate, we can see that the maximum $H_{\rm sh}$ for a given pair of $M_{\rm WD}$ and $B_{\rm p}$ takes place at the same $\dot{m}_{\rm b}$ associated with ${\epsilon_{\rm sh}\sim1}$.
Such a critical $\dot{m}_{\rm b}$ separates the parameter space into two regimes, namely \brem-dominated or cyclotron-dominated cooling.
In other words, for a given triple ${(M_{\rm WD}, B_{\rm p}, \dot{m}_{\rm b})}$, there is always a $\dot{m}_{\rm b}^{\rm crit}$, such that the flow is \brem-dominated when ${\dot{m}_{\rm b}>\dot{m}_{\rm b}^{\rm crit}}$, and cyclotron-dominated otherwise.

We show in Fig.~\ref{FigPSR6.1} the critical specific accretion rate at the location of the shock front $(\dot{m}_{\rm sh}^{\rm crit})$ as a function of $M_{\rm WD}$ and the WD magnetic field intensity at the shock location $(B_{\rm sh})$.
By fixing the cross-section at the shock $S_{\rm sh}=10^{17}$~cm$^2$, according to Eq.~\ref{coolratio}, for a given pair $(M_{\rm WD},B_{\rm sh})$, there is a unique value of $\dot{m}_{\rm sh}$ such that ${\epsilon_{\rm sh}=1}$.
The lines depicted in Fig.~\ref{FigPSR6.1} correspond to values of $\dot{m}_{\rm sh}$ such that $\epsilon_{\rm sh}=1$, i.e. they are values of $\dot{m}_{\rm sh}$ below which cyclotron emission cannot be neglected.
In the figure, for a given value of $\dot{m}_{\rm sh}$, all models defined by $(M_{\rm WD},B_{\rm sh})$ above such a line have non-negligible cyclotron emission.
On the other hand, \brem\,emission always dominates in the bottom region below the same line.

Notice that even though we can neglect cyclotron cooling for several cases of weak magnetic fields, for sufficiently low values of $\dot{m}_{\rm sh}$ (${\lesssim10^{-3}}$~\gscm), even magnetic fields as weak as $\lesssim0.1-1$~MG are enough to play a role in the cooling process.
In particular, for IPs with rather low accretion rates (e.g. those located below the orbital period gap), the assumption of neglecting cooling due to cyclotron, as done in several works, does not seem appropriate.
On the other hand, for specific accretion rates as great as ${\dot{m}_{\rm sh}\gtrsim100}$~\gscm, even a very strong magnetic field is not enough to make cyclotron a relevant process.

We finish this section by saying a few words about the impact of the accretion area at shock position on the above-mentioned results.
Even though we fixed $S_{\rm sh}=10^{17}$~cm$^2$ in Fig.~\ref{FigPSR6.1}, we would like to emphasize that changing $S_{\rm sh}$ will affect only slightly those curves.
Indeed, values greater (smaller) than that will slightly move the curves down (up) in the plane $(M_{\rm WD},B_{\rm sh})$.
This is because the dependence with $S_{\rm sh}$ in Eq.~\ref{coolratio} is much weaker than the dependence with other parameters, which makes the influence of $S_{\rm sh}$ very small.

%% file: TABLES/Table_STANDARD.tex
\begin{table}[tb!] 
%
\begin{flushleft} 
\centering
\caption{Parameters of the \textit{standard} model discussed in this paper.
The non-defined parameters (i.e. $B_{\rm lat}$, $\dot{M}_{\rm WD}$, ${\Delta \varphi}$ and ${\Delta R}$) are set such that $R_{\rm th}=130\,R_{\rm WD}$ and $\dot{m}_{\rm b}=1$~\gscm.
In addition, we assume that the source is at a distance of $100$~pc. See Section~\ref{cyclopscode} for details about these parameters.
}
\label{TabSTAN}
\begin{tabular}{ccccccc}
\hline
\addlinespace[0.085cm]
$i$ &
$B_{\rm long}$ &
$\beta$ &
$M_{\rm WD}$ &
$B_{\rm p}$ &
$\dot{m}_{\rm b}$ &
$R_{\rm th}$ \\
\addlinespace[0.005cm]
($^{\rm o}$) &
($^{\rm o}$) &
($^{\rm o}$) &
(\Msun) &
(MG) &
(\gscm) &
($R_{\rm WD}$)  \\
\addlinespace[0.085cm]
\hline
\addlinespace[0.085cm]
90 &
0 &
5 &
0.8 &
1   & 
1   &
130 \\
%
%
%
%
\hline
\end {tabular}
\end{flushleft}
\end{table}

%% file: CONTENT/Section04.tex
After discussing how the model parameters affect both the balance between \brem\,and cyclotron cooling and the temperature profiles in our PSRs, we can turn to the analysis of their influence on X-ray spectra as well as the unavoidable degeneracy in the parameter space.


To understand the dependencies and degeneracies of the predicted X-ray spectra, one needs to take into account an important correlation between the PSR temperature and the hardness of the X-ray spectra.
The term {\it hard} X-rays corresponds to those photons carrying the highest energies (${\sim10-100}$~keV), while those carrying the lowest energies (${\lesssim10}$~keV) are referred to as the {\it soft} part of the X-rays.
The greater the contribution toward higher energies, the harder the spectrum.

Throughout this paper, we have discussed hydro-thermodynamical aspects of the accretion flow toward the WD surface.
In such a picture, the potential energy of the infalling gas is converted into kinetic energy until reaching the shock, where density and temperature are enhanced.
Then, the greater the gas velocity at the shock, the higher the temperature, and the more energetic the \brem\,emission in the PSR.
Thus, {\it the greater the temperature in the PSR, the harder the X-ray spectrum}, or alternately, {\it the greater the importance of the component associated with the highest energies}.
Keeping this and Figs.~\ref{FigPSR4.1} and \ref{FigPSR4.2} in mind, in what follows we analyze how the X-ray spectrum hardness is affected by the main parameters describing the accretion structure in magnetic CVs.


\begin{figure*}[htb!]
\begin{center}
\includegraphics[width=0.49\linewidth]{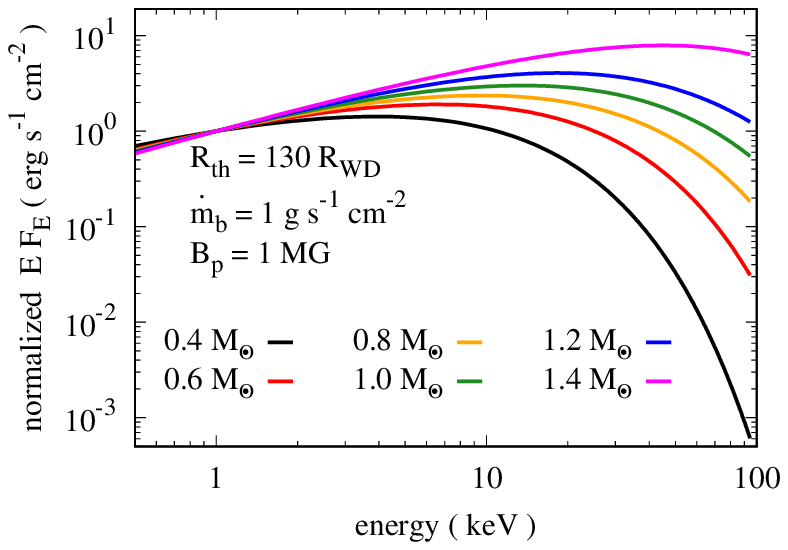}
\includegraphics[width=0.49\linewidth]{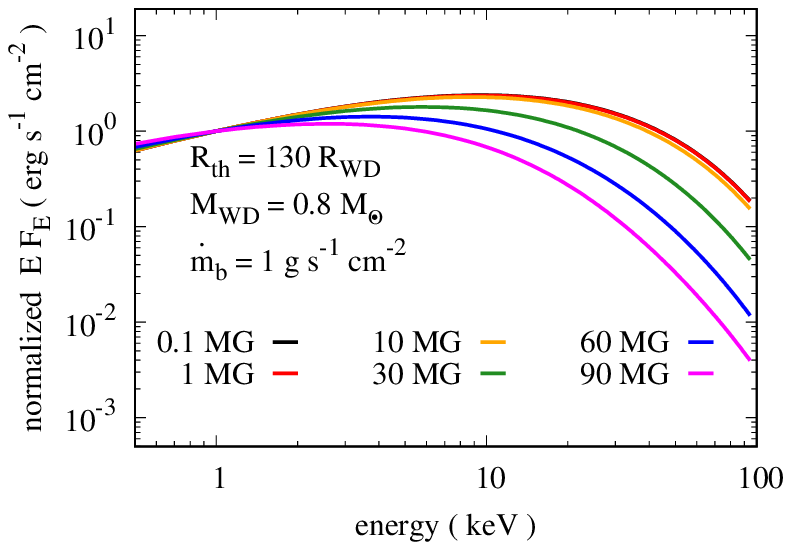}
\includegraphics[width=0.49\linewidth]{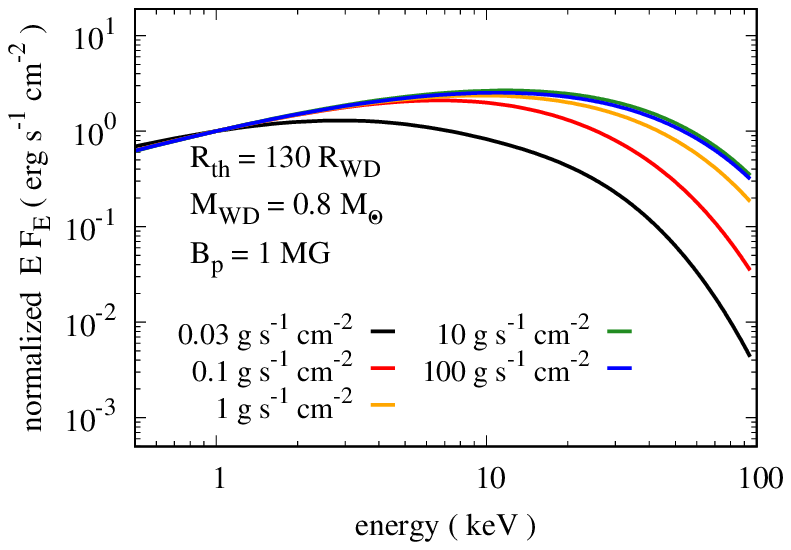}
\includegraphics[width=0.49\linewidth]{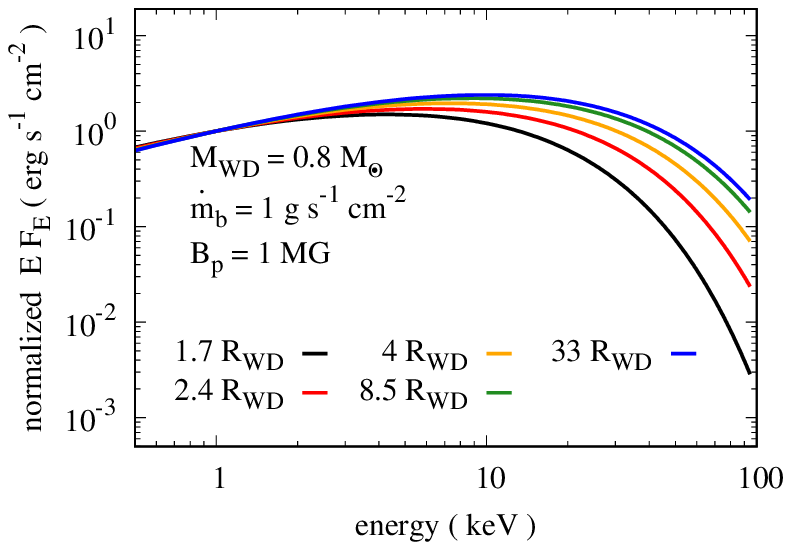}
\end{center}
\caption{Mock X-ray spectra for several different combinations of $M_{\rm WD}$, $B_{\rm p}$, $\dot{m}_{\rm b}$, and $R_{\rm th}$.
In all cases, for simplicity, we set the remaining parameters as in the standard model (Table~\ref{TabSTAN}) and normalized all spectra to their values at $1$~keV.
In all panels, we fixed all parameters but one, namely $M_{\rm WD}$ (top left panel), $B_{\rm p}$  (top right panel), $\dot{m}_{\rm b}$ (bottom left panel) and $R_{\rm th}$ (bottom right panel).
These parameters, when fixed, are indicated in each panel and, when variable, in the keys.
Notice that all four parameters play a key role in shaping the X-ray spectrum, especially $M_{\rm WD}$, after $\sim10$~keV.
In particular, the greater the $M_{\rm WD}$ and/or the smaller the $B_{\rm p}$ and/or the greater the $\dot{m}_{\rm b}$ and/or the greater the $R_{\rm th}$, the harder the spectrum.
See text for more details.
}
\label{FigPSR6.2}
\end{figure*}


\begin{figure}[t!]
\begin{center}
\includegraphics[width=0.99\linewidth]{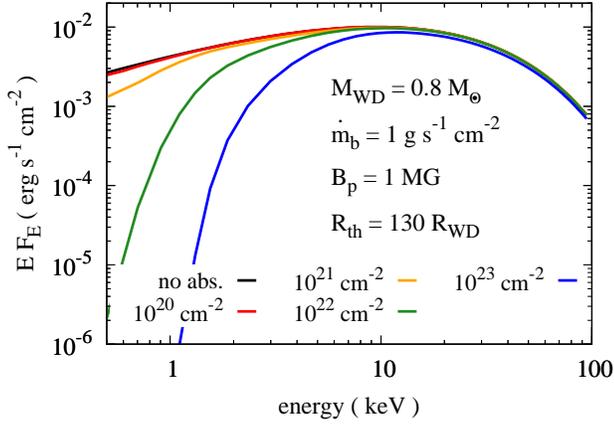}
\end{center}
\caption{Mock X-ray spectra taking into account absorption, for four different values of the hydrogen column density $N_{\rm H}$, namely $10^{20}$, $10^{21}$, $10^{22}$, $10^{23}$, in units of cm$^{-2}$, and a case with no absorption at all, which are indicated in the key.
All parameters are fixed and set as in the standard model (Table~\ref{TabSTAN}).
Notice that absorption is irrelevant for low values of $N_{\rm H}$, i.e. $\lesssim10^{20}$~cm$^{-2}$.
For moderate values, i.e. {$10^{20}$\,cm$^{-2}\,\lesssim\,N_{\rm H}\,\lesssim\,10^{22}$~cm$^{-2}$}, absorption might drastically change the spectrum below $\approx2-3$~keV.
However,  for sufficient large values of $N_{\rm H}$, i.e. $\gtrsim10^{22}$~cm$^{-2}$, absorption plays a significant role in shaping the spectrum beyond $\approx2-3$~keV.
}
\label{FigPSR6.4}
\end{figure}


\begin{figure}[t!]
\begin{center}
\includegraphics[width=0.99\linewidth]{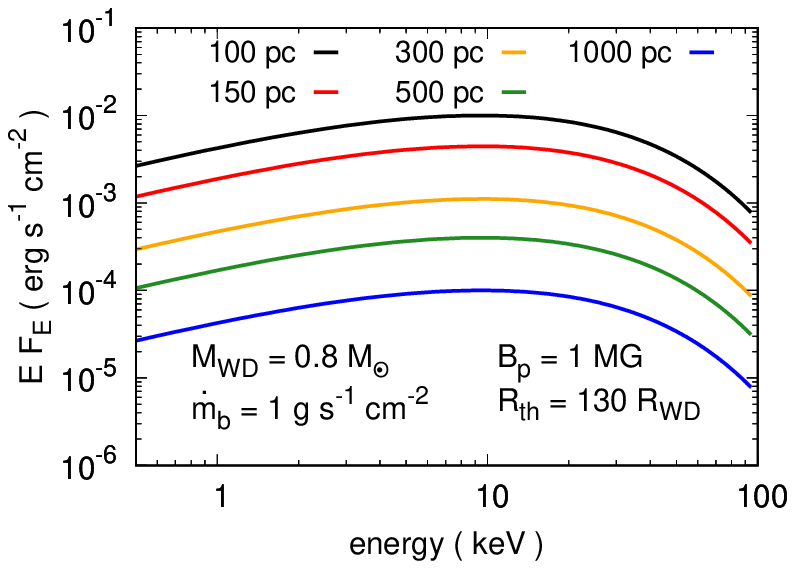}
\end{center}
\caption{Mock X-ray spectra for the standard model (Table~\ref{TabSTAN}), taking into account different distances, namely $100$, $150$, $300$, $500$ and $1000$~pc, which are indicated in the key.
Clearly, as expected, the larger the distance, the lower the X-ray spectrum flux.
Moreover, such a distance-dependent flux is consistent with the widely used \xspec~code (see Appendix~\ref{app_compxspec}).}
\label{FigXRayDIST_NEW}
\end{figure}


\subsection{X-ray spectra}
\label{influXRAY}

An X-ray spectrum in the \cyclops~code is entirely due to \brem~emission (see Section~\ref{cyclopscode} for more details).
That said, readers should bear in mind that other potentially important processes able to affect the hard part of the spectrum are not taken into account in the current version of the \cyclops~code, such as Compton reflection from the WD surface, which is expected to contribute to the emission at energies higher than $\sim10$~keV.
Despite that, X-ray spectra generated with the \cyclops~code are in very good agreement with those generated by the widely used \xspec~code (see Appendix~\ref{app_compxspec} for a detailed comparison between both codes).


We show mock X-ray spectra from the \cyclops~code in Fig.~\ref{FigPSR6.2}.
For a proper comparison, all spectra have been normalized to their fluxes at $1$~keV, which allows us to address the contribution of particular parameters to shape the spectra.
This implies that the y-axis in this figure is somewhat artificial, which does not spoil the analyses, provided that our main goal here is to compare the spectrum shapes.
In addition, no absorption was included, either internal or due to the interstellar medium, as this effect will be addressed separately in other parts of the paper.
In order to generate the mock spectra, in all cases, we allowed four parameters to vary, namely $M_{\rm WD}$ (top left panel), $B_{\rm p}$ (top right panel), $\dot{m}_{\rm b}$  (bottom left panel), and $R_{\rm th}$ (bottom right panel), and fixed the remaining parameters as in the standard model (Table~\ref{TabSTAN}).

%
Starting with the influence of the WD mass, which is shown in the top left panel of Fig.~\ref{FigPSR6.2}, it is clear that the \cyclops~code consistently predicts harder spectra for larger $M_{\rm WD}$.
This is because the greater $M_{\rm WD}$, the higher the temperature in the PSR, and in turn the more enhanced the production of energetic photons.
From the figure, it is also clear that all WD masses produce similar spectra until ${\approx1}$~keV.
At energies greater than that, the lower the $M_{\rm WD}$, the faster the flux falls-off toward higher energies.
Even though there are differences between $\approx1$~and~$\approx10$~keV, they become much more evident at the highest energies.

%
Regarding the influence of the WD magnetic field strength, which is depicted in the top right panel of Fig.~\ref{FigPSR6.2}, we see that for weaker $B_{\rm p}$, the spectra are harder.
The higher the WD magnetic field, the higher the extra cooling (relative to bremstrahlung) causing a smaller average temperatures in the PSRs for stronger $B_{\rm p}$, especially in the region from where the \brem\, radiation comes, i.e. close to the WD surface (see also Fig.~\ref{FigPSR5.1}).
As in the case of the WD mass, spectra due to all values of $B_{\rm p}$ are rather similar until $\approx1$~keV, become different after that until $\approx10$~keV, and become substantially different at higher energies.
We notice that, given the particular choice of other parameters, spectra for $B_{\rm p}\lesssim10$~MG are indistinguishable.

%
With respect to the influence of the specific accretion rate, which is shown in the bottom left panel of Fig.~\ref{FigPSR6.2}, we notice that the larger the $\dot{m}_{\rm b}$, the harder the spectrum.
This is due to the clear correlation between the PSR temperature and the $\dot{m}_{\rm b}$. A lower $\dot{m}_{\rm b}$ leads to a smaller amount of matter and in turn a lower density, which causes a reduction in the cooling rate and in turn makes the PSR taller. As a result, the kinetic energy released at the shock is reduced, causing a reduction in the maximum temperature.
As in the case of $M_{\rm WD}$ and $B_{\rm p}$, the differences among spectra become larger toward higher energies.
In addition, from the choice of other parameters, differences in the spectra for ${\dot{m}_{\rm b}\gtrsim10}$~\gscm~are imperceptible.

%
Regarding the threading region radius, its impact on X-ray spectra is shown in the bottom right panel of Fig.~\ref{FigPSR6.2}.
From the figure, we see that the greater the $R_{\rm th}$, the harder the spectrum.
This is due to the reduction of the available potential energy as $R_{\rm th}$ becomes smaller, which makes the corresponding kinetic energy smaller. 
This implies a reduction of the PSR temperature as shown in Figs.~\ref{FigPSR4.1} and \ref{FigPSR4.2}.
Unlike the other cases discussed above, the differences in the spectra start at energies a bit higher ($\approx3$~keV).
Even though differences become larger as the energy increases, like in other cases, we notice that spectra are more distinguishable
at very high energies ($\gtrsim30$~keV).
Moreover, given the choices of other parameters, the spectra for ${R_{\rm th}\gtrsim8.5}$~\Rwd~are virtually identical.

%
It is important to highlight at this point that the above-mentioned correlations have also been found in other works.
For instance, \citet{Hayashi_2014} found that spectra are softer for lower specific accretion rates.
In addition, \citet{Suleimanov_2016} found that systems with smaller magnetosphere radius produce softer spectra.
We can then conclude that X-ray spectra built with the \cyclops~code present features that are in good agreement with previous works.

%
In the previously shown dependencies, absorption was not taken into account since our goal was to illustrate how each parameter affects the X-ray spectra.
In what follows we will consider interstellar absorption (see Section~\ref{cyclopscode}) and its impact on X-ray spectra, which is illustrated in Fig.~\ref{FigPSR6.4}, where we show spectra for different hydrogen column densities $N_{\rm H}$.
A clear correlation we can see from the figure is that the effect of absorption increases for lower energies.
In addition, the greater the $N_{\rm H}$, the stronger the absorption influence on the hard part of the spectrum.
Indeed, for low values of $N_{\rm H}$ (${\lesssim10^{20}}$~cm$^{-2}$), absorption is negligible throughout the entire energy range of current X-ray instruments.
On the other hand, for moderate values of $N_{\rm H}$, i.e. ${10^{20}~{\rm cm}^{-2}\lesssim N_{\rm H} \lesssim10^{22}~{\rm cm}^{-2}}$, absorption plays a key role in shaping the continuum at energies ${\lesssim2-3}$~keV and is negligible at higher energies.
Finally, for high values of $N_{\rm H}$ (${\gtrsim10^{22}}$~cm$^{-2}$), a larger proportion of the spectrum is subjected to absorption, and the greater the $N_{\rm H}$, the greater the portion of the spectrum affected by absorption.

%
So far we have focused on normalized spectra, since our goal was to investigate how model parameters affect their shape.
However, the observed flux depends on the distance to the source, as illustrated in Fig.~\ref{FigXRayDIST_NEW}, where we considered five distances, namely $100$, $150$, $300$, $500$ and $1000$~pc.
Since the flux depends on the inverse of the squared distance, it is not surprising the correlation seen in the figure, i.e. the larger the distance, the lower the spectrum flux.
Most importantly, the distance-dependent X-ray flux in the \cyclops~code has been properly calibrated using the \xspec~code (see Appendix~\ref{app_compxspec}).

%
Even though we have not discussed the influence of other \cyclops~geometrical parameters, such as the PSR colatitude, the WD magnetic field longitude and orbital inclination, we stress that their impact on shaping X-ray spectra is negligible.
Indeed, bremsstrahlung emission does not depend on the WD magnetic field direction, unlike cyclotron emission, important in optical bands.
What is important for X-ray emission is the magnetic field orientation with respect to the rotation axis, i.e. the WD magnetic field latitude $B_{\rm lat}$ \citep[e.g.,][]{Ferrario_1989}, which is embedded in $R_{\rm th}$ in the analysis we have been performing.

Concerning the orbital inclination, it also has no (or very weak, if at all) impact on the X-ray spectra.
This is because in the \cyclops~code, \brem~is solely responsible for the X-ray emission, and the PSR is optically thin for \brem\,emission so that the direction by which it is seen does not matter. 
In this case, the emission is proportional to the \brem\ emissivity integrated over the PSR volume.
For high inclinations, the PSR is self-eclipsed in some spin phases causing a decrease in the average flux along the WD rotation.
If the self-eclipse is partial, the spectrum shape can slightly change along the WD spin.
As the temperature is not uniform in the PSR, the occulted portion can have different temperatures relative to the visible portion in a phase in which the PSR is partially behind the WD.
And the resulting spectrum can differ from the spectrum produced when the entire PSR is visible.
We would like to emphasize, though, that in the presence of other effects not included here \citep[e.g., Compton humps,][]{Hayashi_2018}, the inclination might be an important parameter to be considered, since such effects might strongly depend on the direction by which the PSR is seen.

\begin{figure*}[htb!]
\begin{center}
\includegraphics[width=0.49\linewidth]{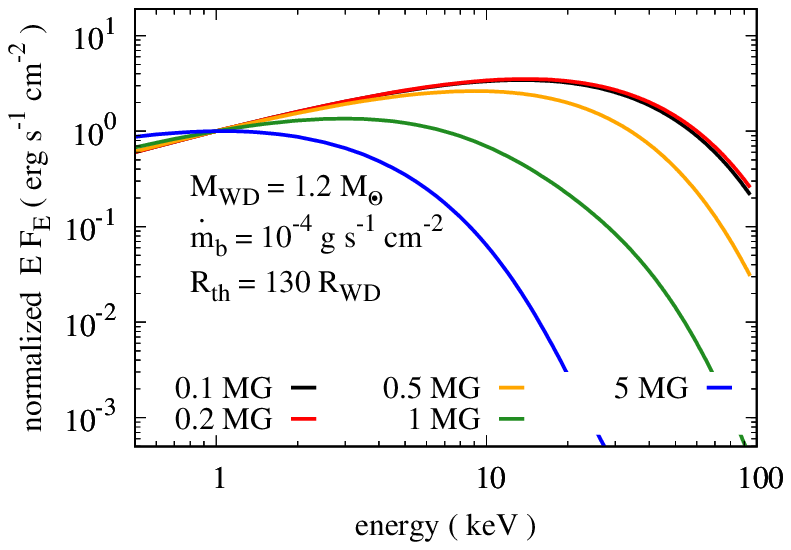}
\includegraphics[width=0.49\linewidth]{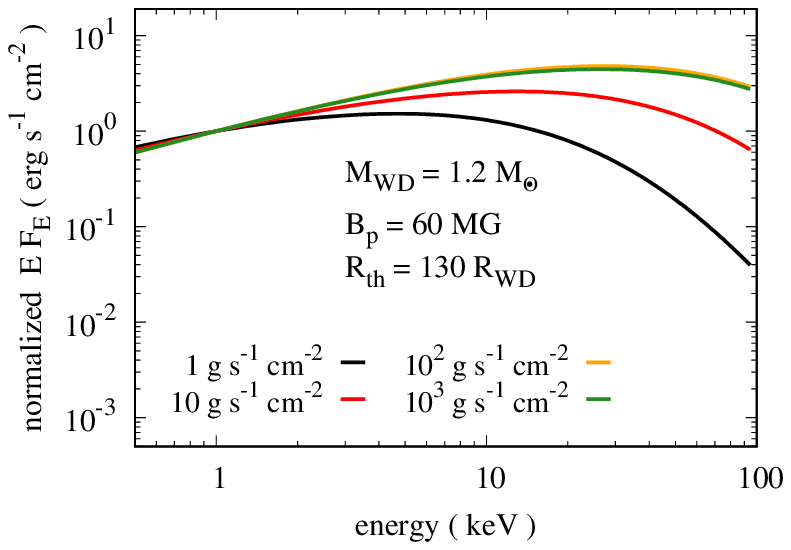}
\end{center}
\caption{Mock X-ray spectra comparing the influence of  $B_{\rm p}$ (left panel) and fixing $M_{\rm WD}=1.2$~\Msun\, and $\dot{m}_{\rm b}=10^{-4}$~\gscm, and comparing the influence of  $\dot{m}_{\rm b}$ (right panel) and fixing $M_{\rm WD}=1.2$~\Msun and $B_{\rm p}=60$~MG.
In both panels, the remaining parameters are set as in the standard model (Table~\ref{TabSTAN}) and all spectra are normalized to their values at $1$~keV.
Notice that the thresholds leading to degeneracy in both panels are different than those in Fig.~\ref{FigPSR6.2}, due to the balance between cyclotron and \brem\,radiative processes, which causes a movement of the thresholds through the parameter space.
}
\label{FigPSR6.3}
\end{figure*}

\subsection{The degeneracy problem}
\label{influDEG}

So far, we have discussed the influence of key parameters in shaping the X-ray spectrum individually.
We will discuss in what follows how correlations among parameters affect the X-ray spectra, especially in creating degeneracies.
This is rather important for the purposes of fitting schemes as one parameter might compensate another one, which implies that rather similar spectra might be built, even for very different combinations of the parameters.

While discussing Fig.~\ref{FigPSR6.2}, we mentioned some thresholds for the parameters above (or below) which the spectra look very alike.
We would like to stress that such thresholds strongly depend on the combination of parameters.
For example, we found that for ${\dot{m}_{\rm b}=1}$~\gscm, ${M_{\rm WD}=0.8}$~\Msun\, and ${R_{\rm th}=130\,R_{\rm WD}}$, there is a degeneracy among spectra for ${B_{\rm p}\lesssim10}$~MG, which is the threshold for this combination of parameters.
However, as discussed in Section~\ref{influCOOL}, due to the balance between cyclotron and \brem\,radiative processes, there is a critical $\dot{m}_{\rm b}$ below which cyclotron cooling dominates over \brem\,cooling.
That said, it is not surprising that this degeneracy might become restrict to smaller and smaller $B_{\rm p}$, as $\dot{m}_{\rm b}$ decreases.
Indeed, for sufficiently low specific accretion rates, effects due to the WD magnetic field cannot be neglected anymore, as argued in Section~\ref{influCOOL}.
This means that the corresponding spectra is also different, even for relatively weak $B_{\rm p}$.

This is illustrated in the left panel of Fig.~\ref{FigPSR6.3}, where we show spectra for different values of weak $B_{\rm p}$, for ${\dot{m}_{\rm b}=10^{-4}}$~\gscm, ${M_{\rm WD}=1.2}$~\Msun, ${R_{\rm th}=130\,R_{\rm WD}}$, and the remaining parameters as in the standard model (Table~\ref{TabSTAN}).
All spectra have been normalized to their fluxes at $1$~keV.
Notice that for such a low $\dot{m}_{\rm b}$, the spectra start becoming degenerated at $B_{\rm p}\sim0.2$~MG, which is the threshold for this combination of parameters.
Thus, there is practically no distinction among spectra when $B_{\rm p}$ is weaker than that.

Another example we discussed is for the combination ${M_{\rm WD}=0.8}$~\Msun, ${B_{\rm p}=1}$~MG and ${R_{\rm th}=130\,R_{\rm WD}}$, which drives a degeneracy among spectra when
${\dot{m}_{\rm b}\gtrsim10}$~\gscm, which is the threshold for this combination of parameters (see Fig.~\ref{FigPSR6.2}, bottom-left panel).
As in the above-discussed case, for sufficiently strong WD magnetic fields and massive WDs, such an alike behavior can be broken when
$10$~\gscm~${\lesssim\dot{m}_{\rm b}\lesssim100}$~\gscm, but still remains for ${\dot{m}_{\rm b}\gtrsim100}$~\gscm.
This is shown in the right panel of Fig.~\ref{FigPSR6.3}, where we depict spectra for different values of high $\dot{m}_{\rm b}$, for ${B_{\rm p}=60}$~MG, ${M_{\rm WD}=1.2}$~\Msun, ${R_{\rm th}=130\,R_{\rm WD}}$ and the remaining parameters as in the standard model (Table~\ref{TabSTAN}).
Notice that for such a different combination of $M_{\rm WD}$ and $B_{\rm p}$, the onset of degeneracy moves from ${\dot{m}_{\rm b}\sim10}$~\gscm\ to ${\dot{m}_{\rm b}\sim100}$~\gscm.

So far, we have provided several examples in which, while fixing all parameters but one, we showed that there is a critical value for this variable parameter above/below which the differences in the resulting X-ray spectra are imperceptible.
We can therefore induce that such thresholds always exist, regardless of the combination of fixed parameters considered and the variable parameter analyzed.
In addition, due to the balance between cyclotron and \brem\,radiative processes, such thresholds move through the parameter space.

The approach we have considered requires a set of four parameters to be fitted, namely the WD mass, the WD magnetic field, the specific accretion rate and the threading region radius, which are the most important parameters shaping the PSR temperature and density profiles, and, in turn, the X-ray spectra.
Thus, we deal with a four-dimensional fitting problem, which can potentially get even more complicated by the addition of more geometrical parameters, such as those related to the footprint of the PSR on the WD surface.
That said, it is actually not surprising that many combinations of these four parameters could lead to X-ray spectra in accordance with an observed one.
This happens because the parameters might compensate one another, leading to a rather similar X-ray spectrum.

For instance, in Section~\ref{influXRAY}, we showed how the spectrum hardness correlates with the parameters.
In particular, we showed that the greater the WD mass, the harder the spectrum.
We also showed that the greater the threading region radius, the harder the spectrum.
In this way, it seems rather probable that a high-mass WD combined with a small threading region radius could lead to a spectrum similar to that of a low-mass WD combined with a large threading region radius.
In a similar fashion, many different combinations of model parameters can naturally lead to spectra that equally well fit an observed one.

In what follows, we will provide more generic examples in which several combinations of the four parameters investigated here lead to virtually identical X-ray spectra.
This is an important issue for fitting schemes in magnetic CV emission modeling and becomes inevitable when only X-ray spectrum is taken into account.
Therefore, the \textit{degeneracy problem} is a physical problem that needs to be consistently addressed in any fitting scheme.
Besides showing more generic examples characterizing the degeneracy problem, we will also discuss ways of solving this problem.
Our approach should also potentially provide tools to investigate additional properties, such as the geometrical parameters as well as internal absorption.

\subsection{Two examples of degenerated spectra}
\label{influDEGmodels}


\begin{figure}[tb!]
\begin{center}
\includegraphics[width=0.99\linewidth]{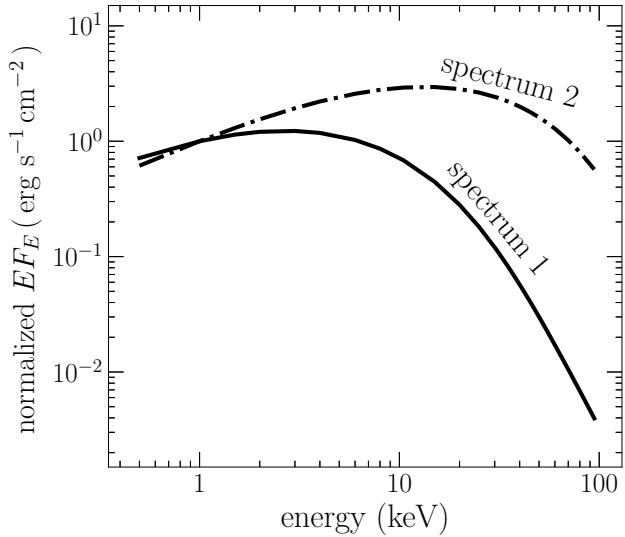}
\end{center}
\caption{X-ray spectra associated with the two sets of models described in Table~\ref{TabMod}, normalized to their values at $1$~keV.
The spectra are very different, the \textit{spectrum 2} being much harder than the \textit{spectrum 1}.
Even though the models in each set are significantly different, they all lead to the same X-ray spectrum, likely because the average temperatures weighted by the squared density are rather similar in each set of models.
}
\label{Fig_2sets_degen_spectra}
\end{figure}


\input{TABLES/Table_Models.tex}


\begin{figure*}[htb!]
\begin{center}
\includegraphics[width=0.4975\linewidth]{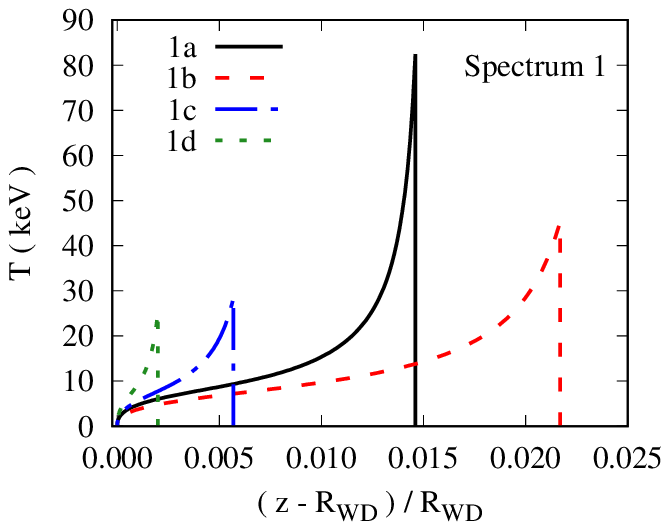}
\includegraphics[width=0.4975\linewidth]{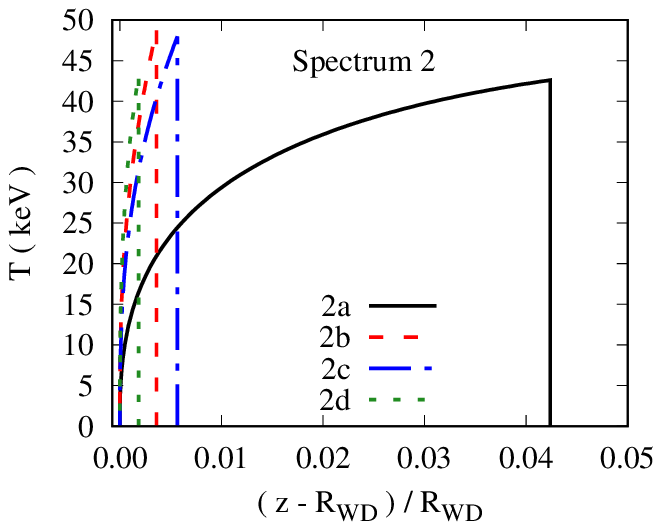}
\end{center}
\caption{PSR temperature distributions of the models listed in Table~\ref{TabMod}, the set \textit{spectrum 1} being shown in the left panels and the set \textit{spectrum 2} in the right panels.
We can clearly see that the profiles of the models are different.
However, the model average temperatures weighted by the squared density, in each set, are rather similar, which helps to explain why they exhibit the same normalized X-ray spectrum.
}
\label{FigPSR7.2}
\end{figure*}


\begin{figure*}[htb!]
\centering
%
%
%
\includegraphics[width=0.99\linewidth]{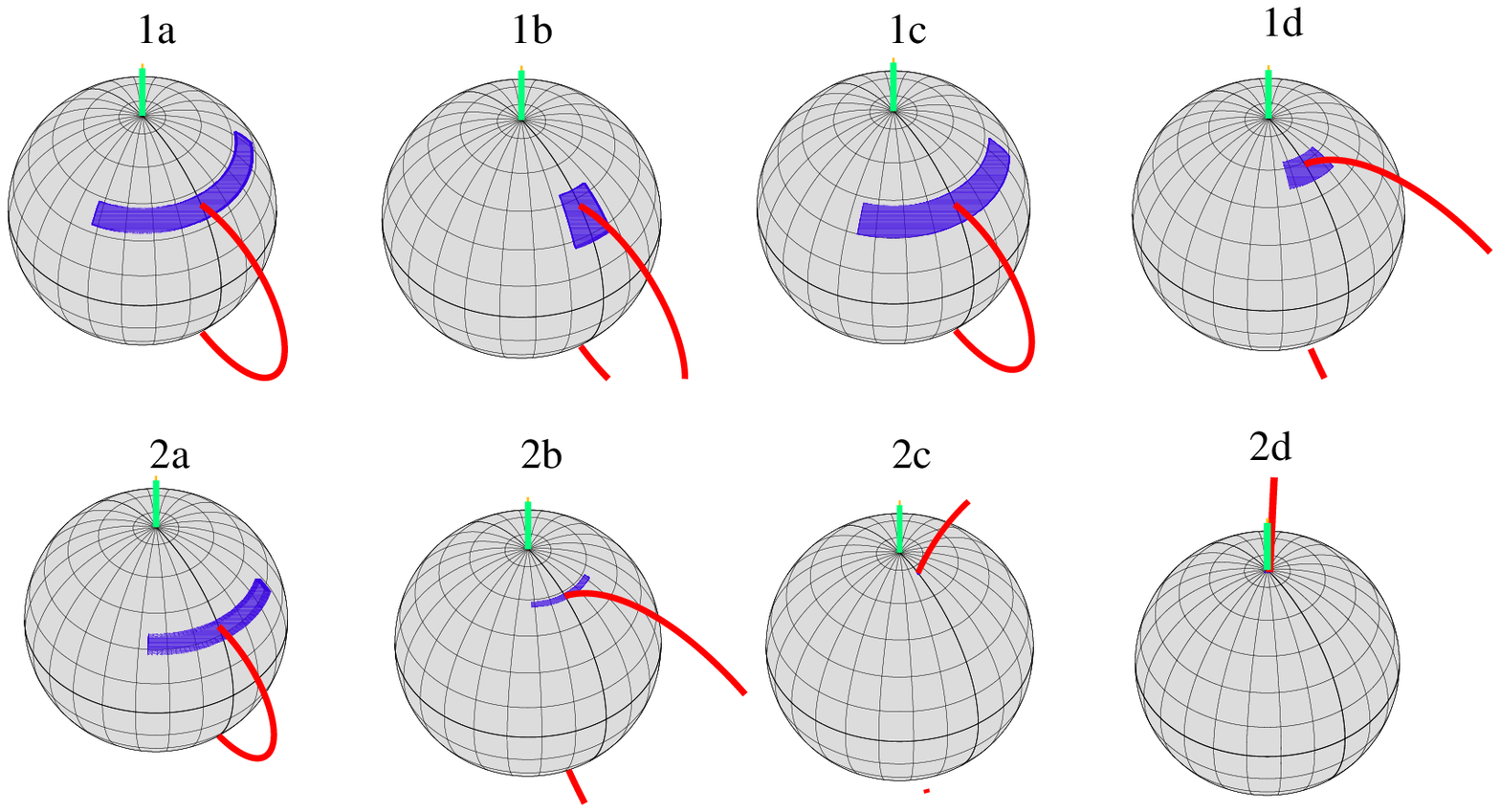}
\caption{Geometry of the models listed in Table~\ref{TabMod}, at the WD rotation phase of 0.1, as seen by the observer.
The PSR is shown in blue and the WD magnetic field axis, which is coincident with the WD rotation axis in these particular models, in green.
The red line indicates the magnetic field line passing through the center of the PSR, which can be thought as the reference field line for the accretion structure.
Notice that the PSR of models 2c and 2d cannot be seen in the figure because their cross-sections are very small.
In addition, all WDs, which have different masses, have the same size.
This does not correctly represent the models, in which the WD radius follows the mass-radius relation.
So this figure is not appropriate to compare the footprint area of the different models.
}
\label{FigModGeometry}
\end{figure*}


In order to illustrate how the degeneracy problem affects X-ray fitting strategies, we consider here two sets of four different models, each set providing the same normalized X-ray continuum spectrum. 
These two spectra are shown in Fig.~\ref{Fig_2sets_degen_spectra} and the properties of the models that produce them are listed in Table~\ref{TabMod}.
We show in Fig.~\ref{FigPSR7.2} the temperature profiles of all these eight models.

By inspecting Fig.~\ref{Fig_2sets_degen_spectra}, we can clearly see that the \textit{spectrum 2} is much harder than the \textit{spectrum 1}.
In this way, we should expect much higher temperatures in the PSRs of the models associated with the \textit{spectrum 2}, which is in fact the case. 
The average temperatures weighted by the squared density in the PSR of the models producing the \textit{spectrum 1} and \textit{spectrum 2} are ${\sim7-8}$ and ${\sim27-30}$~keV, respectively.
In particular, these temperatures are quite similar in each set of models.

Regarding the models 1a, 1b, 1c and 1d, it is quite clear from the left panel of Fig.~\ref{FigPSR7.2} that their profiles are rather different.
An interesting distinction among these models is the shock temperature, which can be different by a factor of four.
Another evident difference is the PSR height, which can vary by a factor of 10.
The very short PSR in model 1d, in particular, is due to the very strong magnetic field strength, which makes the cyclotron cooling substantially enhanced, in comparison with the other models in this set (see Fig.~\ref{FigPSR5.2}).

With respect to the second set of models, the situation is a bit different.
Even though from the right panel of Fig.~\ref{FigPSR7.2} models 2b, 2c and 2d look similar, they are different.
By inspecting Table~\ref{TabMod}, despite the fact that these models have similar shock temperatures, they have considerably different shock heights, specific accretion rates, WD masses and threading region radii.
On the other hand, the threading region in model 2a is much closer to the WD and it has a much smaller specific accretion rate and a much weaker magnetic field, in comparison with models 2b, 2c and 2d, producing a much taller PSR.
Despite that, its shock temperature is comparable to those in the other models, which is likely due to its much higher WD mass.

The geometrical properties of all models listed in Table~\ref{TabMod} are shown in Fig.~\ref{FigModGeometry}, in which the phase 0.1 was chosen.
Despite all these models have the same orbital inclination ($45^{\rm o}$), the same magnetic field latitude ($90^{\rm o}$) and longitude ($0^{\rm o}$), they are geometrically very different.
For instance, the PSR colatitude and the PSR cross-section of all models are substantially different.
The models in the set \textit{spectrum 2} are particularly interesting.
The PSR approaches the magnetic pole, as one moves from model 2a to model 2d. 
This in turn causes the threading region to move farther away from the WD, from model 2a to model 2d.
Interestingly, the PSR cross-sections of models 2c and 2d are so small that we cannot even see those PSRs in the figure.

A hard spectrum like \textit{spectrum 2} can be achieved in several ways.
For instance, spectra associated with high-mass WDs can be very hard, as well as spectra connected with either weak fields or high specific accretion rates.
All models in the set spectrum 2 have relatively high WD masses (${\gtrsim0.9}$~\Msun), relatively high specific accretion rates (${\gtrsim10}$~\gscm), and very diverse magnetic field strengths as well as threading region radii.

Model 2a in this set is particularly interesting, since it has the highest WD mass, the lowest field strength, the lowest specific accretion rate and the smallest threading region radius. These characteristics imply that this model has the tallest and sparsest PSR in this set.
Should this model be more representative of a hypothetical CV exhibiting the spectrum 2, such a CV would be a very peculiar IP, since it would harbor an unusually high-mass WD.

On the other hand, model 2c seems quite close to the properties of most polars, provided its WD mass is consistent with what is found among CVs, its WD hosts a moderately strong magnetic field, it has a low accretion rate and a relatively large threading region radius.
Apart from these few notes, one cannot conclude which of these models better describes a CV exhibiting an X-ray continuum spectrum like the spectrum 2.
To disentangle the models, one would inevitably need to know more properties of such a system and/or to have additional constraints.
In other words, \textit{an X-ray continuum spectrum alone does not tell us much about the properties of a given magnetic CV}.

An example of misleading assumptions in schemes to estimate magnetic CV properties is the modeling of EX~Hya performed by \citet{Luna_2015}, \citet{Suleimanov_2016} and \citet{Suleimanov_2019}.
Given the lack of further constraints, in the former work, the authors assumed the same values for the magnetosphere radius derived in \citet{Revnivtsev_2011} and \citet{Semena_2014}, which is ${\approx2.7}$~\Rwd.
In the first attempt using their break frequency method, \citet{Suleimanov_2016} assumed that this system had a short PSR, given the high specific accretion rates they assumed, and found the WD mass and magnetosphere radius of this system should be ${\approx0.73}$~\Msun~and ${\approx2.6}$~\Rwd, respectively, agreeing in turn with the results achieved by \citet{Revnivtsev_2011} and \citet{Semena_2014}.

However, \citet{Luna_2018} investigated EX~Hya with X-ray data from several satellites and found that it has a tall PSR, with height comparable to its WD radius.
In addition, these authors showed that its magnetosphere is most likely much larger than that predicted by the break frequency method, but smaller than its co-rotation radius.
By knowing that, \citet{Suleimanov_2019} then assumed that EX~Hya has a relatively tall PSR ($0.25$~\Rwd), still smaller than the height estimated from X-ray observations ($\sim$~\Rwd).
In this case, they found a better agreement with the WD mass derived from eclipses, but still significantly smaller.
Their inferred magnetosphere radius is still small (${\approx3}$~\Rwd), though.

Another problem in the analysis of Suleimanov et al., for this particular system, is that the break frequency method used by these authors does not seem to work for EX~Hya.
The Doppler tomograms performed by \citet{Mhlahlo_2007} reveal that this system has a large accretion curtain \citep[see also][]{Norton_2008}, extending to a distance close to the WD Roche lobe radius, while that estimated with the break frequency is at most a few \Rwd.

The example above clearly illustrates how assumptions made in X-ray continuum spectra fitting might become dangerous.
Such assumptions are otherwise needed because of the intrinsic degeneracy problem coupled with the lack of observational constraints, since an X-ray continuum spectrum alone does not provide the information required to properly constrain the parameter space in this sort of fitting scheme.

%% file: TABLES/Table_Models.tex
\begin{table*}[htb!] 
\centering
\begin{center} 
\caption{Parameters of the models discussed in Section~\ref{influDEGmodels}, namely the WD mass $(M_{\rm WD})$, the WD magnetic field strength at the pole  $(B_{\rm p})$, the accretion rate $(\dot{M}_{\rm WD})$, the accretion area at the PSR bottom $(S_{\rm b})$, the specific accretion rate at the PSR bottom $(\dot{m}_{\rm b})$, the PSR colatitude $(\beta)$, the threading region radius $(R_{\rm th})$, the shock height $(H_{\rm sh})$, the shock temperature $(T_{\rm sh})$, the shock density $(\rho_{\rm sh})$ and the average temperature weighted by the squared density $(\langle T \rangle)$. 
For the other \cyclops~parameters, we assumed an orbital inclination of $45^{\rm o}$, a distance of $100$~pc, a magnetic field latitude of $90^{\rm o}$ and a magnetic field longitude of $0^{\rm o}$.
Finally, we assume that the interstellar extinction is negligible.
}
\label{TabMod}
\centering
\setlength\tabcolsep{5.5pt} 
\begin{tabular}{lrrcccrrrccr}
\hline
\addlinespace[0.085cm]
Model & 
\multicolumn{1}{c}{$M_{\rm WD}$} & 
\multicolumn{1}{c}{$B_{\rm p}$} & 
\multicolumn{1}{c}{$\dot{M}_{\rm WD}$} & 
\multicolumn{1}{c}{$S_{\rm b}$} & 
\multicolumn{1}{c}{$\dot{m}_{\rm b}$} & 
\multicolumn{1}{c}{$\beta$}  & 
\multicolumn{1}{c}{$R_{\rm th}$} & 
\multicolumn{1}{c}{$H_{\rm sh}$} & 
\multicolumn{1}{c}{$T_{\rm sh}$}  & 
\multicolumn{1}{c}{$\rho_{\rm sh}$} &
\multicolumn{1}{c}{$\langle T \rangle$} \\
\addlinespace[0.005cm]
 & 
\multicolumn{1}{c}{(\Msun)} & 
\multicolumn{1}{c}{(MG)} & 
\multicolumn{1}{c}{($10^{-10}$\,\Msunyr)} & 
\multicolumn{1}{c}{($10^{16}$\,cm$^{2}$)} & 
\multicolumn{1}{c}{(\gscm)} & 
\multicolumn{1}{c}{($^o$)} & 
\multicolumn{1}{c}{($R_{\rm WD}$)} & 
\multicolumn{1}{c}{($R_{\rm WD}$)} & 
\multicolumn{1}{c}{(keV)}  & 
\multicolumn{1}{c}{($10^{-9}$\,g\,cm$^{-3}$)} &
\multicolumn{1}{c}{(keV)}  \\
\addlinespace[0.085cm]
\hline
\addlinespace[0.2cm]
\multicolumn{11}{c}{\it Spectrum 1} \\
\addlinespace[0.2cm]
1a & $1.35$ &  $29$ &  $~~0.50$ & $1.58$ & $~~0.20$ & $48$ & $1.80$ & $0.0146$ & $83$  &  $~~~~0.92$ & 7.7\\
\addlinespace[0.085cm]
1b & $1.18$ &  $21$ &  $~~0.39$ & $1.66$ & $~~0.15$ & $45$ & $1.99$ & $0.0217$ & $45$  &  $~~~~0.93$ & 7.7\\
\addlinespace[0.085cm]
1c & $1.06$ &  $63$ & $15.85$ & $8.74$ & $~~1.14$ & $50$ & $1.72$ & $0.0057$ & $29$  &  $~~~~9.25$ & 7.4 \\
\addlinespace[0.085cm]
1d & $0.74$ & $141$ & $15.84$ & $4.05$ & $~~2.47$ & $27$ & $4.82$ & $0.0019$ & $25$  & $~~21.46$ & 7.1\\
\addlinespace[0.2cm]
\multicolumn{11}{c}{\it  Spectrum 2}  \\
\addlinespace[0.2cm]
2a & $1.29$ &   $1$ &  $15.85$ & $1.30$ &  $~~7.69$ & $54$ &    $1.52$ & $0.0424$ & $43$  &   $~~45.68$ & 28.9\\
\addlinespace[0.085cm]
2b & $1.03$ &  $68$ &  $63.10$ & $0.68$ & $58.93$ & $28$ &    $4.48$ & $0.0036$ & $49$  &  $365.66$ & 28.5\\
\addlinespace[0.085cm]
2c & $0.94$ &  $26$ &   $~~0.63$ & $0.11$ & $35.54$ & $14$ &   $18.35$ & $0.0057$ & $48$  &  $221.45$ & 29.6\\
\addlinespace[0.085cm]
2d & $0.87$ &  $44$ &   $~~3.98$ & $0.03$ & $95.38$ &  $2$ & $1136.27$ & $0.0019$ & $43$  &  $633.39$ & 27.4\\
\addlinespace[0.085cm]
\hline
\end{tabular}
\end{center}
\end{table*}

%% file: CONTENT/Section05.tex
So far, we have discussed the degeneracy problem, i.e. the degeneracy arising from the amount of parameters to be fitted, in a very general fashion.
Similarly important is to provide possible solutions that can potentially help to break the degeneracy in the parameter space.
The only way to solve the degeneracy problem is by introducing additional constraints to the fitting scheme, so that the models could be distinguished.
We discuss in this section approaches using X-ray data that can break the degeneracy, by focusing on the two sets of four different models introduced in Section~\ref{influDEGmodels}.

\subsection{Breaking the degeneracy with emission lines}
\label{sec_degen_line}

A way of increasing the constraints for particular systems is the inclusion of emission lines in the X-ray spectra, since they provide extra constraints for the parameters of the PSR.
For instance, \citet{Hayashi_2014} showed that the the ratio of the hydrogen-like to the helium-like iron $K\alpha$ lines changes with respect to the specific accretion rate.
This and other line complexes might help to break the degeneracy, since the line emission spectrum strongly depends on the PSR temperature and density distributions \citep[e.g.,][]{Fujimoto_1997,Ezuka_1999,Ishida_1999}.

In addition, one could extract useful information about the shock height from the iron fluorescent line and the Compton hump.
Examples of objects in which this sort of estimate was possible include EX~Hya \citep[][]{Luna_2018} and V1223~Sgr \citep[][]{Hayashi_2011}.
Knowing the PSR height can, to some degree, help in breaking the degeneracy.
This information, for instance, could help to distinguish the models in the set \textit{spectrum~1} and eventually the model 2a from the models 2b--d in the set \textit{spectrum~2} (see Fig.~\ref{FigPSR7.2}).

Therefore, even though incorporating emission lines to the fitting strategy should help, this effect alone seems unable to solve the degeneracy problem.
This is specially true if we take into account that line emission also depends on the elemental abundances, which introduce an additional parameter to the fitting scheme.
Unfortunately, addressing to which extent emission lines help cannot be currently done with the \cyclops~code, since it cannot handle emission lines, but should be verified in future works.

\subsection{Breaking the degeneracy with consistent break frequency estimate}
\label{sec_degen_breakfreq}

For some IPs, accurate hard X-ray observations are or will become available \citep[e.g., through the \nustar~Legacy Survey program,][]{Shaw_2020}.
A property that can sometimes be extracted from the power spectrum in these cases is the break frequency \citep[e.g.,][]{Suleimanov_2016,Suleimanov_2019}, which corresponds to the Keplerian frequency at the magnetosphere boundary \citep[e.g.,][]{Revnivtsev_2009,Revnivtsev_2010,Revnivtsev_2011}.
From that, the magnetosphere radius can be estimated, which decreases the number of free parameters in the fitting scheme.

The main difficulty of this approach is that the break frequency cannot be usually/easily extracted from the data.
For instance, \citet{Shaw_2020} analyzed a sample of 19 magnetic CVs with \nustar~and could estimate the break frequency for only one of them.
For the remaining systems, these authors assumed either spin equilibrium, in which the magnetosphere radius is the co-rotation radius, for the other 16 IPs in their sample, or that the magnetosphere radius was $10^3$~$R_{\rm WD}$, in the case of the two asynchronous polars they analyzed.

We can then conclude by now that estimating the break frequency is not an easy task.
Even worse, it might not necessarily provide an accurate estimate of the magnetosphere radius \citep[see][for a discussion related to EX~Hya]{Luna_2018}.
Therefore, despite the fact that it can be useful, wherever it is possible to apply this method, it is most likely not enough to solve the degeneracy problem, especially considering the multidimensional parameter space involved in magnetic CV modeling, like the one investigated here.

\begin{figure}[tb!]
\begin{center}
\includegraphics[width=0.99\linewidth]{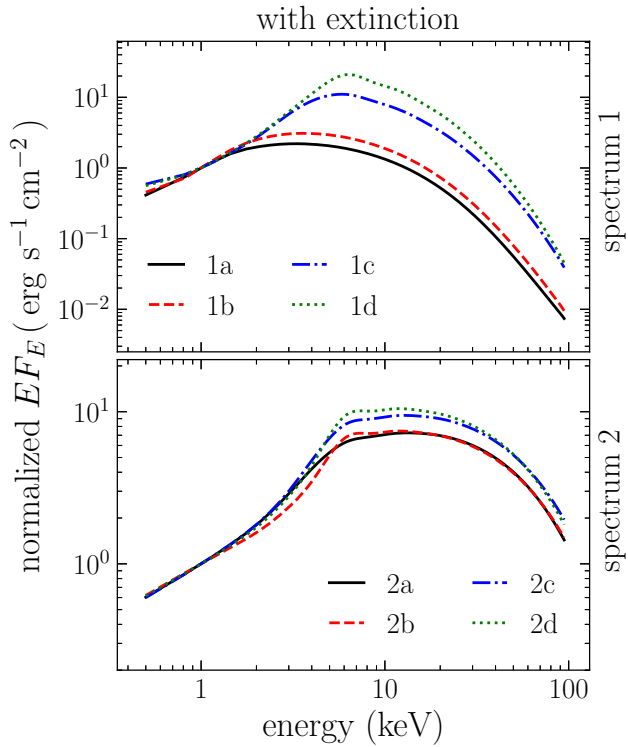}
\end{center}
\caption{Effect of the pre-shock region extinction on the X-ray spectra (normalized to their values at $1$~keV) of the models in the sets \textit{spectrum 1} and \textit{spectrum 2}, which parameters are listed in Table \ref{TabMod}.
The changes in comparison with Fig.~\ref{Fig_2sets_degen_spectra} are due to photoelectric absorption and Compton scattering of a partially ionized material.
The changes in the set \textit{spectrum 1} are significant for the pair of models 1c and 1d, in comparison with the pair of models 1a and 1b.
However, the changes in all models of the set \textit{spectrum 2} are very small.
This suggests that, even though incorporating only pre-shock region extinction to a fitting scheme may help, it is unlikely able to solve the degeneracy problem.
} 
\label{fig_spectra_preshock_extinction}
\end{figure}

\subsection{Breaking the degeneracy with consistent pre-shock region modeling}
\label{sec_degen_preshock}

A spectrum in the \cyclops~code corresponds to the combination of spectra in different rotational phases.
In each phase, the extinction of the pre-shock region is calculated according to the model accretion geometry and inclination.
Therefore, the PSR emission and pre-shock extinction are consistently calculated for a given model.
In particular, \cyclops\ can produce the spectrum resulting of a partial absorption from the pre-shock region, similarly to what is done by the \pcfabs~model of \xspec, which is widely used to reproduce magnetic CV spectra.
However, \cyclops\ has an approach consistent with the 3D adopted geometry for the entire accretion structure (PSR and pre-shock region), which is required to properly model the phase-dependent flux modulation of magnetic CVs.

We illustrate in Fig.~\ref{fig_spectra_preshock_extinction}  how the spectra shown in Fig.~\ref{Fig_2sets_degen_spectra} can be modified when the pre-shock region extinction
is taken into account.
As described in Section~\ref{sec_extinction}, this region is expected to extinguish the PSR emission by photoelectric absorption and Compton scattering. 
The spectra, which are normalized to their values at $1$~keV, were calculated considering a partially ionized pre-shock region that has a constant density equals to one fourth of the density at the shock front.
Regardless of the model, we can clearly see that the pre-shock region can produce non-negligible changes in the spectrum shape, specially for energies below a few keV.
This is not surprising and can be explained by the dependence of the photoabsorption cross-section on the energy, since absorption more strongly affects the soft region of spectrum.
On the other hand, even though Compton scattering, which affects the hard part of the spectrum, does not usually play a significant role in shaping the spectra of the models, it is sufficiently important in models 2b and 2d, reducing in turn the flux in these models.

Regarding the models in the set \textit{spectrum 1}, there are two pairs of models with very similar spectra.
While the spectra of models 1c and 1d are strongly affected by pre-shock region photoabsorption, models 1a and 1b are only moderately affected by this effect.
Most importantly, given the intrinsic uncertainties involved in observational data, it seems very unlikely that the models in each pair could be distinguished, which does not help much in breaking the degeneracy in the set \textit{spectrum 1}.

The spectra of the models in the set \textit{spectrum 2} are very similar, irrespective of the energy range, which means that despite the fact that these models are different, with different accretion geometries, pre-shock region extinction shapes their spectra in a similar way.
In addition, it seems unlikely that, even with high-quality observational data, these models could be unambiguously distinguished.
Therefore, even though incorporating this effect makes the fitting more physically appealing, it does not provide a way to solve the degeneracy problem in this set of models.

We finish this discussion about incorporating pre-shock region extinction in the modeling by emphasizing that it alone most likely cannot break the degeneracy while fitting X-ray continuum spectra.
Indeed, from Fig.~\ref{fig_spectra_preshock_extinction}, despite the fact that absorption and scattering change the spectrum shapes, we still end up with alike spectra, since the role played by these processes can be similar.
For instance, models 1a and 1b have virtually the same spectra, which also happens for models 1c and 1d, and for all models in the set \textit{spectrum 2}.

We can conclude by now that adding pre-shock region extinction as an extra ingredient to a fitting strategy, even though being physically required, does not seem enough to solve the degeneracy problem.
Therefore, additional constraints are most likely still needed in order to unambiguously distinguish the models.

\begin{figure*}[tb!]
\begin{center}
\includegraphics[width=0.99\linewidth]{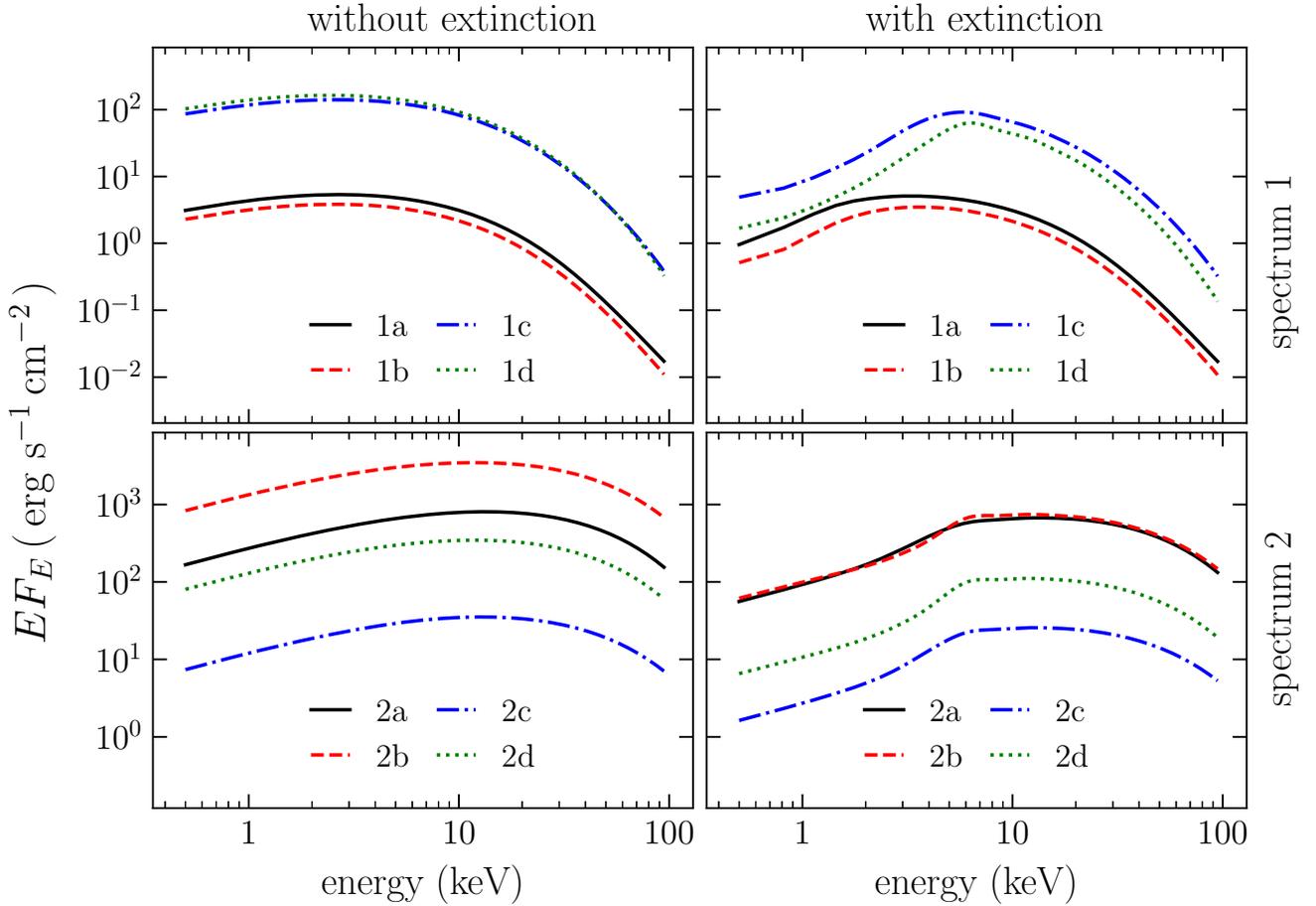}
\end{center}
\caption{Effect of distance on the X-ray spectra of the models in the set \textit{spectrum 1} (top panels) and \textit{spectrum 2} (bottom panels).
In the spectra, which were obtained using the parameters listed in Table \ref{TabMod} and assuming a distance equals to $100$~pc, we considered two cases, namely including pre-shock region extinction (right panels) and completely neglecting this effect (left panels).
We can see that the degeneracy in the set \textit{spectrum 1} most likely cannot be broken, since the pair of models 1a and 1b, as well as the pair of models 1c and 1d, still have alike spectra, regardless of whether we incorporate pre-shock region extinction or not.
Regarding the models in the set \textit{spectrum 2}, the fluxes are different in the case of no pre-shock region extinction, but when this effect is included, the pair of models 2a and 2b have virtually identical spectra.
These examples illustrate that even though incorporating \emph{Gaia} distances and consistent pre-shock region modeling to a fitting scheme helps to some extent to break the degeneracy in some cases, they are likely not enough to unambiguously solve the degeneracy problem.
}
\label{fig_spectra_absolute_flux}
\end{figure*}

\subsection{Breaking the degeneracy using accurate distance}
\label{sec_degen_distance}

Along the PSR, the \brem~emissivity depends on the electron number density and the temperature.
In the optically thin case, which is valid for the X-ray emission of the PSR, the total emission is the integral of the emissivity over the PSR volume.
This implies that the `observed' \brem~flux of a model at a fixed distance depends entirely on its PSR properties (see Eqs.~\ref{NormXSPEC} and \ref{NormCYCLOPS}).
This means that, for models having different PSRs, such as the models in the sets \textit{spectrum 1} and \textit{spectrum 2} (see Fig.~\ref{FigPSR7.2}), even though they exhibit virtually identical normalized X-ray spectra, the X-ray fluxes considering the PSR at a same fixed distance could be in principle different.
We would like to emphasize that, as shown in Appendix~\ref{app_compxspec}, the fluxes provided by the \cyclops\ code are in very good agreement with those from \xspec.

In the context of magnetic CV fitting strategies, by knowing the distance to the investigated system, one can put all analyzed models at the same distance and compare their fluxes.
In other words, if an accurate distance estimate for the investigated system exists, e.g. using the accurate parallaxes from the \emph{Gaia} satellite, the non-normalized theoretical spectra can be directly compared with the observed one.
That said, even though many models can have very similar spectrum shapes, i.e., virtually identical normalized spectra, such spectra could be separated when the distance is incorporated to the fitting scheme.

We shall add to the discussion that, with the accurate distances from the \emph{Gaia} satellite, deriving X-ray luminosities is possible, which in turn allows us to estimate the accretion rate.
For example, \citet{Suleimanov_2019} managed to estimate accretion rates for a large fraction of IPs and found that  most systems accrete at  rates of ${\sim10^{-9}}$~\Msun~yr$^{-1}$, which is consistent with what is expected from CV secular evolution modeling, provided their orbital periods (usually ${\gtrsim3}$~hr) and their (usually) negligible WD magnetic moments.

We show in Fig.~\ref{fig_spectra_absolute_flux} the non-normalized X-ray spectra of the models in the sets \textit{spectrum 1} and \textit{spectrum 2}, assuming a distance of $100$~pc for all models, and taking into account or not the pre-shock region extinction.
Regarding the models in the set \textit{spectrum 1}, irrespective of whether we take into account or not the pre-shock region extinction, models 1a and 1b as well as models 1c and 1d have virtually identical spectra.
This is likely because the accretion rates and the accretion area, and in turn, the specific accretion rate of the models in each of these two sets are similar (see Table~\ref{TabMod}).
In particular, in models 1a and 1b, ${\dot{m}_{\rm b}=0.20}$ and $0.15$~\gscm, respectively, while in models 1c and 1d, ${\dot{m}_{\rm b}=1.14}$ and $2.47$~\gscm, respectively.
This implies that models 1c and 1d have denser PSRs than models 1a and 1b. Moreover, the four models have similar volumes, so the different emission levels can be understood only by the different densities, since the \brem~emissivity depends on the squared density.

Comparing now the case in which the pre-shock region extinction is included with the case in which this effect is ignored, we can clearly see that the hard part of the spectra in this set of models is barely affected.
On the other hand, as already stated before, the soft flux is reduced quite a bit, especially for models 1c and 1d, which is likely because these models have denser pre-shock regions, in comparison with models 1a and 1b.

The models in the set \textit{spectrum 2} are more affected when we incorporate not only distance, but also pre-shock region extinction.
Starting with the case without pre-shock region extinction, we can see that, despite the fact that the shape of the spectra of all the models in this set are virtually identical, they can be separated when they are put at the same distance.
Unlike the models in the other set, those in this set have very different accretion rates and accretion areas, which implies that they are characterized by different specific accretion rates.
Then, their PSRs have different density distributions, although they have similar temperature distributions.
This difference in the density distribution combined with very different emitting volumes is likely the reason why they emit in a different way when put at the same distance.

The model 2a is particularly interesting and illustrates another dependence of the luminosity of the PSR.
This model has the tallest and sparsest PSR as well as the smallest specific accretion rate in this set of models, and at the same time emits more than models 2c and 2d.
This happens because, even though the specific accretion rate might be an important indicator of the X-ray emission intensity, it is not the only parameter affecting the flux.
The volume of the emitting region is also relevant and most likely explains why this model emits more than the others, but model 2b.

Regarding the impact of the pre-shock region extinction on the models of the set \textit{spectrum 2}, as already stated previously, given the dependence of the photoabsorption cross-section on the energy, the soft part of all those models are moderately affected by pre-shock region absoption.
However, unlike for the other set of models, in which the pre-shock region extinction does not significantly affect the flux in the hard part, the situation for this set is different, the hard flux in models 2b and 2d being strongly reduced due to Compton scattering.

We shall finish this discussion about the distance by emphasizing that by incorporating it alone to the fitting scheme, or even coupled with consistent pre-shock region modeling, most likely cannot break the degeneracy while fitting an X-ray continuum spectrum.
We have shown in Fig.~\ref{fig_spectra_absolute_flux} that, even though we could distinguish the models emitting different flux in some cases, we can still have in the end virtually identical spectra.
In the particular set of models investigated here, models 1a and 1b have very alike spectra, as well as models 1c and 1d, irrespective of whether pre-shock region extinction is taken into account or not.
In addition, when this effect is considered, models 2a and 2b end up with virtually identical spectra.

We can then conclude that including only accurate distance estimates as well as pre-shock region modeling in a fitting strategy is most likely not enough to undoubtedly solve the degeneracy problem in a general situation.
Therefore, similarly to what we have already found, additional constraints other than distance and pre-shock region extinction are most likely still needed in order to unambiguously distinguish the models.
In what follows, we discuss how incorporating X-ray light curves, in different energy ranges, can potentially solve the degeneracy problem, provided consistent pre-shock region extinction is also included in the modeling.

\subsection{Breaking the degeneracy with X-ray light curves}
\label{sec_degen_lx}

Given the characteristics of X-ray observations, the same data set can be used to produce spectra and light curves.
Hence, the useful constraints provided by light curves does not require any additional observing time, only additional data reduction.
Of course, this is valid if the data set can provide light curves with enough signal-to-noise ratio.

The zero phase of the light curves corresponds to the WD spin phase in which the meridian plane that contains the center of PSR also contains the observer's direction, i.e., the radial vector passing through the center of the PSR and the line-of-sight define a plane perpendicular to the plane of the sky.
In addition, for the sake of simplicity, throughout this section all light curves have been normalized to their maximum fluxes.
Moreover, while discussing the light curves, we adopt the following definition of pulsed fraction

\begin{equation}
{\rm PF} \ = \ \frac{F_{\rm max} \ - \ F_{\rm min}}{F_{\rm max}} \ = \ 1 \ - \ \frac{F_{\rm min}}{F_{\rm max}} \,,
\end{equation}
\

\noindent
where $F_{\rm max}$ and $F_{\rm min}$ are the maximum and minimum fluxes of the light curve, respectively.

The modulation of the X-ray flux with the WD rotation in magnetic CVs can have two origins, namely \textit{self-eclipse} and \textit{phase-dependent pre-shock region extinction}.
The former might happen for many different geometrical configurations because the PSR can be partially or fully hidden by the WD at several rotation phases, which causes a drop in the observed flux at those phases.
Hereafter, we use the term \textit{self-eclipse} to represent full or partial PSR occultation by the WD. 
On the other hand, depending on the geometrical properties of the system, the PSR can be partially or fully eclipsed by the pre-shock region, which in turn may strongly extinguish the flux from the PSR.
We start our discussion by focusing on the first effect, i.e., we initially ignore the existence of the pre-shock region.

\subsubsection{Impact of self-eclipse}

From the above-mentioned definition of phase zero adopted in this work, the maximum occultation must occurs at phase $0.5$.
In addition, the shapes and depths of such minima in the light curve are mainly dependent on the system geometrical properties such as the orbital inclination, the PSR colatitude and the PSR cross-section.
Finally, the flux must be constant (PF~${\sim0}$~\%) when the PSR can be fully seen by the observer in all rotation phases.

\begin{figure*}[htb!]
\begin{center}
\includegraphics[width=0.99\linewidth]{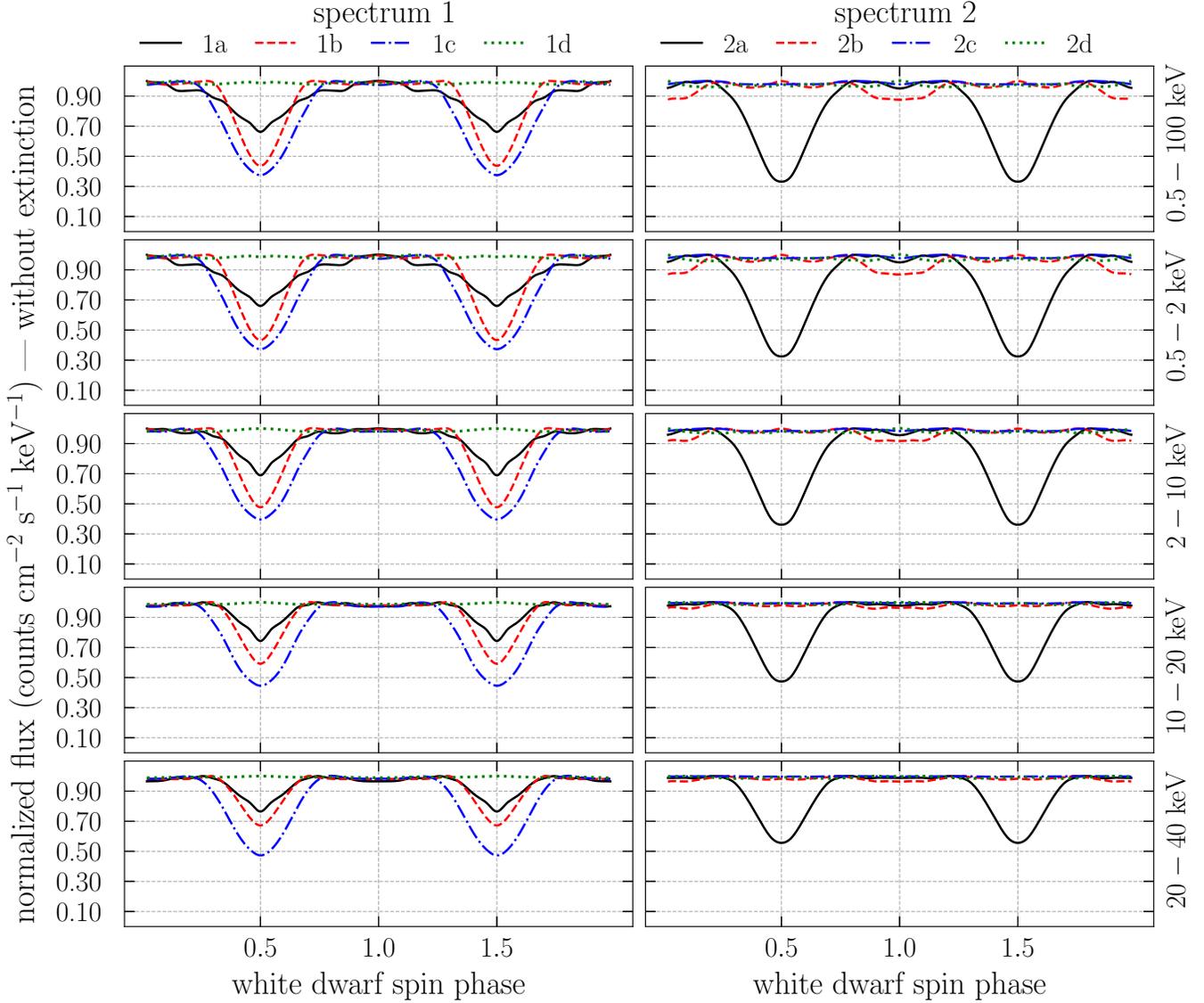}
\end{center}
\caption{X-ray light curves, normalized to their maximum fluxes, including only the effect of the partial self-eclipse of the PSR by the WD from the two sets of models listed in Table~\ref{TabMod}, the set \textit{spectrum 1} shown in the left column and the set \textit{spectrum 2} in the right column, in five energy bands (in keV), namely $0.5-100$ (first row), $0.5-2$ (second row), $2-10$ (third row), $10-20$ (fourth row) and $20-40$ (fifth row).
Despite the fact that no pre-shock region extinction is considered here, we can see that the light curves of each model in the set \textit{spectrum 1} are different from one another, which helps to distinguish them.
On the other hand, the light curves of the models in the set \textit{spectrum 2} exhibits no modulation, except for model 2a.
These different features happen because of the different geometrical characteristics of all these eight models, which lead to different modulations due to self-eclipse.
Given the absence of modulation in half of the models investigated here, we can then conclude that X-ray light curves, without consistent pre-shock region modeling, is likely unable to break the degeneracy existent when only the X-ray spectrum is taken into account.
}
\label{Fig_xr_lc_se}
\end{figure*}

The self-eclipse light curves of the models listed in Table~\ref{TabMod} are shown in Fig~\ref{Fig_xr_lc_se}.
We present the light curves in five energy intervals, namely $0.5-100$ (integrated energy range), $0.5-2$, $2-10$, $10-20$, and $20-40$~keV. 
Models 1a, 1b and 1c in the set \textit{spectrum 1} exhibit strong modulation due to partial self-eclipses (see Fig.~\ref{FigModGeometry}, which illustrates each model geometry).
The PF varies from ${\sim0}$~\% (model 1d) to ${\sim60}$~\% (model 1c), which implies that no model in this set presents total self-eclipse.
Model 1d is a typical example of systems not undergoing self-eclipse, which have a light curve characterized by a constant flux (PF~${\sim0}$~\%) irrespective of the rotation phase and energy range.

Regarding the set \textit{spectrum 2}, only model 2a exhibits partial self-eclipse, while all other models have PF~${\sim0}$~\%.
This happens because the PSRs of models 2b -- 2d are located very near the WD rotation pole and the inclination is low enough to avoid occultation of the PSR at any WD spin phase.
We would like to draw the readers attention to the fact that the slight oscillations of the fluxes of models 1a, 1d, and 2b -- 2d are not real.
This happens because the desirable level of precision in the \cyclops~code is not always readily available, due to the finite spatial resolution of its 3D grid.

\begin{figure*}[htb!]
\begin{center}
\includegraphics[width=0.99\linewidth]{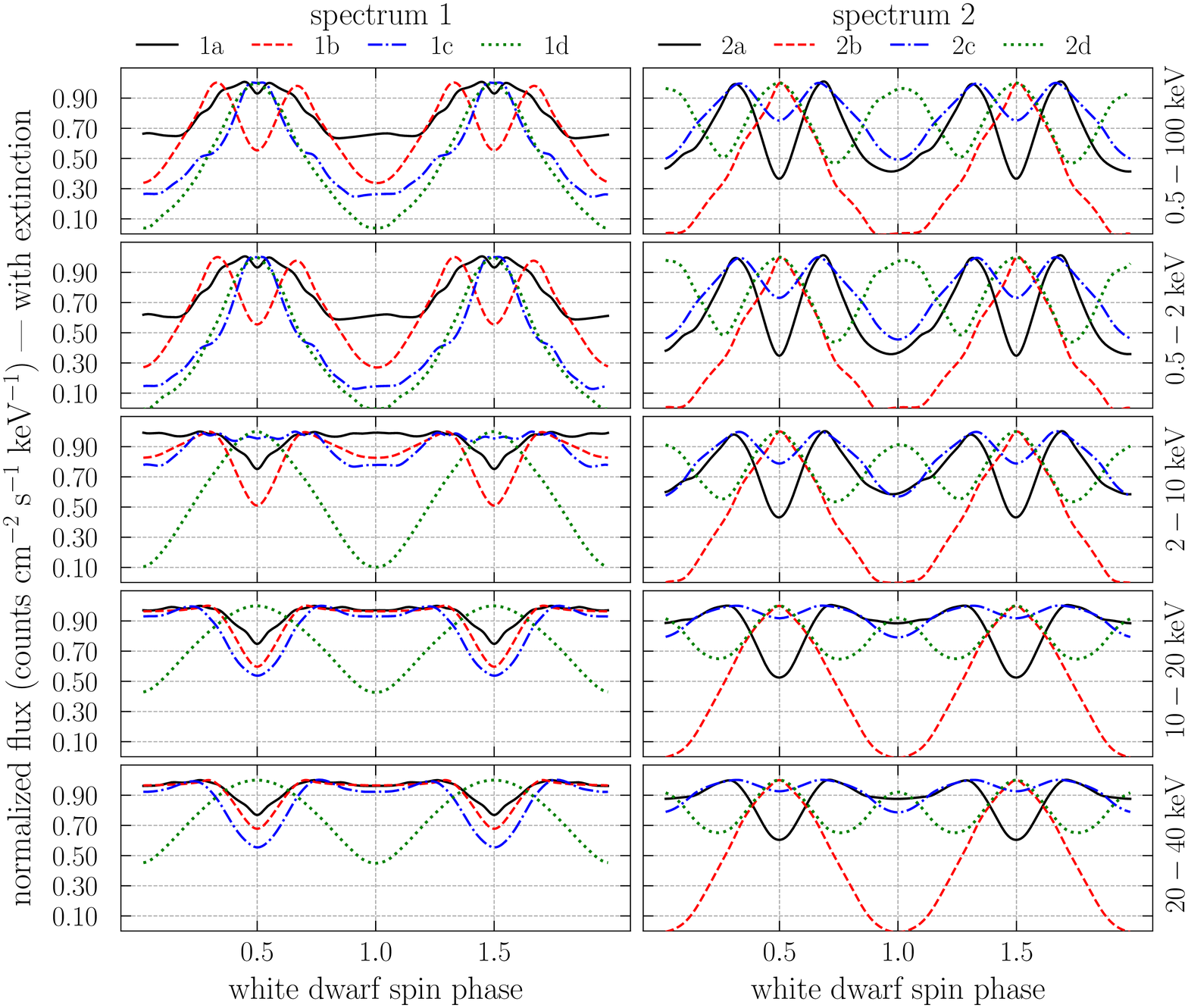}
\end{center}
\caption{X-ray light curves, normalized to their maximum fluxes, including the partial self-eclipse of PSR by the WD from and the pre-shock extinction for the two sets of models listed in Table~\ref{TabMod}, the set \textit{spectrum 1} being shown in the left column and the \textit{spectrum 2} in the right column, in five energy bands (in keV), namely $0.5-100$ (first row), $0.5-2$ (second row), $2-10$ (third row), $10-20$ (fourth row) and $20-40$ (fifth row).
Even though some models have very similar light curves in some energy ranges, clearly, when several energy intervals are considered simultaneously, the light curves of each model investigated here are different from one another.
This then allows us to unambiguously distinguish the models and, in turn, solve the degeneracy problem that arises when only the X-ray spectrum is taken into account.
}
\label{Fig_xr_lc_ext}
\end{figure*}

\subsubsection{Impact of the pre-shock extinction}

After discussing how self-eclipse can modulate X-ray light curves, we now investigate the impact of incorporating the pre-shock region extinction into the X-rays light curves. 
Before proceeding further, we shall remind the readers about a couple of important aspects related to the pre-shock region extinction (see Section~\ref{sec_extinction} for more details on how this is treated in \cyclops).

First, in the models investigated here, i.e., those described in Table~\ref{TabMod}, we assume that the material is considered half neutral and half ionized, and the pre-shock region has a constant density equals to one fourth of the density at the shock position (see Eq.~\ref{rhoff}), which definitely maximizes the effect of the extinction of the pre-shock region on the observed emission.
Second, the pre-shock region extinction is proportional to the density integrated along the line-of-sight, and can be optically thick for the processes and parameters considered here, unlike the \brem~emission, which is always optically thin.

We show in Fig.~\ref{Fig_xr_lc_ext} how the X-ray light curves shown in Fig.~\ref{Fig_xr_lc_se} change by the inclusion of the pre-shock region extinction.
While the self-eclipse causes the minimum fluxes to occur at phase $0.5$, the minima due to the pre-shock region extinction happens at phase $0.0$, except for model 2d. 
As explained earlier, the self-eclipe minimum is expected at phase $0.5$ because at this phase the PSR is just behind the WD, as seen by the observer, which causes the largest possible occultation of the PSR by the WD.
On the other hand, the pre-shock-region-extinction mininum is expected to occur when the pre-shock region is just in front of the WD at phase $0.0$, i.e.,  between the WD and the observer, since this configuration leads to the largest possible extinction in most cases.

The double-trough feature caused by a combination of self-eclipse and pre-shock region extinction can be seen in the soft light curves of two models in which partial self-eclipse occurs, namely models 1b and 2a.
Interestingly, this feature is very similar to what was found by \citet[][their fig.~1]{Hoogerwerf_2006} in the EX~Hya light curve for Mg XI.
On the other hand, the self-eclipse minima in the soft light curves of models 1a and 1c cannot be clearly seen in Fig.~\ref{Fig_xr_lc_ext}.
However, in the hard light curves of the four above-mentioned models, we mainly see the minima caused by self-eclipse, since the extinction of their pre-shock regions is virtually negligible in the hard part.
This implies that in these cases there is a change of $0.5$ cycle in the light curve minimum as the energy increases, since the minima moved from phase $0.5$ to phase $0.0$ by the inclusion of the pre-shock region extinction.
Moreover, the light curves of models 1d and 2b are characterized by a single-trough feature, which is caused by the pre-shock region extinction.

Even though there is no partial self-eclipse in model 2c, since its PSR can in principle be always seen by the observer (see Fig.~\ref{FigModGeometry}), its light curve exhibits a double-trough feature.
This happens because the PSR is seen by the observer behind the pre-shock region in all phases.
That said, the additional trough at phase $0.5$ seen in the light curve of model 2c is a direct consequence of the phase-dependent pre-shock region extinction.
In addition, if the extinction is large enough, the self-eclipse modulation can be washed out from the light curves, which is the case for models 1a and 1c.
Finally, we shall mention that pre-shock region extinction can modify the depth of the minima caused by self-eclipe in double-trough light curves, like those of models 1b and 2a.

Regarding the minimum shapes, we first note that the larger the PSR footprint, the larger the geometrical cross-section of the entire accretion structure, which also regulates the X-ray flux modulation with the WD rotation phase due to the pre-shock-region extinction.
For example, models 1a and 1c have a relatively wider PSR than model 1b and 1d, and consequently, flatter minima. 
In addition, for model 2a, the minimum exclusively due to the pre-shock region extinction is flatter than that mainly caused by self-eclipse.

\begin{figure*}[htb!]
\begin{center}
\includegraphics[width=0.99\linewidth]{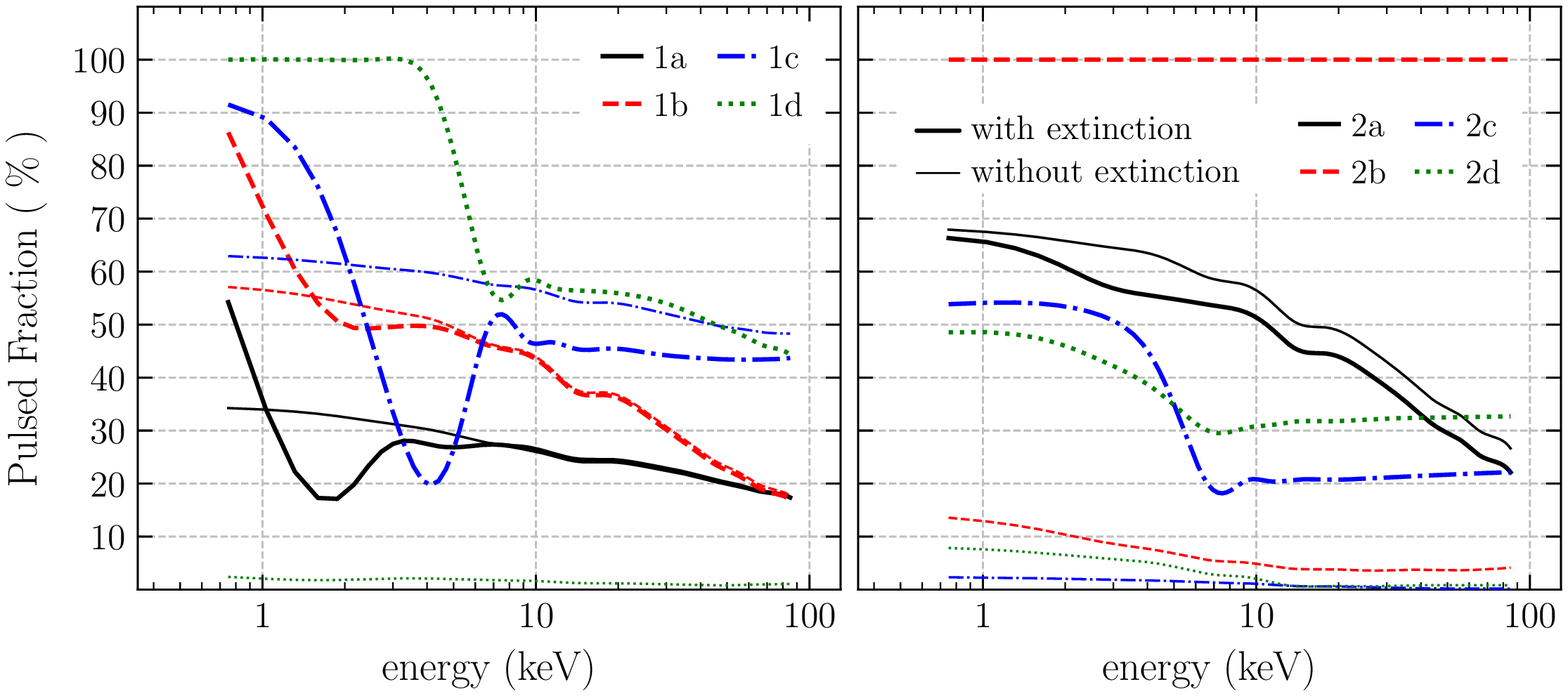}
\end{center}
\caption{Dependence with energy of the pulsed fraction obtained from the X-ray light curves, for the set of models \textit{spectrum 1} (left panel) and \textit{spectrum 2} (right panel), which are listed in Table~\ref{TabMod}, for the case in which the extinction of the pre-shock region is considered (thick curves) and that in which this effect is ignored (thin curves).
Clearly, similarly to what we found while analyzing the X-ray light curves in only five energy intervals, the pulsed fraction in all models differently varies with energy when the entire X-ray range is considered.
In addition, when the pre-shock region is ignored, the pulsed fraction always decreases as the energy increases, since the modulation in the light curves in this case is exclusively due to self-eclipse. 
However, when this effect is taken into account, irrespective of the energy, the pulsed fraction is always greater than zero, which implies that the X-ray light curves are modulated in the entire X-ray wavelength range.
This further strengthen our claim that X-ray light curves, in different energy intervals, are most likely enough to unambiguously solve the degeneracy problem in X-ray spectral modeling, provided the extinction of the pre-shock region is properly modeled.
}
\label{Fig_xr_pf}
\end{figure*}

\subsubsection{Variation of the pulsed fraction with energy}

So far, while discussing on the X-ray light curves, we have not considered in details how they vary with energy.
When different energy ranges are taken into account, the flux modulation is expected to be very different in the soft and hard parts.
A way of looking at this is by means of how the pulsed fraction varies with energy, considering tens of narrow energy intervals, which is done in Fig.~\ref{Fig_xr_pf}, for the models in the sets \textit{spectrum 1} and \textit{spectrum 2}.
Before proceeding further, we shall remind the readers that the non-null pulsed fractions seen in models 1d and 2b--d for the case in which the pre-shock region is ignored are not real.
This is because the X-ray light curves of these models are not modulated by self-eclipse.
Such non-null pulsed fractions are due to finite spatial resolution of the 3D grid in the \cyclops~code.

Starting with the self-eclipse case, we can see in Figs.~\ref{Fig_xr_lc_se} and \ref{Fig_xr_pf} (thin lines) that there is a decrease in the pulsed fraction as the energy increases.
For instance, for model~2a, while the pulsed fraction is ${\sim65-70}$~\% for energies below $2$~keV, it is reduced to ${\sim30-40}$~\% for energies above $20$~keV.
This result can seem weird at first glance, since self-eclipse is a geometrical effect.
However, it is an expected result because the temperature in the PSR is highest at the shock position and drops toward the WD surface (see Fig.~\ref{FigPSR7.2}).
This implies that the region contributing more to the soft X-ray emission is closer to the WD, and therefore can be occulted even in situations in which the top of the PSR is visible by the observer. 
Therefore, for a given geometry in which partial self-eclipse occurs, the lower the energy, the larger the hidden portion of the PSR contributing to that particular energy.

When we incorporate pre-shock region extinction, the situation can easily become far more complicated, as illustrated in Fig.~\ref{Fig_xr_lc_ext} and \ref{Fig_xr_pf} (thick lines).
Even though the above-described dependence with energy is still expected in most models, it is clear from the figures that it can be broken in the presence of pre-shock region extinction.
For instance, model 2b does not exhibit self-eclipse and its extinction is total, i.e. its pulsed fraction is ~${\sim100}$~\%, regardless of the energy range, which means that its pre-shock region is optically thick even at hard energies.
In any event, except for this model, we can see in Fig.~\ref{Fig_xr_pf} that the pulsed fraction of all other models is clearly larger in the soft part than in the hard part.

Regarding the models in which self-eclipse occurs (i.e., 1a, 1b, 1c and 2a), the pre-shock region extinction strongly affects their light curves at energies lower than a few keV.
In addition, the pulsed fraction curves of models 1a and 1c exhibit a trough below $10$~keV.
As the energy increases, after reaching such a trough, the pulsed fraction increases until reaching a peak, and finally decreases again.
On the other hand, for models 1b and 2a, the pulsed fraction smoothly decreases with energy, without a trough at energies lower than $10$~keV.

As previously explained, the pre-shock region extinction in these models is very large at low energies, which somehow blurs or even erases the self-eclipse modulation.
However, when the impact of the extinction becomes much less pronounced (models 1c and 2a) or negligible (models 1a and 1b) at high energies, the self-eclipse modulation dominates, explaining why the pulsed fraction, when extinction is included, is much closer to the pulsed fraction originated from self-eclipse.
For models 1a and 1b, in particular, the extinction of the pre-shock region is virtually negligible at energies higher than a few keV, which makes their pulsed fractions due to self-eclipse and extinction to converge in this energy range.
On the other hand, the pre-shock region extinction of models 1c and 2a is still sufficiently important to affect the light curves in the entire X-ray range, which decreases the pulsed fraction in comparison with the self-eclipse case, irrespective of the energy.

Finally, we note that Fig.~\ref{Fig_xr_lc_ext} shows that this energy-dependent behavior of the pulsed fraction is accompanied by a change of $0.5$ cycle in the light curve minimum, for models 1a and 1c.
On the other hand, for models 1b and 2a, as the energy increases, the light curves lose the double-trough feature, which is caused by a combination of self-eclipse and pre-shock region extinction, and becomes characterized by a single trough, which is due to self-eclipse.

We now discuss the above results in the context of observations of magnetic CVs.
The pulsed fraction in these systems is constant or decreases with energy (e.g., \citealt{CotiZelati_2016}; \citealt{Tomsick_2016}).
The standard explanation for this behavior is the presence (or not) of photoelectric absorption in the mass flow between the stars.
Despite not having explored the entire parameter space, the few examples we have shown here are enough to demonstrate that the self-eclipse of the PSR can also produce this behavior.
Specifically, the partial self-eclipse of the PSR produces a steady and smooth decrease of the pulsed fraction with energy. 
More importantly, this effect is intrinsic to a 3D PSR, depending only on the viewing angle of the system, and can be correctly estimated considering the exact same parameters used to calculate the PSR emission itself.

On the other hand, to calculate the pre-shock extinction it is necessary to adopt a mass distribution in the pre-shock region, which is a far less understood subject.
This effect also produces a decrease in the pulsed fraction with increasing energy, but with a more complex behavior. 
In particular, the \chandra~and \nustar~pulsed fraction behavior of EX~Hya \citep{Luna_2018} resembles the curves that consider the pre-shock extinction shown in Fig.~\ref{Fig_xr_pf}.
In summary, both effects (i.e., self-eclipe and pre-shock region) should be consistently taken into account to correctly understand the amplitude of X-ray light curves of magnetic CVs.

%% file: CONTENT/Section06.tex
We have discussed throughout Section~\ref{degeneracymethods} the required data and the methods able to help to break the degeneracy in the parameter space of magnetic CVs.
In particular, we have shown that even though all of them are expected to help to solve the degeneracy problem, some of them alone do not seem enough to do the job in a more general situation.
For instance, even though incorporating either consistent pre-shock region modeling or accurate distance estimates helps, several models in the sets \textit{spectrum 1} and \textit{spectrum 2} still end up with very similar X-ray spectra.
In addition, very specific constraints required in any attempt of breaking the degeneracy are not always available, such as the PSR height or the break frequency.

However, X-ray light curves correspond to a very fruitful way out of this problem.
We have shown in Figs.~\ref{Fig_xr_lc_ext} and \ref{Fig_xr_pf} that the X-ray light curves of the four models in each set investigated here are substantially different.
In particular, their modulations and location/amplitude of their maxima and minima are substantially different from one another.
This implies that X-ray light curves could be promisingly used to solve the degeneracy problem.
Additionally, we showed in these figures normalized X-ray light curves to better visualization of their shapes.
With the \emph{Gaia} distances, we could disentangle the models even further by comparing their fluxes from the same distance, which is important for models having alike light curve shapes.

The method we propose that is most likely able to solve the degeneracy problem is based on the utilization of only X-ray data in the fitting scheme.
From the same X-ray data, not only the continuum spectrum can be obtained, but also light curves in different energy ranges.
More specifically, by incorporating several soft and hard light curves to the fitting scheme, one should be able to disentangle models with similar continuum spectra as well as similar light curves in some energy ranges.
In addition to the continuum spectrum and the light curves, the inclusion of emission lines should help even further to solve the degeneracy problem.
That said, the information required to break the degeneracy is most likely hidden in the complete X-ray data, which includes the spectrum (continuum plus emission lines) coupled with the light curves in different energy ranges.

However, while comparing Figs.~\ref{Fig_xr_lc_se} and \ref{Fig_xr_lc_ext}, it is quite clear that this method is expected to work only if the pre-shock region is properly modeled, i.e., if the pre-shock region extinction is properly taken into account, since this region plays a key role in shaping the light curves not only in the soft part but also in the hard part.
It is very likely that many models not characterized by self-eclipse could provide X-ray spectra consistent with the observed one.
Therefore, those models could not be disentangled since no modulation would show up in their light curves.
Another problem is that not always the signal-to-noise ratio is high enough so that light curves could be built from the X-ray data.
In those cases, unfortunately, light curves do not help and additional constraints other than those provided by the X-ray data would be most likely still required (e.g., optical light and polarization curves).

The  above-described method we propose here seems the most suitable for a more general approach, provided it is relatively easy to use X-ray light curves in magnetic CV fitting strategies, since they could be easily obtained from the same data used to generate the spectrum (i.e., no additional data is required).
In addition, none of the other available methods can break the degeneracy in a parameter space composed of more than a few parameters.
Therefore, currently our approach appears the most adequate way out of this issue, given its simplicity and power.

We would like to draw the readers attention to one important fact.
X-ray spectra are mainly related to the physical properties in the PSR, i.e. its temperature and density distributions, which could be combined in many different ways to provide virtually identical spectra.
On the other hand, even though the geometrical properties of the PSR are less relevant for determining the emitted X-ray spectrum, they strongly affects the light curves, irrespective of the wavelength.
On top of that, the overall magnetic CV emission modeling not only involves the physical and the geometrical properties of the PSR but also the parameters of the system itself, such as the WD mass and the accretion rate.

Therefore, to warrant correct understanding of a particular magnetic CV, an essential ingredient of any fitting scheme is to properly model the 3D PSR and its impact on the main observables, such as X-ray spectrum and light curves.
To the best of our knowledge, the \cyclops~code is the only one currently available that is able to handle this kind of analysis, which makes it a powerful and sophisticated tool to explore magnetic CV emission.

%% file: CONTENT/Section07.tex
An updated version of \cyclops~code \citep{Costa_Rodrigues_2009,Silva_2013} to model X-ray spectra and multiwavelength light curves of magnetic CVs was presented.
Despite the power of the \cyclops~code in fitting magnetic CV observables as well as the many physical processes it takes into account, we would like to stress that some potentially important issues are not included in this version of \cyclops~and further developments of the code could address them.


Regarding the PSR model, our currently approach assumes one single fluid with equipartition between ions and electrons.
Despite this is a reasonable assumption in many cases, there are at least two situations, for massive WDs, in which this assumption is no longer plausible, i.e. in which the electron-ion equipartition time-scale is longer than the total cooling time-scale.
For very low specific accretion rates, the low density within the PSR can easily make energy exchange between electrons and ions inefficient \citep[e.g.,][]{Hayashi_2014}, which would imply relatively long electron-ion equipartition time-scales.
In addition, for very strong magnetic fields, the cooling due to cyclotron emission is enhanced, which might cause a strong reduction in the total cooling time-scale \citep[e.g.,][]{Sarty_2008}.
Such potentially important cases cannot be properly investigated with the current version of the \cyclops~code.
Therefore, by upgrading the code from a one-fluid to a two-fluid approach would make \cyclops~capable of investigating virtually all types of PSRs, regardless of their parameters.


With respect to the boundary conditions (Eqs.~\ref{vff}, \ref{rhoff}, \ref{BVPCon1}, \ref{BVPCon2}), our model assumes the standard strong shock conditions coupled with the flow reaching zero temperature at the WD surface. 
The latter assumption is most likely not correct and a more realistic approach would be assuming that the flow converges to the photospheric temperature, or even to that of a locally heated photosphere \citep{Cropper_1998}.
This is definitely problematic with respect to the emission line complex originating from regions very close to the WD photosphere, where the temperature is much lower than in the remaining PSR.
That said, a way to improve our current model is by relaxing the boundary conditions so that they better match the real physical conditions expected in magnetic CV PSRs.


Regarding the PSR cooling processes (Eq.~\ref{cool}), we neglect cooling due to Compton scattering and electron conduction as well as cooling from emission lines.
Compton scattering is expected to change only very slightly the temperature and density profiles in the PSR \citep[e.g.,][their fig. 3]{Suleimanov_2008}.
On the other hand, electron conduction may be important at the PSR bottom \citep[e.g.,][]{Wu_2000}.
Finally, cooling due to emission lines is not negligible at the very bottom of the PSR, where the temperature is low (${\lesssim10^6}$~K).
For systems with low WD masses, the shock temperature and the post-shock gas temperature are low enough that the emission lines can actually contribute to a significant fraction of the total cooling. 
A fully consistent treatment including all these processes in the hydrodynamic calculation is non-trivial.
In addition, the current version of the \cyclops~code does not handle radiative transfer with either Compton scattering or electron conduction, which makes our choice for the cooling processes consistent within the code.
Having said that, future versions of the \cyclops~code could include these effects, not only in the cooling function, but also in the radiative transfer scheme. 


Other possible improvements of the \cyclops\ code are related to a better description of the pre-shock region.
For instance, the abundances of the material responsible by the photoelectric absorption should be allowed to vary and, at the same time, be consistent with estimates for the system being studied, which can be obtained from the emission lines.
In addition, a more complex upgrade would correspond to estimating the ionization fraction in the accretion structure.


Another ingredient not included in the current version of the \cyclops~code is the reflection from the WD surface.
This process is responsible for the iron K$\alpha$ line at  ${\approx6.4}$~keV and the Compton hump between ${\sim10}$ and ${\sim50}$~keV \citep{George_1991}.
In addition, based on \emph{NuSTAR} observations, \citet{Mukai_2015} and \citet{Luna_2018} showed that reflection is an important ingredient to be included in magnetic CV parameter estimation, especially for IPs, which can potentially provide further constraints. 
A reasonable model for the reflection was developed by \citet{Hayashi_2018}, who also investigated the influence of the parameters on the iron complex and Compton hump.
A model like that could be incorporated in future versions of the \cyclops~code and complementary analysis be performed (e.g. inferring the influence of the magnetic field vector).


Finally, the thermo-hydrodynamic differential equations for post-shock flows in the \cyclops~code are not solved simultaneously with the radiative transfer for \brem~and cyclotron radiations, which is a more realistic approach \citep[e.g.,][]{Fischer_2001}.
Indeed, we first model the PSR based on the 1D approach described in Section~\ref{app_cyclopspsr}.
Then, we perform the 3D phase-dependent and line-of-sight-dependent radiative transfer throughout the PSR as described in \citet{Costa_Rodrigues_2009} and \citet{Silva_2013}.

%% file: CONTENT/Section08.tex

We have used an updated version of \cyclops~code 
to model X-ray spectra and multiwavelength light curves of magnetic CVs, which includes accurate PSR modeling, by solving the stationary one-dimensional hydro-thermodynamic equations describing the accreting plasma, where bremsstrahlung and cyclotron radiative processes play a role in cooling the gas from the shock until the WD surface.
We have adopted a quasi-dipolar geometry, have taken into account the WD gravitational potential and a finite size of magnetosphere radius, and have allowed the magnetic field to decay as the distance from the WD surface increases.
While comparing our approach with previous works, we showed that our model agrees reasonably well with previous investigations and showed how the temperature profiles are affected by geometrical and physical parameters.
In particular, we discussed a four-dimensional fitting scheme in modeling magnetic CVs, whose parameters are the WD mass, the WD magnetic field, the specific accretion rate and the threading region radius.

\vspace{0.2cm}
Our main results can be summarized as follows.\\


i) We showed that a balance between cyclotron and \brem~radiative processes always exists in the presence of non-negligible cyclotron cooling, as expected.
In other words, there is always a combination of parameters such that the importance of these emission processes in shaping the PSR profiles are comparable.\\

ii) Due to the balance between \brem~and cyclotron processes, there is always a critical specific accretion rate above which \brem~radiation governs the cooling, and below which cyclotron radiation is the main mechanism driving the cooling.
In particular, the greater the WD magnetic field at the PSR bottom, the greater such a critical specific accretion rate, i.e. influence of cyclotron radiation takes place for a larger range of specific accretion rates.\\

iii) Due to the balance between \brem~and cyclotron processes, the previously found anti-correlation between the shock height and the specific accretion rate is broken in the presence of non-negligible cyclotron cooling.
Additionally, the above-mentioned critical specific accretion rate sets the region in the parameter space where such an anti-correlation is broken.\\


iv) We found that, for intermediate polars with extremely low accretion rates (e.g. due to instability in the accretion disk or low state), the cyclotron cooling cannot usually be neglected.
This is due to the above-mentioned balance between \brem\,and cyclotron processes, which occurs at low specific accretion rates in systems with weak magnetic fields.\\


v) With respect to X-ray spectra, we found that the greater the WD mass and/or the smaller  the WD magnetic field and/or the greater the specific accretion rate and/or the greater the threading region radius, the harder the spectrum, as expected.
This is intrinsically connected with the role of such parameters in shaping the PSR temperature and density profiles.\\

vi) While adopting PSRs with uniform temperature and density distributions, we showed that the \cyclops~X-ray spectra are in very good agreement with those generated by the \xspec~code, which is a widely used X-ray package to fit X-ray spectra, not only with respect to distance-dependent \brem~emission but also regarding the impact of the interstellar extinction.\\

vii) We showed that there exist thresholds in the parameter space below/above which X-ray spectra become indistinguishable, resulting in a degeneracy among the parameters.
In addition, such thresholds strongly depend on the combination of parameters.
In other words, due to the balance between \brem~and cyclotron radiative processes, the interplay between the WD mass, the WD magnetic field and the specific accretion rate strongly affects the thresholds mentioned above.\\


viii) We argued that in our four-dimensional modeling scheme, there is always a degeneracy, regardless of the region  of the parameter space one explores, which needs to be addressed in any meaningful fitting scheme.\\

ix) We further argued that any attempt of solving the degeneracy problem requires at least more constraints, i.e. having only the X-ray continuum spectrum for a given source is not enough to look for a global minimum used to uniquely estimate the main magnetic CV parameters.\\

x) We showed that even though several methods seem to some extent useful to break the degeneracy in the parameter space, some of them alone are most likely not enough to unambiguously solve the degeneracy problem.
In particular, despite the fact that incorporating either consistent pre-shock region modeling or accurate distance estimates may help, most likely in a fitting scheme one would still end up with many different models having very similar X-ray spectra.
Moreover, some very specific constraints are not always available, such as the PSR height or the break frequency.
However, incorporating light curves to the fitting scheme looks like a promising way out of the degeneracy problem and, based on that, we proposed a method that seems able to solve this problem.\\


xi) The method we proposed corresponds to incorporating X-ray light curves in different energy ranges to the fitting scheme, in case there is enough signal-to-noise ratio for that, potentially together with accurate \gaia~distance estimates.
This is an appealing method since X-ray data already comprise spectra and light curves, and no extra observational effort is required.
However, this method is expected to work only if the pre-shock region is properly modeled, since this region plays a key role in shaping X-ray light curves.\\

xii) While investigating the energy-dependent pulsed fraction in the X-ray light curves modeling, we have shown that the partial self-eclipse of the PSR should not be neglected, since it can produce pulsed fractions that smoothly decrease with energy.
This implies that the observed decrease of the pulsed fraction with energy in some magnetic CVs is not necessarily due to photoeletric absorption of the pre-shock region alone, and could perhaps be at least partially explained by self-eclipse, depending on the geometrical properties of system.
In addition, this combination of self-eclipse and pre-shock region absorption could lead to energy-dependent light curves, in the sense that the location or existence of the minima present in the X-ray light curves could depend on the energy.\\


xiii) Finally, we showed that the \cyclops~code, which is a 3D emission modeling code, is a sophisticated tool that can and should be used by the community in future magnetic CV emission modeling, not only with respect to X-ray data, but also regarding data in virtually any frequency range, including polarized emission.
In addition, to the best of our knowledge, \cyclops~is the only code available to the community able to consistently model the 3D pre-shock region, which is an essential ingredient to properly deal with X-ray emission from magnetic CVs.\\

\section*{Availability of the \cyclops~code}

The \cyclops~code is written in {\sc idl} and is freely available under request.
In order to  obtain a copy of the \cyclops~code described in this paper, please write to \href{mailto:claudia.rodrigues@inpe.br}{claudia.rodrigues@inpe.br}.

%% file: Acknowledgements.tex
We would like to thank an anonymous referee for the comments and suggestions that helped to improve this paper.
We thank Alfredo N. Iusem for pleasantly explaining many issues associated with the Newton-Raphson root finding method to solve boundary value problems.
We also thank the MCTIC/FINEP (CT-INFRA grant 0112052700) and the Embrace Space Weather Program for the computing facilities at the National Institute for Space Research, Brazil.
D.B. was supported by S\~ao Paulo Research Foundation (FAPESP, grant \#{2017/14289-3}) and partially supported by ESO/Gobierno de Chile.
C.V.R. thanks the grant 303444/2018-5 from CNPq
and the grant \mbox{\#{2017/14289-3}} from S\~ao Paulo Research Foundation (FAPESP). 
%
M.R.S. acknowledges support from the Millennium Nucleus for Planet Formation (NPF) and Fondecyt (grant 1181404).
M.C. thanks the grant \#2015/25972-0 by FAPESP.
%
%
%
%
G.J.M.L. is a member of the CIC-CONICET (Argentina) and acknowledge support from grant ANPCYT-PICT 0901/2017.
A.S.O acknowledges São Paulo Research Foundation (FAPESP) for financial support under grant \#2017/20309-7.
S.G.P. acknowledges the support of a Science and Technology Facilities Council (STFC) Ernest Rutherford Fellowship.
%
%
P.E.S. acknowledges FAPESP for financial support under grant \#2017/13551-6.
%
%
M.Z. acknowledges support from CONICYT PAI (Concurso Nacional de Inserci\'on en la Academia 2017, Folio 79170121) and CONICYT/FONDECYT (Programa de Iniciaci\'on, Folio 11170559).

%% file: CONTENT/AppendixA.tex
\subsection{Basic Assumptions and Geometry}
\label{psrmod}

In our modeling, the PSR properties are obtained from the solution of the stationary one-dimensional hydro-thermodynamic differential equations describing the accreting plasma.
We further assume a dipole-like geometry \citep[i.e. cubic cross-section variation, e.g.,][]{Hayashi_2014,Suleimanov_2016} and equipartition between ions and electrons \citep[e.g.,][]{Wu_1994,Cropper_1999,Som_2018}.
We would like to draw the readers attention to the fact that, unlike the PSR modeling, a three-dimensional geometry is adopted in the remaining parts of the \cyclops~code, e.g., the radiative transport and pre-shock region.
Hereafter, we will only discuss the PSR modeling, although in Section~\ref{cyclopscode} we provide more details about the other aspects of the \cyclops~code.

A stationary model with equipartition is a reasonable assumption provided that the characteristic time-scale of energy exchanged between electrons and ions (i.e. electron-ion equipartition time-scale) is much shorter than the cooling time-scale. 
There are two particular situations in which such an assumption does not hold.
First, for very low specific accretion rates and massive WDs, the non-equipartition area in the PSR might be well extended (${\gtrsim80}$\,\%), which means that the equipartition approach is no longer valid \citep[e.g.,][]{Hayashi_2014}.
This is because of the low density within the PSR, which makes energy exchange between electrons and ions inefficient.
However, \citet{Hayashi_2014b} applied their model considering non-equipartition between ions and electron to the IPs EX~Hya and V1223~Sgr and found that even for a relatively small specific accretion rate ($\sim0.05$ g s$^{-1}$ cm$^{-2}$), the equipartition between ions and electrons is quickly achieved close 
to the shock.
Second, for very strong magnetic fields and massive WDs, models assuming non-equipartition predict considerably different shock heights and qualitatively different density and temperature distributions in regions far from the WD surface than models assuming equipartition \citep[e.g.,][]{Wu_2000,Saxton_2007,Sarty_2008}.

With respect to the cross-section variation, a dipolar geometry is a more realistic approach than standard cylindrical models in which the cross-section is assumed to be constant.
Indeed, \citet{Saxton_2007} showed that the inclusion of a dipolar accretion geometry provides a harder continuum and in turn lowers estimates for WD masses when compared with masses inferred from models excluding this effect.
In addition, \citet{Hayashi_2014} showed that differences between both approaches become important when the shock height is ${\gtrsim20}$\,\% the WD radius \citep[see also][]{Canalle_2005}.
This is particularly true for accretion columns in IPs, in which the magnetic field is only moderately strong and the shock position might be relatively distant from the WD surface so that changes in the cross-section might play a significant role.

In Fig.~\ref{FigPSR1} we illustrate the geometrical and physical aspects of our PSR model. 
In what follows, we describe the differential equations, boundary conditions and the numerical scheme to solve these equations.

\subsection{Boundary Value Problem}
\label{secbvp}

The steady state equations presented below are for a plasma that is restricted to flow along the WD magnetic field lines.
We assumed here that the plasma is a fully ionized gas respecting the ideal gas law

\begin{equation}
P \ = \  n \, k_B \, T \, ,
\label{BVPEqn0}
\end{equation}
\

\noindent
where 
$P$ is the average pressure,
$T$ is the average temperature,
$k_B=1.380658\times10^{-16}$~erg~K$^{-1}$ 
is the Boltzmann constant,
and
$n$ in the average number density given by

\begin{equation}
n \ = \ \frac{\rho}{\mu \, m_H} \, ,
\label{nave1}
\end{equation}
\

\noindent
where
$\rho$ is the mass density,
${\mu=0.615}$ is the mean molecular mass of the fully ionized gas with solar metallicity,
and
${m_H=1.673525\times10^{-24}}$~g is the 
atomic hydrogen mass.

We further assume that the flow is symmetric with respect to the magnetic line passing through the PSR center, variations in the cross-section mimic a dipolar geometry \citep[as suggested by][]{Hayashi_2014} and that the hydro-thermodynamic quantities are functions of only $z$ (see Fig. \ref{FigPSR1}) within the PSR.
Under such assumptions, the differential equations related to the mass continuity, the momentum conservation and the energy conservation are \citep[e.g.,][]{Hayashi_2014}:

\begin{equation}
\der{}{z} \left(S \rho v \right) \ = \ 0 ,
\label{BVPEqn1}
\end{equation}

\begin{equation}
\der{}{z} \left( \rho v^2 + P \right) +
\frac{\rho v^2}{S} \der{S}{z} +
\rho \, g_{\rm WD}  = \ 0 ,
\label{BVPEqn2}
\end{equation}

\begin{equation}
v\der{P}{z} + \gamma P \der{v}{z} + 
(\gamma - 1) 
\left( \Lambda - \frac{\rho v^3}{2S} \der{S}{z} \right)
= \ 0 ,
\label{BVPEqn3}
\end{equation}
\

\noindent
where 
$\gamma=5/3$ is the adiabatic index,
$v(z)$ is the flow velocity, 
$S(z)$ is the cross-section,
$\Lambda(z)$ is the cooling function, which is described in Section \ref{seccoolfunc}, 
and 
$g_{\rm WD}(z)$ is the WD gravitational field given by

\begin{equation}
g_{\rm WD}(z) \ = \ \frac{GM_{\rm WD}}{z^2} ,
\end{equation}
\

\noindent
where 
$G=6.67259\times10^{-8}$~cm$^3$~g$^{-1}$~s$^{-2}$ is the gravitational constant 
and 
$M_{\rm WD}$ is the WD mass.

Note that Equation \ref{BVPEqn1} contains an integral of the form 

\begin{equation}
S(z) \, \rho(z) \, v(z) \ = \ S(z) \, \dot{m}(z) \ = \ \dot{M}_{\rm WD}
\end{equation}
\

\noindent
where 
$\dot{M}_{\rm WD}$ is the accretion rate 
and
$\dot{m}(z)$ is the specific accretion rate, i.e. the accretion rate per unit area.
Notice that $\dot{M}_{\rm WD}$ is constant but both $S$ and $\dot{m}$ vary throughout the PSR such that their product is constant and equals to $\dot{M}_{\rm WD}$.
We follow here the formulation by \citet{Hayashi_2014} and define the cross-section as

\begin{equation}
S(z)  \ = \ S_{\rm b} \, \left( \frac{z}{R_{\rm WD}} \right)^{n},
\label{EqnS}
\end{equation}
\

\noindent
where 
$S_{\rm b}$ is the accretion area at the bottom of the PSR (see Fig. \ref{FigPSR1}),
$n=0$ corresponds to the cylindrical geometry 
and 
$n=3$ to the dipolar geometry.
The bottom of the PSR corresponds to ${z=R_{\rm WD}}$ and ${S=S_{\rm b}}$, the latter coming directly from the adopted magnetic accretion geometry.

The cross-section derivative through the PSR is then given by

\begin{equation}
\der{S}{z}   \ = \  n \, 
\left( \frac{S_{\rm b}}{R_{\rm WD}} \right)\, 
\left( \frac{z}{R_{\rm WD}} \right)^{n-1}.
\label{EqnDS}
\end{equation}
\

In order to facilitate the solution of the PSR structure, instead of solving Equations \ref{BVPEqn1}, \ref{BVPEqn2} and \ref{BVPEqn3} in the variable $z$ we utilize a different independent variable.
The variable $z$ varies from $z=z_{\rm sh}=R_{\rm WD}+H_{\rm sh}$ until $z=R_{\rm WD}$, where $z_{\rm sh}$ is the shock position (i.e. $z_{\rm sh} = R_{\rm WD} + H_{\rm sh}$), $H_{\rm sh}$ is the shock height with respect to the WD surface (see Fig.~\ref{FigPSR1}),
and $R_{\rm WD}$ is the WD radius given by \citep{Nauenberg_1972}

\begin{equation}
\frac{R_{\rm WD}}{{\rm cm}} \ = \ 7.8\times10^8 \, \sqrt{ \left( \frac{M_{\rm WD}}{1.44 \, {\rm M}_\odot} \right)^{-2/3} \
- \ \left( \frac{M_{\rm WD}}{1.44 \, {\rm M}_\odot} \right)^{2/3}}.
\end{equation}
\

We define a variable $z'$ such that it goes from $0$ to $H_s$ \citep[e.g.,][]{Suleimanov_2005}, i.e.

\begin{equation*}
z' \ := \ z_{\rm sh} - z.
\end{equation*}
\

Equations \ref{BVPEqn1}, \ref{BVPEqn2} and \ref{BVPEqn3} can be translated into the following two-dimensional system of ordinary differential equations \citep[][]{Hayashi_2014}, which has $z'$ as the independent variable:

\begin{equation}
\der{P}{z'}  = 
\frac{\left( \gamma - 1 \right) 
\dot{m} \left( \Lambda - \frac{\rho v^3}{2S} \der{S}{z'}\right)
\ + \ 
\left( \dot{m} / v \right) \gamma P g_{\rm WD}}{\gamma P \ - \ \dot{m} v } ,
\label{BVPEqnP}
\end{equation}

\begin{equation}
\der{v}{z'}  =  \frac{g_{\rm WD}}{v} - \frac{1}{\dot{m}} \, \der{P}{z'}  .
 \label{BVPEqnV}
\end{equation}
\

The system variables are the pressure $(P)$ and the velocity $(v)$ and they depend on $z'$, which is the independent variable.
The system of ordinary differential equations expressed by Equations \ref{BVPEqnP} and \ref{BVPEqnV} is associated with boundary conditions (Section \ref{secbc}), which all together correspond to the boundary value problem defining the PSR.

\subsection{Boundary Conditions}
\label{secbc}

The system described by Equations \ref{BVPEqnP} and \ref{BVPEqnV} is associated with a boundary value problem, in which the boundary conditions are given at the shock position (i.e. $z' = z_{\rm sh} - z = 0$) and at the WD surface (i.e. $z' = z_{\rm sh} - z = H_{\rm sh}$). 
At the shock position we assume the Rankine-Hugoniot condition and at the WD surface we impose that the velocity is null.

The pre- and the post-shock values of the hydrodynamic variables are related by the Rankine-Hugoniot strong-shock jump conditions.
At the shock (just above it), the plasma falls with its free-fall velocity $(v_{\rm ff})$. The conditions for the jump are such that, at the shock, the density increases by a factor of four and the velocity decreases by a factor of four, i.e.

\begin{equation}
v_{\rm post} \ = \ \frac{v_{\rm pre}}{4} \ = \ 
\frac{v_{\rm ff}}{4} = 
\frac{1}{4}\,\sqrt{2GM_{\rm WD}\,\left( \frac{1}{z_{\rm sh}} - \frac{1}{R_{\rm th}} \right) },
\label{vff}
\end{equation}

\begin{equation}
\rho_{\rm post} \ = \ 4\rho_{\rm pre} \ = \ 
4\frac{\dot{m}}{v_{\rm pre}} \ = \ 
4\frac{\dot{m}}{v_{\rm ff}} \ = \ \frac{\dot{m}}{v_{\rm post}},
\label{rhoff}
\end{equation}
\

\noindent
where $R_{\rm th}$ is the threading region position relative to the WD center (see Fig.~\ref{FigPSR0}) and corresponds to the point where the magnetic pressure first dominates the gas ram pressure.
This position can also be understood as the magnetosphere limit.

As already pointed out by \citet{Suleimanov_2016}, the assumption of a finite magnetosphere is more accurate than assuming a free-fall velocity from the infinity.
This is rather important for systems with small magnetospheres (a few $R_{\rm WD}$), where matter reaches considerably smaller velocities compared to the case of accretion from infinity \citep[see also][]{Luna_2018}.

The boundary conditions at the shock (i.e. ${z'=0}$), which are indicated with sub-index `${\rm sh}$', are then given by \citep[e.g.,][]{Suleimanov_2005} 

\begin{equation}
\left\{
\begin{array}{l}
v_{\rm sh} \ = \ 0.25 \, \sqrt{2GM_{\rm WD}\,\left( 1/z_{\rm sh} - 1/R_{\rm th} \right)}, \\ 
 \\
\rho_{\rm sh} \ = \ \dot{m}_{\rm sh} \, / \, v_{\rm sh} , \\
 \\
P_{\rm sh} \ = \ 3 \, \dot{m}_{\rm sh} \, v_{\rm sh},\\
 \\
T_{\rm sh} \ = \ 3 \, \mu \, m_H \, v^2_{\rm sh} \, / \, k_B,\\
\end{array}
\right. \label{BVPCon1}
\end{equation}
\

\noindent
and the condition at the bottom, i.e., at the WD surface is

\begin{equation}
v_{\rm b} \ = \ v(z'=H_{\rm sh}) \ = \ 0. \label{BVPCon2}
\end{equation}
\

Before describing the numerical scheme to solve the boundary value problem associated with the PSR (Section \ref{numsche}), we discuss the cooling function adopted here.

\subsection{Cooling Function}
\label{seccoolfunc}

The cooling function $(\Lambda)$ adopted here includes effects of both bremsstrahlung and the cyclotron processes, i.e. 

\begin{equation}
\Lambda \ = \ \Lambda_{\rm brem} + \Lambda_{\rm cyc}. 
\label{cool}
\end{equation}

The optically thin bremsstrahlung cooling is given by \citep{Som_2018}

\begin{equation}
\frac{\Lambda_{\rm brem}}{{\rm erg \, cm^{-3} \, s^{-1}}} \ = \ 5.01\times10^6 \, 
\left( \frac{\rho}{10^{-9} \, {\rm g \, cm^{-3}}} \right)^2 \, 
\left( \frac{T}{10^8 \, {\rm K}} \right)^{1/2}.
\label{coolbrem}
\end{equation}
\

To evaluate which one of these two radiative processes dominates the cooling, one usually utilizes the {\it cooling ratio  ($\epsilon_{\rm sh}$)}, which is defined as the ratio between the bremsstrahlung and the cyclotron cooling time-scales at the shock position, given by \citep{Som_2018}

\begin{equation}
\begin{multlined}
\epsilon_{\rm sh} \ = \ 4.32 \,
\left( \frac{B_{\rm sh}}{10 \, {\rm MG}} \right)^{2.85} 
\left( \frac{\dot{m}_{\rm sh}}{ {\rm g\,s^{-1}\,cm^{-2}}} \right)^{-1.85} 
\left( \frac{M_{\rm WD}}{0.8 \, {\rm M_\odot}} \right)^{2.925} 
\left( \frac{R_{\rm WD}}{7\times10^8 \, {\rm cm}} \right)^{-2.925} 
\left( \frac{S_{\rm sh}}{10^{15} \, {\rm cm^2}} \right)^{-0.425},
\end{multlined}
\label{coolratio}
\end{equation}
\

\noindent
where values with sub-index `${\rm sh}$' are quantities at the shock position.
Bremsstrahlung dominates in cases where ${\epsilon_{\rm sh}<1}$ and cyclotron governs otherwise (see Section \ref{influPSR} for more details).

The optically thick cyclotron cooling is given by \citep[e.g.,][]{Canalle_2005,Saxton_2007}

\begin{equation}
\begin{multlined}
\Lambda_{\rm cyc} \ = \ 
\Lambda_{\rm brem} \,
\epsilon_{\rm sh}
\left( \frac{P}{P_{\rm sh}} \right)^2
\left( \frac{\rho}{\rho_{\rm sh}} \right)^{-3.85} 
\left( \frac{S}{S_{\rm sh}} \right)^{-0.425} 
\left( \frac{B}{B_{\rm sh}} \right)^{2.85},
\end{multlined}
\label{coolcyc}
\end{equation}
\

\noindent
where the quantities $P$, $\rho$, $S$ and $B$ are not fixed and vary through the PSR and again quantities with sub-index `${\rm sh}$' are given at the shock position.
In particular, the cross-section variation is given by Eq. \ref{EqnS} and the WD magnetic field assumed here is a dipole, which implies that its variation with $z$ is given by

\begin{equation}
B(z)  \ = \ B_{\rm b} \, \left( \frac{z}{R_{\rm WD}} \right)^{-3},
\label{EqnB1}
\end{equation}
\

\noindent
where $B_{\rm b}$ is the magnetic field at the PSR bottom given by

\begin{equation}
B_{\rm b}  \ = \ \frac{B_{\rm p}}{2} \, \sqrt{4 - 3\cos^2\varrho},
\label{EqnB2}
\end{equation}
\

\noindent
where $B_{\rm p}$ is the magnetic field intensity at the pole, and $\varrho$ is a function of $\beta$, $B_{\rm lat}$ and $B_{\rm long}$.
We would like to stress that in the \cyclops~code, the parameter is $B_{\rm p}$ (i.e. the magnetic field intensity at the pole), while $B_{\rm b}$ (i.e. the magnetic field intensity at the PSR bottom) is computed accordingly provided the geometry of the model (i.e. the PSR center colatitude $\beta$ and the  magnetic field axis latitude $B_{\rm lat}$ and longitude $B_{\rm long}$).
See Fig.~\ref{FigPSR0} for more details on the geometrical aspects of the \cyclops~code.
Finally, for $B_{\rm lat}=90^o$ and $B_{\rm long}=0^o$, i.e. in the case the magnetic axis coincides with the WD rotation axis and the PSR and magnetic axis are in the same plane (like in Fig.~\ref{FigPSR0}), Eq.~\ref{EqnB2} reduces to an expression consistent with eq.~45 in \citet{Canalle_2005}.

\subsection{Numerical Scheme}
\label{numsche}

The boundary value problem is given by the system of ordinary differential equations defined by the Eqs.\,\ref{BVPEqnP} and \ref{BVPEqnV}, coupled with the conditions \ref{BVPCon1} and \ref{BVPCon2}, where the quantity to be determined is the shock height $H_{\rm sh}$.
To solve such a boundary value problem, we iteratively solve the ordinary differential equations.
To do so, we guess shock heights and assume the conditions \ref{BVPCon1} as the initial conditions, until we satisfy the condition \ref{BVPCon2}, at the WD surface, i.e. null velocity at the PSR bottom.

Since the derivatives of the system tend to be huge as the velocity tends to zero (due to the asymptotic behavior of the solutions while approaching the PSR bottom), we solve the initial value problem with a 4--5th-order Runge-Kutta-Fehlberg integrator with adaptive step-size control, in order to reach very small velocities (i.e. with a precision of ${\sim10^{-8}}$).
Indeed, given the non-smooth behavior of the system when ${v\rightarrow0}$, monitoring the local truncation error here is the best way to ensure accuracy, by adjusting the step-size.

We solve the whole boundary value problem with a simple shooting algorithm based on the Newton-Raphson root finding method.
We start with a trial integration that satisfies the boundary conditions at the shock position, from ${v=v_{\rm sh}}$ to ${v=10^{-8}}$~cm~s$^{-1}$.
The final position (i.e. PSR bottom) is then compared with the WD surface position (i.e. ${z'=H_{\rm sh}}$) and the discrepancy is used to adjust the guessing $H_{\rm sh}$, until the boundary condition at the PSR bottom is eventually satisfied, with a precision of ${\sim10^{-8}}$.

%% file: CONTENT/AppendixB.tex
\begin{figure*}[htb!]
\begin{center}
\includegraphics[width=0.497\linewidth]{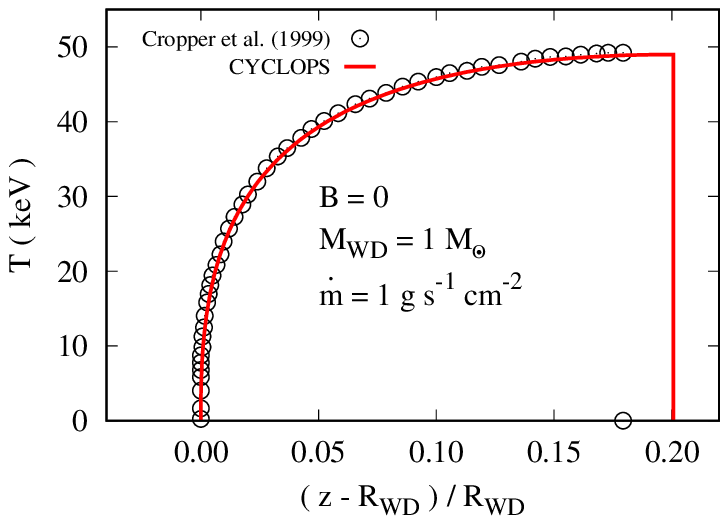}
\includegraphics[width=0.497\linewidth]{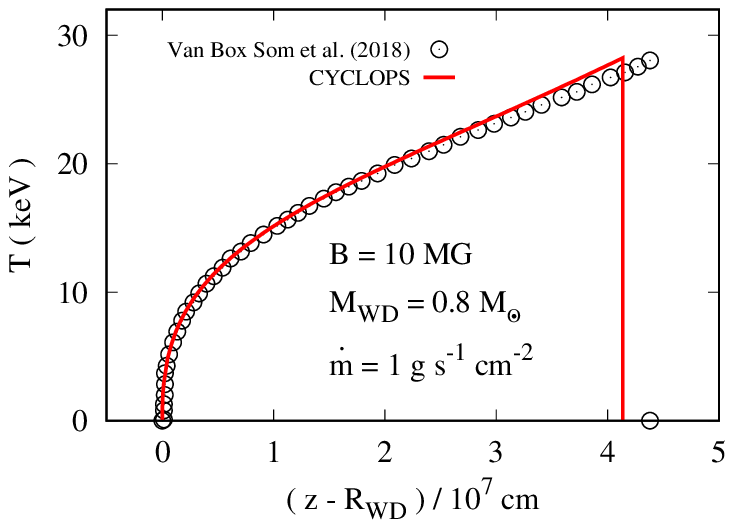} 
\includegraphics[width=0.497\linewidth]{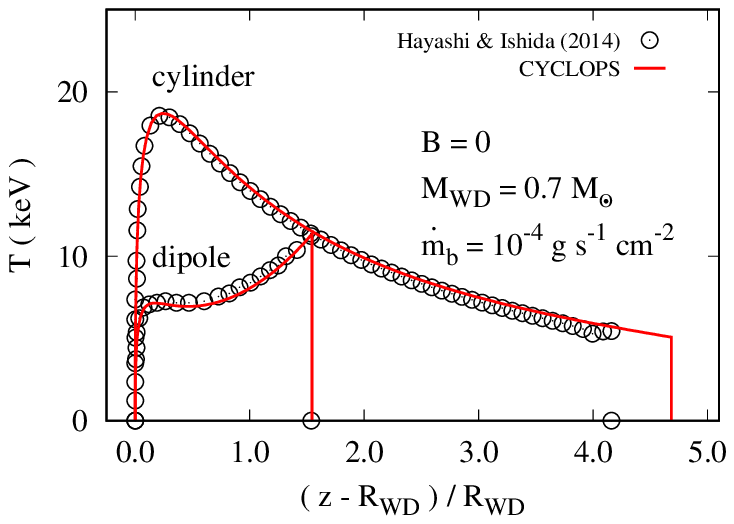} 
\includegraphics[width=0.497\linewidth]{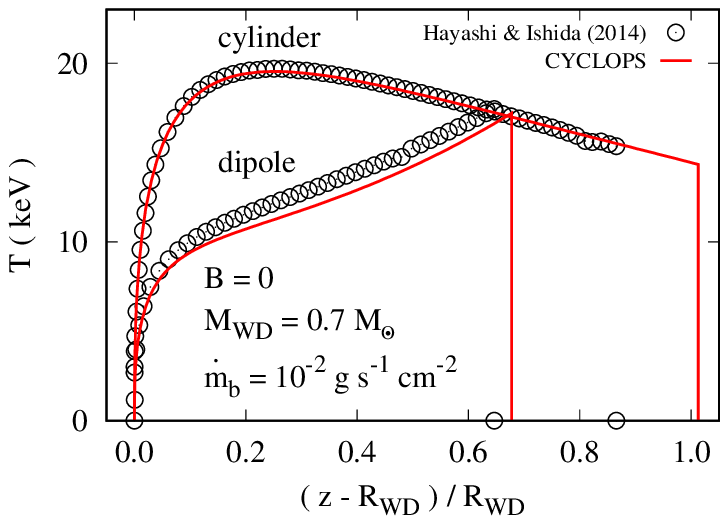}
\includegraphics[width=0.497\linewidth]{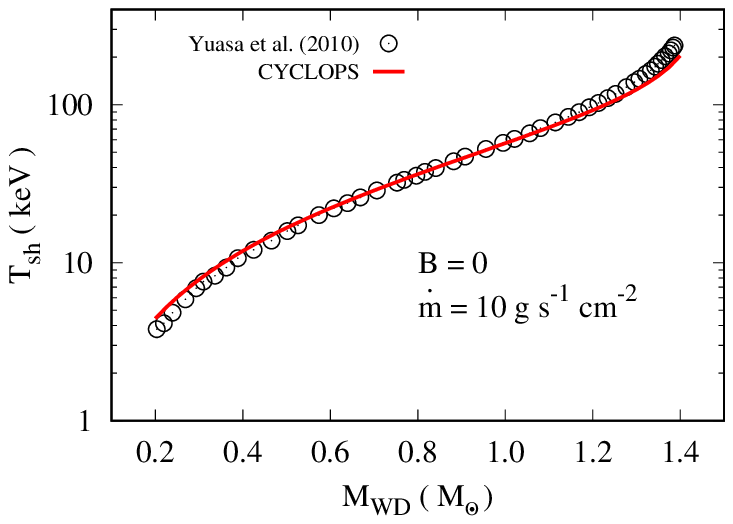}
\includegraphics[width=0.497\linewidth]{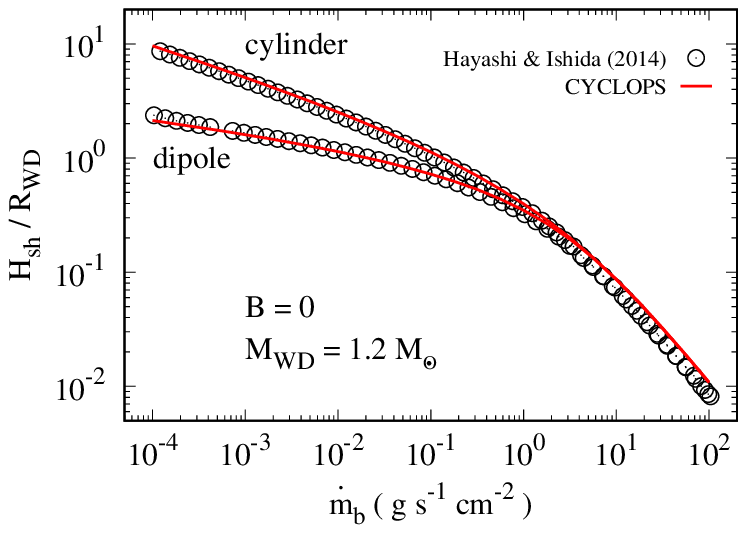}
\end{center}
\caption{Comparison of different \cyclops~solutions with previous works.
In the top and middle rows, we compare our temperature profiles with \citet[][top left panel]{Cropper_1999}, \citet[][top right panel]{Som_2018} and \citet[][middle left and right panels]{Hayashi_2014}.
In the bottom left panel, we show the temperature at the shock $(T_{\rm sh})$ as a function of the WD mass $(M_{\rm WD})$ and compare with \citet[][]{Yuasa_2010}.
In the bottom right panel, we provide the shock height in units of the WD radius $(H_{\rm sh}/R_{\rm WD})$ as a function of the specific accretion rate at the WD surface $(\dot{m}_{\rm b})$ and compare with \citet[][]{Hayashi_2014}.
Our profiles are in very good agreement with these previous works, especially the temperature at the shock and the profile shape as a whole.
Small differences in the shock height ${(\lesssim\,10\,\%)}$ are likely due to different cooling prescriptions as well as numerical techniques to solve the system of ordinary differential equations defining the PSR structure.
}
\label{FigPSR2_NEW}
\end{figure*}


With the aim of showing that our modeling provide consistent results, we compare in this section our approach with several previous works with similar methods.
Starting with our PSR modeling, we show, in the top and middle panels of Fig.~\ref{FigPSR2_NEW}, temperature profiles built with parameters consistent with \citet[][top left panel]{Cropper_1999}, \citet[][top right panel]{Som_2018} and \citet[][middle panels]{Hayashi_2014}.
In the comparisons, a cylindrical model is defined by setting $n=0$ in Eqs.~\ref{EqnS} and \ref{EqnDS}, while a dipolar model is defined by $n=3$.

In the comparison with \citet{Cropper_1999}, we assumed a cylindrical model and set 
${R_{\rm th}\gg R_{\rm WD}}$, 
${B=0}$, 
${\dot{m}=1~{\rm g\,s^{-1}\,cm^{-2}}}$, 
${M_{\rm WD}=1.0~{\rm M_\odot}}$ 
and 
${\mu=0.615}$. 
In the comparison with \citet{Som_2018}, we also assumed the cylindrical model and set 
${R_{\rm th} \gg R_{\rm WD}}$, 
${B=10}$~MG constant throughout the PSR, 
${\dot{m}=1~{\rm g\,s^{-1}\,cm^{-2}}}$,
${M_{\rm WD}=0.8~{\rm M}_\odot}$ 
and 
${\mu=0.5}$. 
Finally, in the comparison with \citet{Hayashi_2014}, we assumed both cylindrical and dipolar models and set
${R_{\rm th} \gg R_{\rm WD}}$, 
${B=0}$, 
${M_{\rm WD}=0.7~{\rm M}_\odot}$ 
and 
${\mu=0.615}$. 
We then considered two low values for $\dot{m}_{\rm b}$, namely 
${10^{-4}\,{\rm g\,s^{-1}\,cm^{-2}}}$
and 
${10^{-2}\,{\rm g\,s^{-1}\,cm^{-2}}}$.

Notice that our profiles are in good agreement with results from the above-mentioned works and the different shock heights $(\lesssim\,10\,\%)$ are likely due to different cooling prescriptions and numerical methods. 
For instance, \citet{Hayashi_2014} adopted the collisional ionization equilibrium cooling function calculated by \spex~package \citep{Schure_2009}, in which cooling due to emission lines, which becomes more efficient for temperatures below a few $10^6$~K, is included.
On the other hand, \cyclops~does not currently handle line cooling and only \brem~cooling contributes to the cooling function in the middle panels of Fig.~\ref{FigPSR2_NEW}.
This explains why the shock heights in \cyclops~are slightly greater than those from \citet{Hayashi_2014} since the inclusion of line contribution to the cooling function slightly enhances the cooling, resulting in turn in slightly shorter PSRs.


Moving forward, we compare our results with \citet{Yuasa_2010} and \citet{Hayashi_2014}, for a set of solutions, in which we assumed parameters consistent with those used by these authors,  i.e. 
${R_{\rm th} \gg R_{\rm WD}}$, 
${B=0}$ 
and 
${\mu=0.615}$.
In the comparison with \citet{Yuasa_2010}, we adopted the cylindrical model, fixed  
${\dot{m}=10~{\rm g\,s^{-1}\,cm^{-2}}}$, 
and let $M_{\rm WD}$ vary from $0.2$ to $1.4$~\Msun. 
In the comparison with \citet{Hayashi_2014}, we adopted the cylindrical and dipolar models, fixed
${M_{\rm WD}=1.2~{\rm M}_\odot}$ 
and let $\dot{m}_{\rm b}$ vary from 
$10^{-4}$~to~${10^2~{\rm g\,s^{-1}\,cm^{-2}}}$.

The comparisons are shown in the bottom row of Fig.~\ref{FigPSR2_NEW}, where in the left panel we show the shock temperature $T_{\rm sh}$ against $M_{\rm WD}$ similar to fig.~11 in \citet{Yuasa_2010}, and in the right panel we show the shock height $H_{\rm sh}$ from the WD surface against $\dot{m}_{\rm b}$ similar to fig.~5 in \citet{Hayashi_2014}.
Notice that we also find a good agreement between our results and theirs. Again, small differences found in these plots are connected with different \brem~cooling prescriptions and numerical techniques to solve the boundary value problem associated with the PSR structure.
In particular, we predict slightly higher (lower) shock temperatures for low-mass (high-mass) WDs, in comparison with \citet{Yuasa_2010}.
Additionally, we predict slightly larger shock height, especially for high 
${\dot{m}_{\rm b}~(\gtrsim1~{\rm g\,s^{-1}\,cm^{-2}})}$, 
in comparison with \citet{Hayashi_2014}, because line cooling is not included in the \cyclops~code, which makes the overall cooling less efficient in our case.

The correlation and anti-correlation shown in the bottom row of Fig.~\ref{FigPSR2_NEW} are interesting and not difficult to understand.
With regards to the correlation between the shock temperature $(T_{\rm sh})$ and $M_{\rm WD}$, we notice that $T_{\rm sh}$ strongly depends on the gas velocity at the shock position $(v_{\rm sh})$, which in turn depends on $M_{\rm WD}$, as $v_{\rm sh}$ is a result of the potential energy converted into kinetic energy.
In this way, the greater $M_{\rm WD}$, the greater $v_{\rm sh}$, an in turn the greater $T_{\rm sh}$.

Now, the anti-correlation between $H_{\rm sh}$ and $\dot{m}_{\rm b}$ exists for negligible cyclotron cooling cases and is directly related to the \brem~cooling efficiency in the PSR. 
Indeed, the greater $\dot{m}_{\rm b}$, 
the greater the density in the PSR. 
As the \brem~cooling rate strongly depends on the density (see Eq. \ref{coolbrem}), the greater the density, the greater the cooling rate, and in turn the smaller the PSR height.
This implies then that the smaller $\dot{m}_{\rm b}$, the greater $H_{\rm sh}$.
As discussed in Section~\ref{influCOOL}, such an anti-correlation is not always present in the case of non-negligible cyclotron radiation.
In particular, there is a critical specific accretion rate above which there is an anti-correlation and below which there is a correlation.
Such a critical specific accretion rate strongly depends on the WD mass and the WD magnetic field.

\begin{figure*}[htb!]
\begin{center}
    \includegraphics[width=0.497\linewidth]{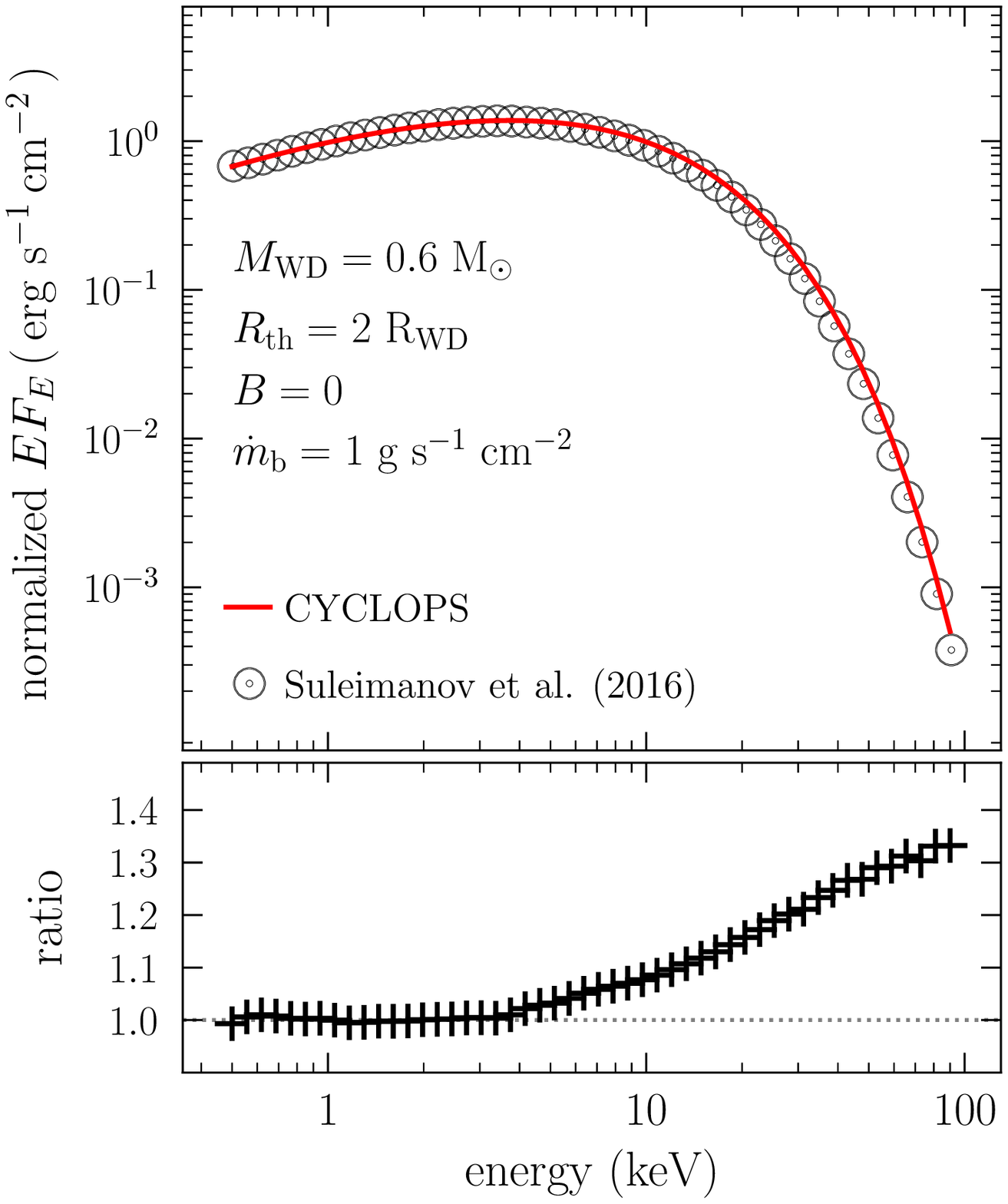}
    \includegraphics[width=0.497\linewidth]{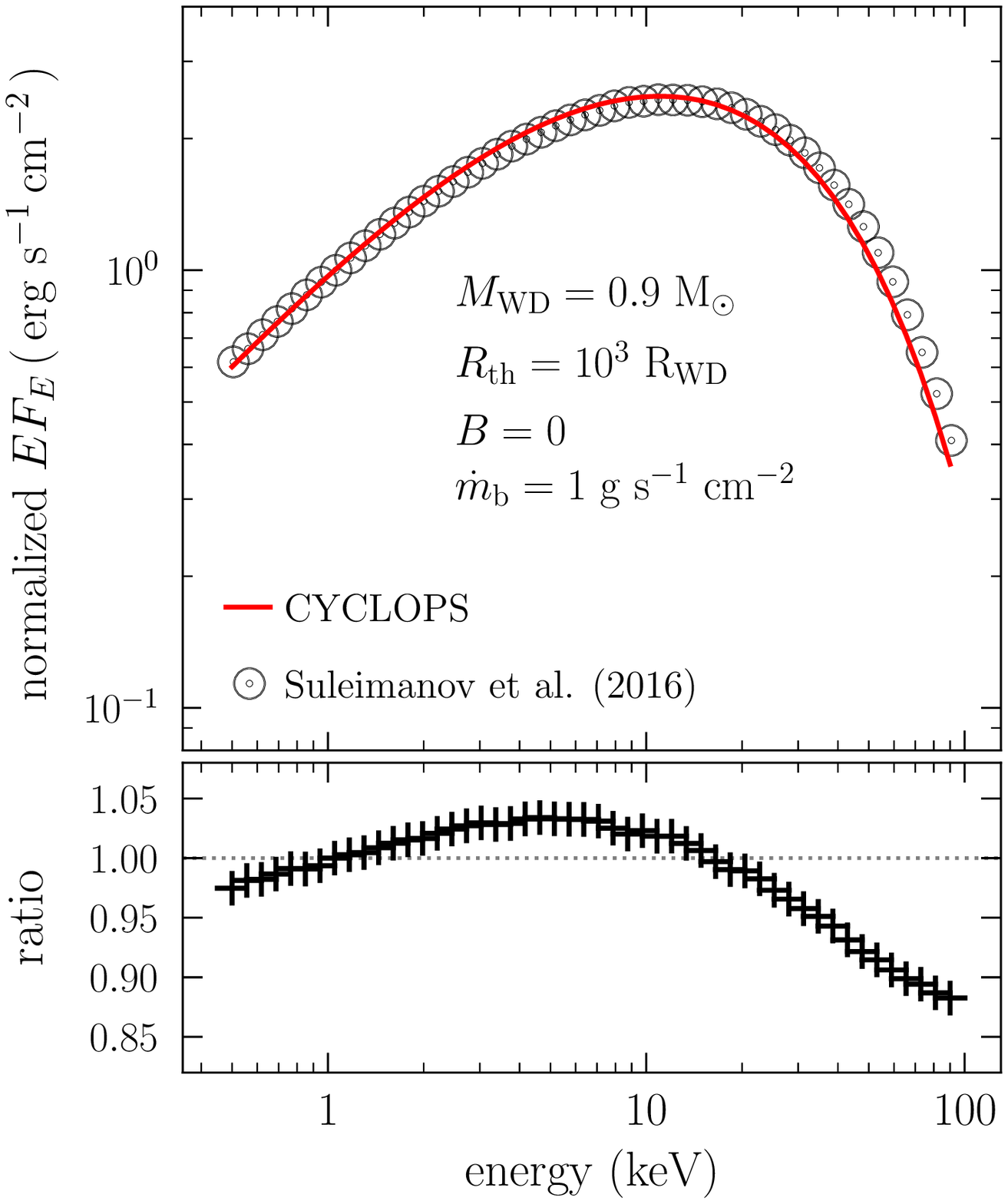} 
\end{center}
\caption{Comparison of two different \cyclops~X-ray spectra with \citet[][]{Suleimanov_2016}.
In both panels, the spectra are normalized by their values at $1$~keV.
Our spectra are in very good agreement with \citet[][]{Suleimanov_2016} since the differences are very small for energies below ${\sim20}$~keV ($\lesssim10$~\%) and still small for energies greater than that (${\lesssim30}$~\%).
Such small difference between both works are likely due to slightly different \brem~prescriptions as well as PSR modeling, which have more impact on the hard part of the spectrum.
}
\label{FigCompXraySulHI}
\end{figure*}

Now, moving to the X-ray modeling, we show in Fig.~\ref{FigCompXraySulHI} two X-ray spectra built with parameters consistent with \citet[][]{Suleimanov_2016}.
In both cases, we assumed negligible cyclotron cooling,
${\dot{m}=1~{\rm g\,s^{-1}\,cm^{-2}}}$,
and
${\mu=0.615}$.
In addition, while in the left panel we adopted
${R_{\rm th}=2\,R_{\rm WD}}$ and
${M_{\rm WD}=0.6~{\rm M}_\odot}$,
in the right panel we set
${R_{\rm th}=1000\,R_{\rm WD}}$ and
${M_{\rm WD}=0.9~{\rm M}_\odot}$. 
Notice that our X-ray spectra are in good agreement with the results by \citet{Suleimanov_2016}.
In particular, the spectra agree reasonably well for energies below ${\sim20}$~keV since differences are usually smaller than ${\sim10}$~\%.
However, for energies greater than that, differences can be a bit higher (${\sim20-30}$~\%).

This is likely a consequence of the different PSR modeling and \brem~prescription in both cases (compare our Section~\ref{cyclopscode} and Appendix~\ref{app_cyclopspsr} with section 3 in \citealt{Suleimanov_2016}), which affect more the hard part than the soft one.
For instance, the boundary value problem in \citet[][see their eq. 21]{Suleimanov_2016} is slightly different from ours, which follows \citet{Hayashi_2014}.
In addition, their cooling function is computed by \apec~using the AtomDB atomic database, which is different from our approach (see Equations \ref{cool}, \ref{coolbrem} and \ref{coolcyc}).
Moreover, the calculation of the X-ray spectrum in \citet[][see their eq. 29]{Suleimanov_2016} is different from what is done in the \cyclops~code \citep[see section 2.2 in][]{Silva_2013}.
Despite the fact that there are differences in the PSR modeling and \brem~emission in both works, provided that the differences between \cyclops~and \citet{Suleimanov_2016} are small in most portions of the X-ray energy range, we can conclude that both results are consistent with one another.

%% file: CONTENT/AppendixC.tex
\begin{figure*}[t!]
\begin{center}
\includegraphics[width=0.47\linewidth]{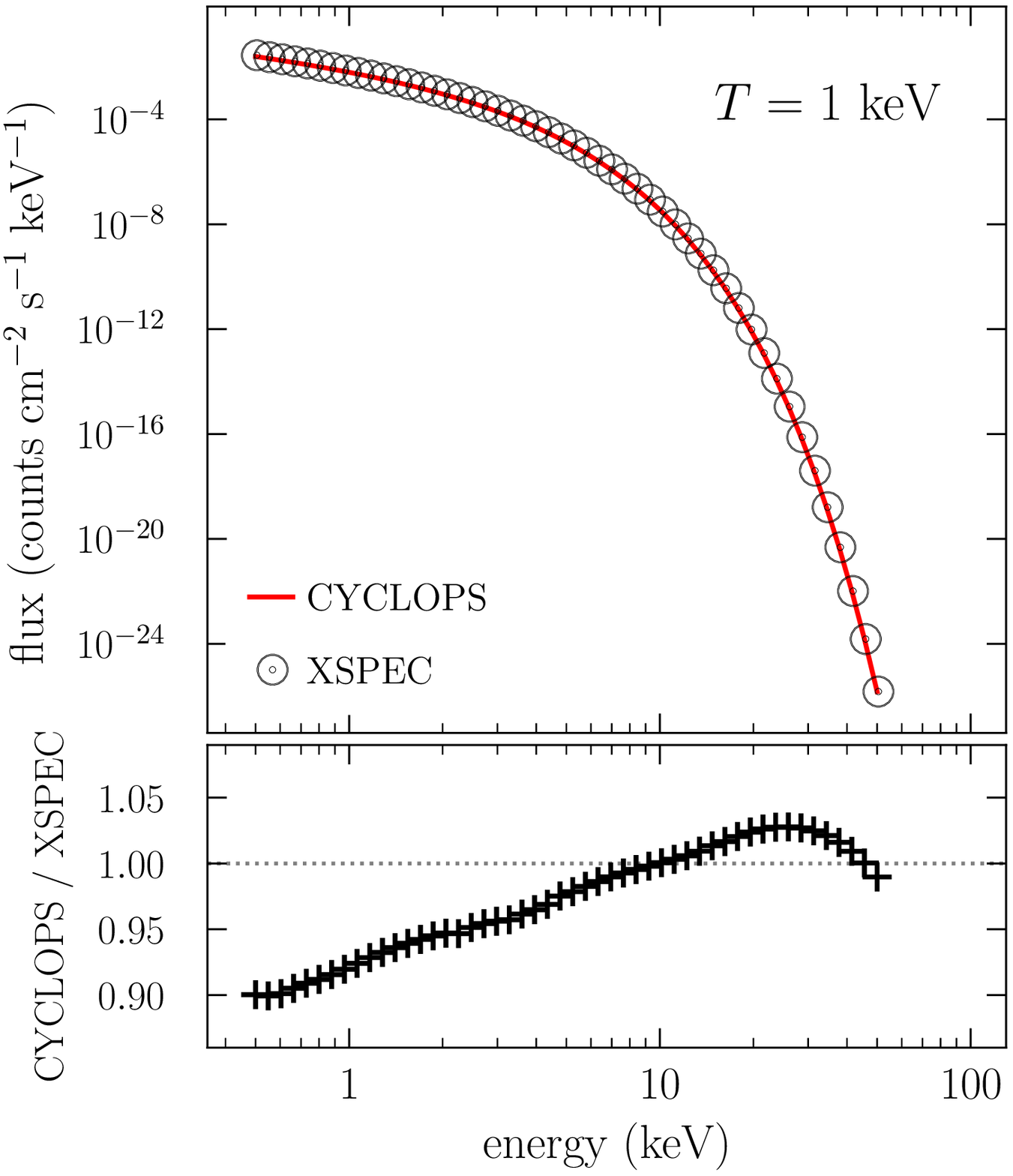}
\hspace{0.2cm}
\includegraphics[width=0.47\linewidth]{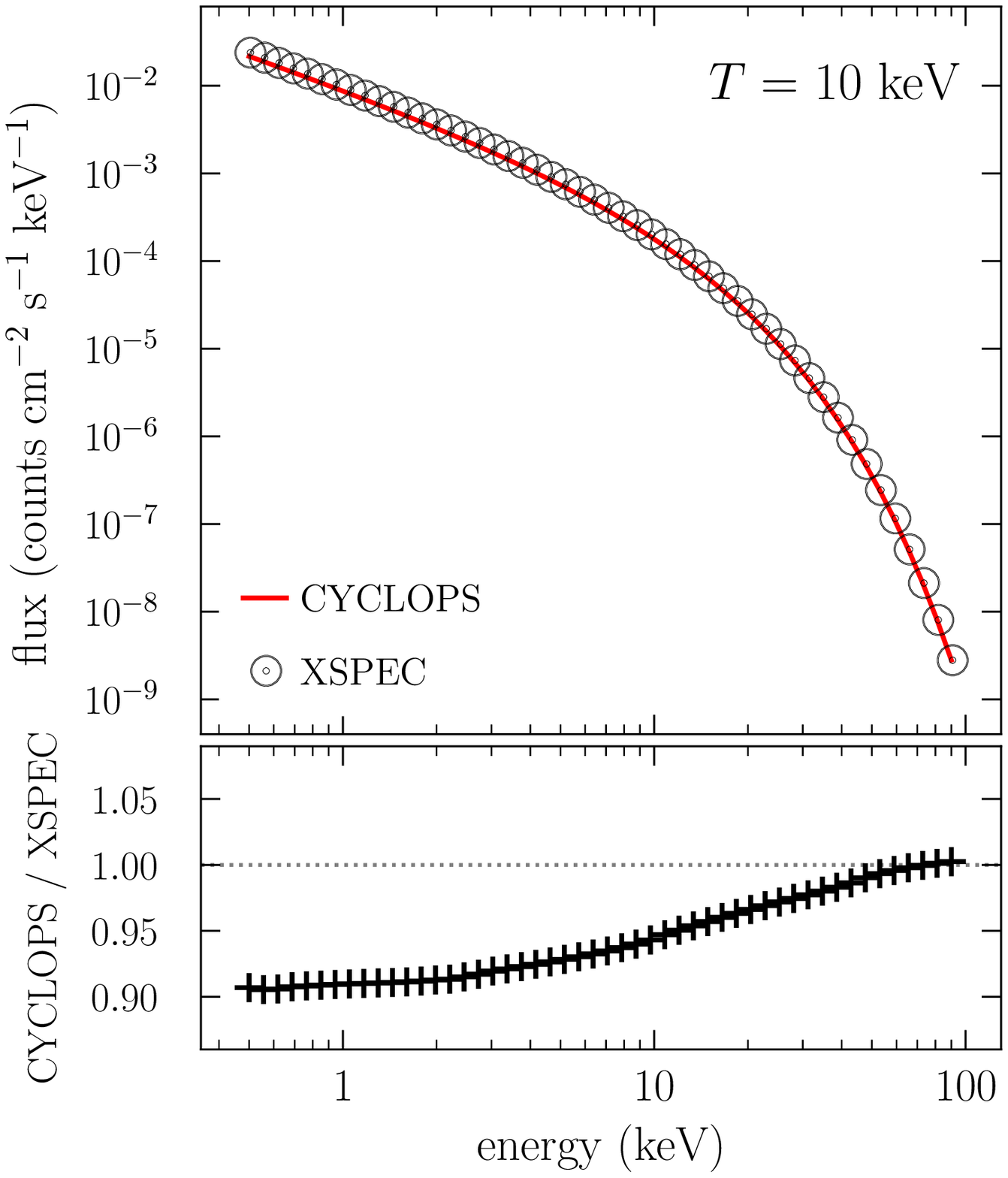}
\includegraphics[width=0.47\linewidth]{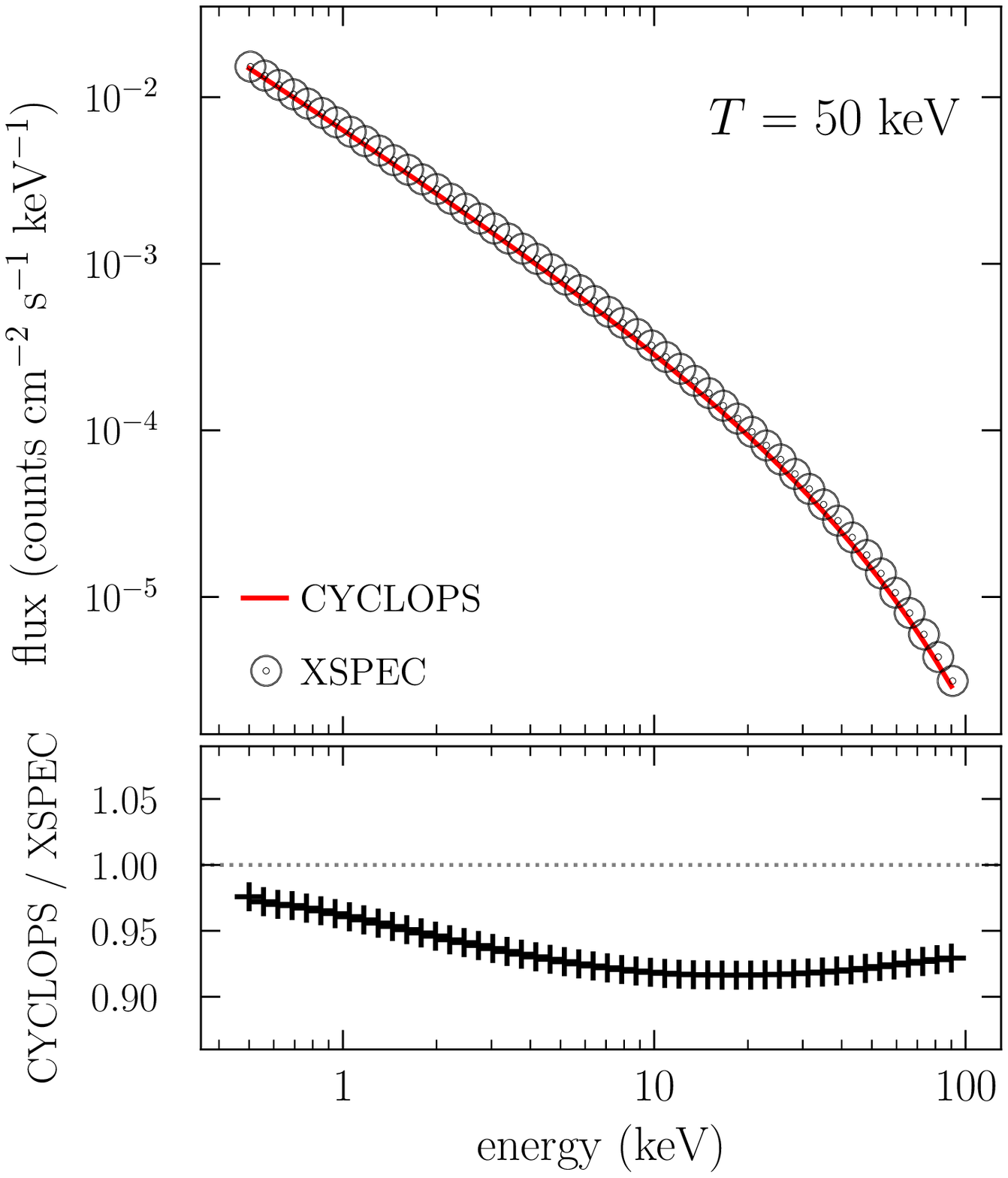}
\hspace{0.2cm}
\includegraphics[width=0.47\linewidth]{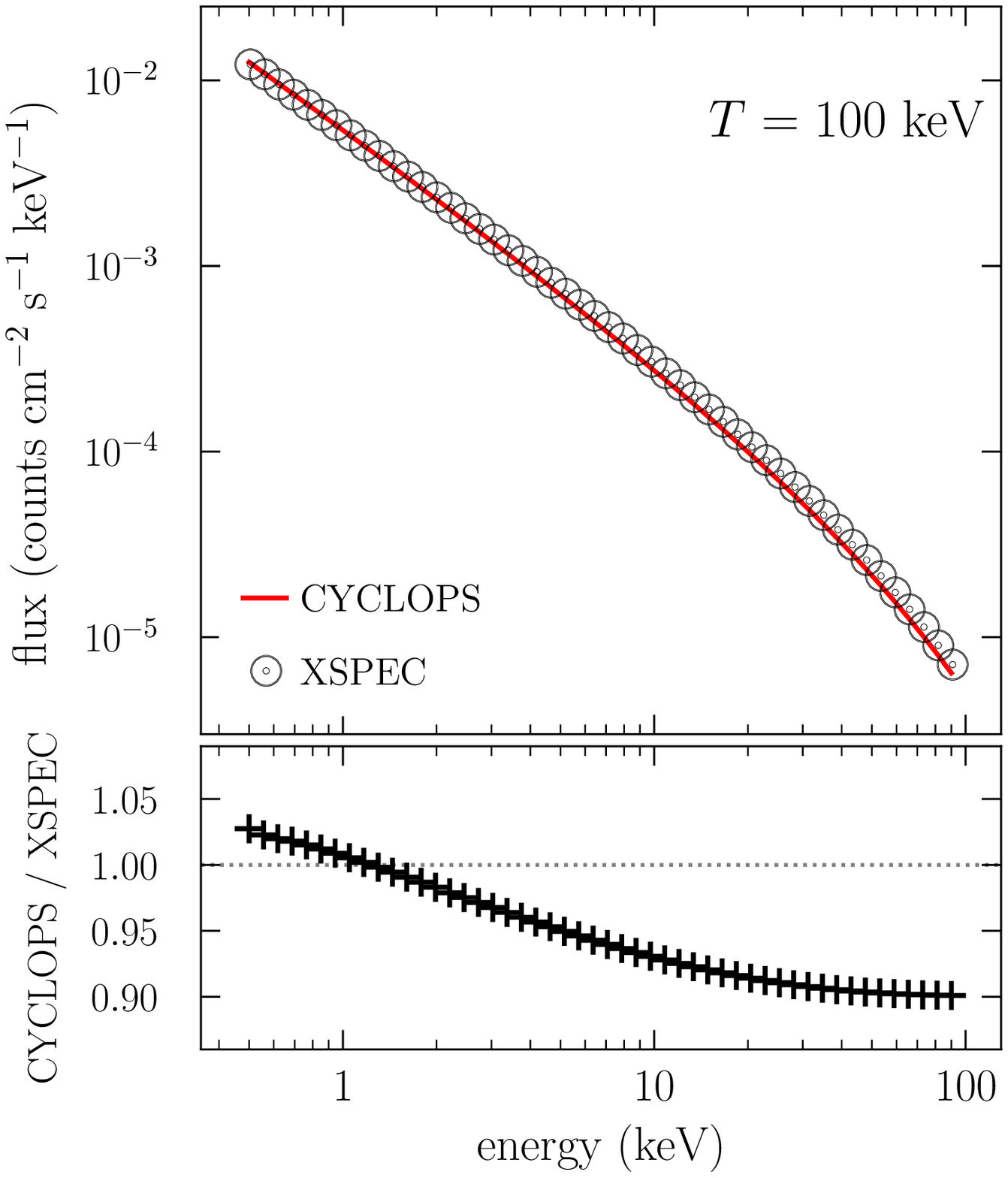}
\end{center}
\caption{Comparison between X-ray spectra from \cyclops~and \xspec~(model \bremss), for four different values of the temperature, namely $1$ (top left panel), $10$ (top right panel), $50$ (bottom left panel) and $100$~keV (bottom right panel).
In all models, the electronic density is $10^{16}$~cm$^{-3}$ and the distance is $100$~pc.
Clearly, the spectra from \cyclops~are in excellent agreement with those from \xspec, provided that the differences are usually smaller than $10$~\%, irrespective of the energy.
We can then conclude that the \cyclops~code properly deals with non-normalized spectra, i.e., it can take into account the distance while fitting an object.
}
\label{FigAppC_distance}
\end{figure*}

Despite the fact that the PSR modeling in the \cyclops~code includes the solution of the stationary one-dimensional hydro-thermodynamic differential equations describing the accreting plasma as described in Appendix~\ref{app_cyclopspsr}, it can also handle PSRs with uniform temperature and density distributions, i.e., assuming that the temperature and densities are the same in the entire PSR.
This allows us to compare its \brem~emission modeling with that of the \xspec~code \citep{Arnaud_1996,Dorman_Arnaud_2001}.
Such a comparison is an essential step in order to demonstrate that the distance-dependent flux in the \cyclops~code is correctly calculated and the interstellar extinction is properly taken into account.

With the goal of showing that \cyclops~provides consistent results, throughout this section we compare it with \xspec~for several cases of homogeneous PSRs.
Before that, we shall emphasize that \cyclops~and \xspec~can only be compared in the optically thin regime, because this is the only case currently implemented in the \xspec~code, although for X-rays this assumption is not problematic.
The \cyclops~code, on the other hand, solves the radiative transfer without any approximation in this regard, since it also deals with the usually optically thick cyclotron emission, which happens in optical wavelengths.

In the following comparison, we utilized the model \bremss~\citep{Karzas_1961} in the \xspec~package.
Even though a slightly different \brem~prescription is adopted in this model with respect to \cyclops, it corresponds to the most suitable model to perform such a comparison.
Other models such as \mekal~\citep{Mewe_1985,Mewe_1986,Kaastra_1992,Liedahl_1995} have \brem~recipes more similar to that of \cyclops~and could in principle be better for our purpose.
However, these models include emission lines of many  elements, which would not result in meaningful comparisons, provided that currently there is no prescription for line emission in the \cyclops~code.
While discussing interstellar extinction, we adopted the \phabs~model in \xspec, which calculates the required photoelectric absorption cross-sections.
In addition, since our primary goal is to show that not only the shape but also the flux of the \cyclops~spectra are consistent with those of \xspec, we assume in all cases  a distance of $100$~pc.
Moreover, we fix the electronic density to $10^{16}$~cm$^{-3}$ and focus here on the impact of the PSR temperature in shaping the X-ray spectra as well as in determining their X-ray flux.

In order to properly compare \cyclops~with \xspec, we need first to make sure that the emitting source is the same in both cases.
For that we utilize the normalization factor for the \bremss~model adopted in \xspec~given by

\begin{equation}
{\rm Norm} \ = \ \frac{3.02\times10^{-15}}{4\,\pi\,D^2} \, \int \, n_e \, n_I \, {\rm d}V
\label{NormXSPEC}
\end{equation}
\
 
\noindent
where $D$ is the distance to the source, $n_e$, $n_I$ are the electron and ion number densities, and $V$ is the volume.
For our purposes here, we assume that $n_e=n_I$, which implies that Eq.~\ref{NormXSPEC} can be expressed as

\begin{equation}
{\rm Norm} \ \simeq \ 2.524\times10^{-53} \, 
\left( \frac{D}{\rm cm} \right)^{-2} \, 
\left( \frac{n_e}{\rm cm^{-3}} \right)^2 \, 
\left( \frac{S_{\rm b}}{\rm cm^2} \right) \, 
\left( \frac{H_{\rm sh}}{\rm cm} \right)
\label{NormCYCLOPS}
\end{equation}
\
 
\noindent
where $S_{\rm b}$ is the accretion area at the PSR bottom and $H_{\rm sh}$ is the PSR height (see Fig.~\ref{FigPSR1}).
This normalization factor directly implies that the `observed' \brem~flux coming out of a model depends not only on the distance but also on the PSR properties.

In order to compare the spectra from \cyclops~and \xspec, we first calculated the PSR of a generic model with the formalism described in Appendix~\ref{app_cyclopspsr}.
That procedure provides a non-homogeneous PSR, i.e., the temperature and density vary along the PSR.
However, as mentioned before in this section, we assume here that the temperature and density are constant throughout the PSR.
To achieve that, we kept the PSR geometrical properties and simply forced the temperature and density to be constant.
Thus, having the PSR cross-section and height from this generic model and fixing the distance, we can compute the normalization factor with Eq.~\ref{NormCYCLOPS}.
This normalization factor is then used to make sure we have the same emitting region properties, i.e., volume and distance, in \cyclops~and \xspec, guarantying in turn that we can compare the fluxes.
Finally, the \xspec~spectra have been calculated using the standard procedure for the \bremss~model.

We show in Fig.~\ref{FigAppC_distance} the X-ray spectra from \cyclops~and \xspec, considering four different temperatures, namely $1$, $10$, $50$ and $100$~keV.
We can clearly see from the figure that the differences between the \bremss~model in \xspec~and \cyclops~are usually smaller than $10$~\%, irrespective of the energy and assumed temperature.
Given that \cyclops~and \xspec~are different codes, it is not surprising that their spectra are not identical.
However, they are very similar, being consistent with one another.
We can then conclude that the \cyclops~code provides X-ray spectra in excellent agreement with the widely used \xspec~code.
Moreover, the precision of the analytical expressions of the free-free Gaunt factor adopted in \xspec~and \cyclops~is around $10$~\% \citep[see, e.g.,][]{Mewe_1986}.
So the differences between the two codes are within the precision we have in the bremsstrahlung emissivities.

We would like to draw the readers attention to the fact that \cyclops~X-ray flux is independent of the assumed spatial resolution in the plane of the sky, which implies that the area of each and every cell in the 3D grid is correctly calculated.
This is because such a resolution defines the size of each cell in the grid and therefore its solid angle, which also depends on the distance.
In addition, to obtain the flux in each line-of-sight that represents the PSR, the emerging intensity from the cell closer to the observer, which results from the radiative transport along the entire line-of-sight, is multiplied by the cell area (in steradian).
Therefore, since the flux is not affected by variations in the spatial resolution, we can conclude that the angular areas of the cells are correctly calculated in \cyclops.

\begin{figure*}[t!]
\begin{center}
\includegraphics[width=0.98\linewidth]{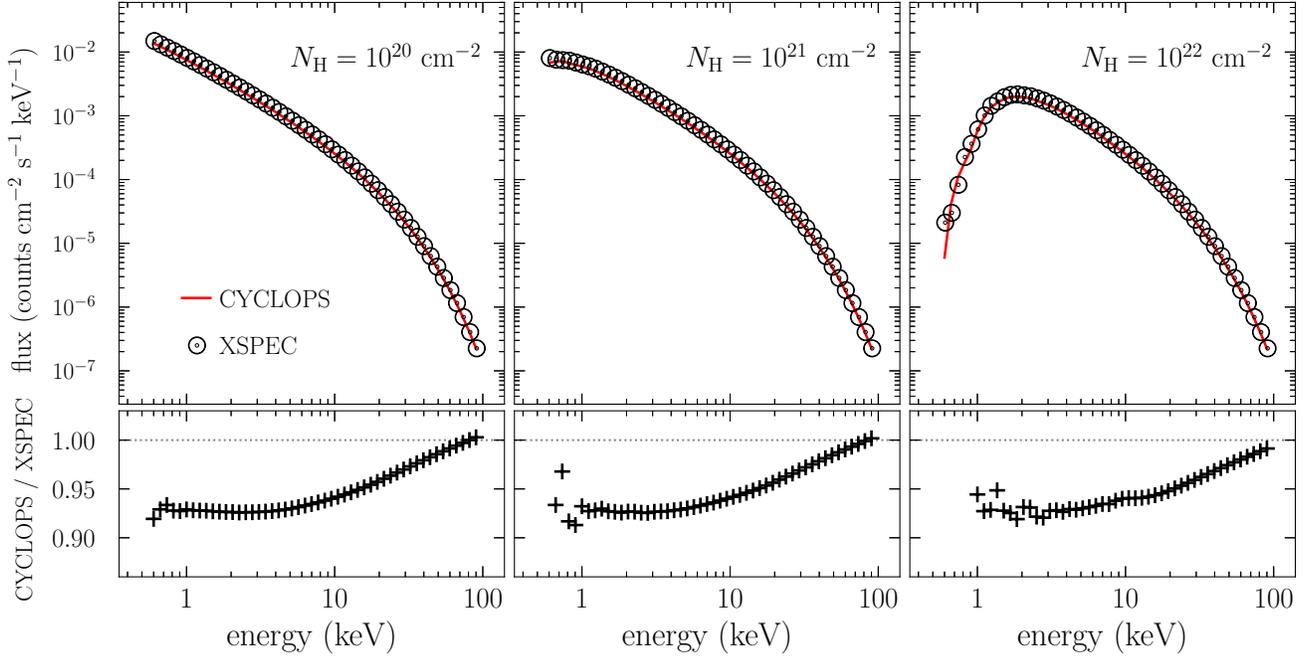}
\end{center}
\caption{Comparison between X-ray spectra from \cyclops~and \xspec~(combination of models \bremss~and \phabs), taking into account photoelectric absorption, for three different values of the hydrogen column density $N_{\rm H}$, namely $10^{20}$ (left panel), $10^{21}$ (middle panel), and $10^{22}$~cm$^{-2}$ (right panel).
In all models, the temperature is $20$~keV, the electronic density is $10^{16}$~cm$^{-3}$, and the distance is $100$~pc.
As expected, absorption is negligible for low values of $N_{\rm H}$, but becomes relevant for ${N_{\rm H}>10^{20}}$~cm$^{-2}$, at energies below a few keV.
Most importantly, the impact of absorption in \cyclops~and \xspec~is very similar (differences are smaller than $10$~\%), except for energies below $1$~keV.
We can then conclude that both codes are consistent with one another.
}
\label{FigAppC_extinction}
\end{figure*}

The last aspect of the X-ray modeling we compare with \xspec~is the interstellar extinction.
We show in Fig.~\ref{FigAppC_extinction} how the photoelectric absorption affects the spectra, by assuming a temperature of $20$~keV and three different values of hydrogen column density $N_{\rm H}$, namely $10^{20}$, $10^{21}$, and $10^{22}$~cm$^{-2}$.
We can clearly see that the impact of absorption in both codes is rather similar, irrespective of $N_{\rm H}$.
On the other hand, for energies below $1$~keV, models in the \cyclops~code are clearly much more absorbed in comparison with what the combination of \phabs~and \bremss~models provide.
In any event, like in the previous case in which we varied the temperature, differences between both codes are smaller than $10$~\%, except for the above-mentioned discrepancy at the softer part.
Therefore, as before, we can conclude that spectra in \cyclops~and \xspec~are in excellent agreement.